\documentclass[fleqn,10pt]{wlscirep}
\usepackage[utf8]{inputenc}
\usepackage[T1]{fontenc}
\usepackage{newunicodechar}

\usepackage{amsmath,amssymb}
\usepackage{graphicx}
\usepackage{booktabs}
\usepackage{multirow}
\usepackage{siunitx}
\usepackage{hyperref}
\usepackage{cleveref}
\usepackage{chato-notes}
\makeatletter
\renewcommand{\mynote}[2][]{\setkeys{chato}{#1}{\color{chato@clrpurple}\sffamily\footnotesize[\ifx\chato@author\@undefined Note\else\chato@author\fi: #2]}}
\makeatother

\usepackage[normalem]{ulem}
\definecolor{chkorange}{rgb}{0.85,0.45,0.0}

\newunicodechar{–}{--}
\newunicodechar{“}{``}
\newunicodechar{”}{''}
\newunicodechar{’}{'}
\newunicodechar{‐}{-}

\usepackage{caption}
\title{How coalition size shapes bill cosponsorship in the U.S. Congress}

\author[1,*]{Bernardo Pereira}
\author[2]{Giulia Preti}
\author[3]{Andrea Santoro}
\author[2]{Gianmarco De Francisci Morales}
\author[4]{Giovanni Petri}
\affil[1]{Dipartimento di Scienze Matematiche, Politecnico di Torino, Turin, 10129, Italy}
\affil[2]{Intesa Sanpaolo Innovation Center, Turin, 10138, Italy}
\affil[3]{ISI Foundation, Turin, 10126, Italy}
\affil[4]{Network Science Institute, Northeastern University London, London, UK}
\affil[*]{Corresponding author: \texttt{bernardo.pereira@polito.it}}

\begin{abstract}
Legislation is built by coalitions, yet studied through pairwise ties.
Using five decades of U.S. House and Senate cosponsorship (1973--2025), we represent each bill as a group.
Prior ties to current members predict who joins better than ties to the sponsor, and group history---earlier co-appearance with pairs of those members---produces a larger likelihood gain than sponsor ties alone.
Coalition formation therefore retains a social memory of earlier groups: cooperation is organized around relational histories embedded in the coalition already assembled, which sponsor--cosponsor ties alone cannot recover.
Completed coalitions also recur and share subgroups more often than degree- and size-preserving nulls predict.
Party sorting peaks in coalitions of a few dozen members; as Congress polarized, small coalitions retained their cross-party share while mid-size coalitions became more partisan.
Women are over-represented chiefly in coalitions led by Democratic women, although same-gender sponsor matching is close to zero after adjustment.
\end{abstract}

\begin{document}

\flushbottom
\maketitle
\thispagestyle{fancy} %

\section*{Introduction}

Over the past five decades, roll-call voting in the U.S. Congress has polarized sharply, yet legislators have kept signing on to each other's bills across party lines~\cite{harbridge_bipartisanship_2015}.
Among the observable traces of political cooperation, bill sponsorship and cosponsorship have proven to be particularly informative proxies for legislative relationships~\cite{victor_legislative_2016, kessler1996dynamics}, signaling positions to constituents and agenda setters~\cite{koger_position_2003, mayhew2004congress}.
Early work in political science established that the decision to cosponsor a bill is far from random: it reflects strategic calculations about constituency interests, ideological proximity, and reputational signaling~\cite{campbell_cosponsoring_1982, wilson_cosponsorship_1997}.
Building on this foundation, Fowler~\cite{fowler_legislative_2006, fowler_connecting_2006} brought bill cosponsorships into the domain of network science, demonstrating that robust measures of legislative influence (and political cooperation) can be extracted from cosponsorship data.
More broadly, whether legislators form broad, cross-partisan coalitions or narrow, ideologically homogeneous ones matters well beyond legislative studies: it speaks to questions ranging from political polarization~\cite{waugh_party_2011, neal_sign_2020} and normative concerns in democratic theory about how majorities form and minorities are included~\cite{guinier1995tyranny} to the science of collective action~\cite{olson1965logic}, where group size and composition shape both the stability and outcomes of cooperative arrangements~\cite{battiston_higher-order_2025}.
A cosponsorship, however, is less a tie between two legislators than a decision to join a group.
That group can range from three members to more than four hundred, has its own partisan and gender mix, and often brings together legislators who have worked together before.
\begin{figure}[!ht]
\centering
\includegraphics[width=0.98\linewidth]{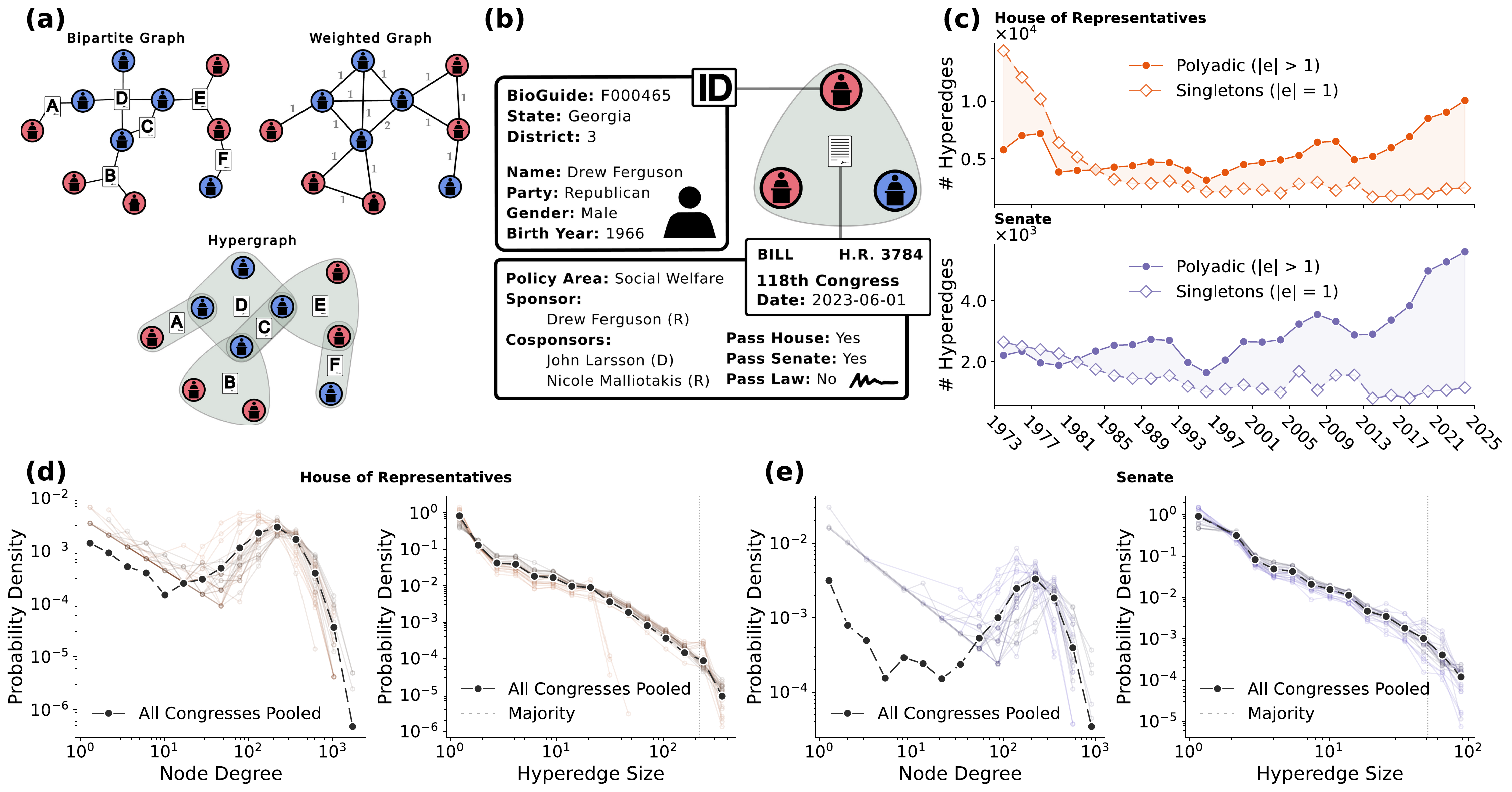}
\caption{\textbf{Hypergraph representation of bill cosponsorship in the U.S. Congress and its structural properties.}
\textbf{(a)} Comparison between network representations. \textit{Top left}: bipartite legislator--bill network. \textit{Top right}: projected pairwise network where higher-order information is reduced to weighted edges. \textit{Bottom}: hypergraph where each bill links a group of legislators through a single hyperedge. Retaining bill membership distinguishes group events that share the same projected pairwise ties.
\textbf{(b)} Schematic illustration of the data structure, where each legislator is represented as a node with available metadata (age, gender, and party affiliation), and each bill defines an annotated hyperedge of all its cosponsors. Unlike pairwise network projections, this representation preserves the higher-order group interactions underlying legislative collaboration.
\textbf{(c)} Evolution of cosponsored bills over time in the House (\textit{top}) and in the Senate (\textit{bottom}). Total number of polyadic hyperedges (bills with at least one cosponsor beyond the original sponsor) and singleton hyperedges (bills supported only by their sponsor) across Congresses.
\textbf{(d)-(e)} Empirical distributions of node degrees (number of sponsored/cosponsored bills per legislator, left subplots) and hyperedge sizes (number of legislators per bill, including the primary sponsor, right subplots) for the House of Representatives (d) and Senate (e). Background colored curves represent the distributions of each congress term, darker shades corresponding to more recent terms. Vertical dashed lines indicate chamber majorities.}
\label{fig:ho_description}
\end{figure}

The size of a coalition affects what joining it means.
Simmel observed that adding a third member transforms a dyad~\cite{simmel_number_1902}, and small-group research finds the sharpest change between dyads and triads~\cite{moreland_are_2010}.
For instance, two legislators co-signing a bill with three total sponsors may imply tighter, deliberate coordination, whereas the same two legislators appearing together on a bill with thirty sponsors may often reflect a broad, symbolic endorsement.
Weighted projections (\Cref{fig:ho_description}\textbf{a} \textit{top right}) of cosponsorship~\cite{fowler_connecting_2006, bratton_networks_2011, neal_sign_2020} collapse this difference, and networks aggregated over a session can obscure variation at finer levels of measurement~\cite{kirkland_measurement_2014}.
Well-established regularities have therefore been documented without asking whether they hold for small and large coalitions alike, or how a coalition comes to have the members it has. These include party agenda control~\cite{Cox_McCubbins_2005}, persistent bipartisanship~\cite{harbridge_bipartisanship_2015}, women's coalition building~\cite{jackie_jill_2011, when_women_2013, lawless_nice_2018} and the electoral calendar~\cite{mayhew2004congress}.

Following the need for grounded definitions of group interactions~\cite{st-onge_defining_2026}, we anchor our higher-order representation in the legislative act of cosponsorship itself, a natural but ephemeral grouping that may reflect diverse social, ideological, or strategic motivations. 
We treat each bill as a group, and model House and Senate cosponsorship across 26 Congresses (1973--2025) as annotated hypergraphs (\Cref{fig:ho_description}\textbf{a} \textit{bottom}).
Whereas previous foundational releases extend only through the 108th Congress (until 2005)~\cite{benson_simplicial_2018, landry_xgi_2023}, our dataset covers up to the 118th Congress (until 2025) and incorporates extensive legislator- and bill-level metadata, including demographics, policy areas, and legislative outcomes (\Cref{fig:ho_description}\textbf{b}).

Bipartite legislator--bill networks (\Cref{fig:ho_description}\textbf{a} \textit{top left}) preserve bills as distinct entities~\cite{breiger_duality_1974}, but make overlaps among their cosponsor sets less transparent; a hypergraph keeps coalition size and overlap explicit~\cite{battiston_higher-order_2025}.
We ask three questions.
Who coalesces with whom, and does it depend on coalition size?
Are coalitions assembled from groups that have worked together before, and at what point in a Congress?
And when a bill is open for support, who joins next: legislators tied to its sponsor, or legislators tied to the members already on it~\cite{backstrom_group_2006, guimera_team_2005}?

We find that party shapes coalitions of every size, most strongly those of a few dozen members, and that small coalitions kept their cross-party share as Congress polarized while mid-size coalitions became markedly more partisan.
Women are over-represented chiefly in coalitions led by Democratic women, yet when individual joining decisions are modeled, candidates show close to no preference for sponsors of their own gender.
Completed coalitions recur, and share overlapping subgroups, more often than expected under a null model that preserves legislators' activity and coalition sizes; large coalitions, moreover, are introduced early in a Congress, while bills introduced late gain cosponsors faster but end with smaller coalitions.
Finally, whether a legislator joins a forming coalition is predicted more strongly by prior ties to its current members than by ties to its sponsor, and is further predicted by group collaborations with those members.
Together, these results suggest that legislative support is better understood as a dynamical process of group formation than as an aggregate of individual or pairwise endorsements.

\section*{Results}

\subsection*{Cosponsorship coalitions span two orders of magnitude}

Our dataset covers 26 Congresses of the U.S. House of Representatives, hereafter the House, and the Senate, spanning the 93rd through the 118th Congress (1973--2025). A coalition comprises a bill's primary sponsor and all its recorded cosponsors; coalition size refers to the completed coalition, except in the recruitment analysis, which uses the coalition present before each joining event.

\Cref{fig:ho_description}\textbf{c} summarizes the number of bills (hyperedges) recorded in each chamber across the full period covered by our dataset.
The House hypergraphs are considerably larger, both in nodes and hyperedges, reflecting the greater number of seats and the higher volume of legislative activity in the lower chamber.
Variation in the number of bills introduced across Congresses is closely linked to the broader socio-political context of each legislative period.
After the 95th Congress (1977--79), the number of bills introduced in the House declined steeply following the lifting of the 25-cosponsor cap per bill, which ended the introduction of many duplicate bills that served as political endorsements of previously introduced ones \cite{congressgov2026sponsorship}.
The lowest number of cosponsored bills falls in the 104th Congress (1995--97), after the Republican Party won control of the House, and a local maximum of bill-introduction rates is observed around the 2008 financial crisis (Supplementary Section~\ref{si:descriptive}).
Over this period cosponsorship became the norm. The pattern of evolution for singleton bills versus polyadic ones is analogous in both chambers. After an initial period where singleton bills dominated legislative activity, there was an inflection, during 1981--83 in the Senate and 1983--85 in the House, and the preponderance of polyadic bills has risen ever since, consistent with the slow adoption of cosponsorship as a routine signal of support~\cite{campbell_cosponsoring_1982, koger_position_2003}.

\Cref{fig:ho_description}\textbf{d}-\textbf{e} (\textit{left}) depicts the distribution of node degrees, i.e., the number of bills in which each legislator participates across all Congresses.
The distribution is asymmetric, with most legislators concentrated around a single peak of a few hundred bills, consistent with and extending the findings of Fowler~\cite{fowler_legislative_2006}, who reported analogous patterns in earlier data.
It differs from the right-skewed, heavy-tailed degree distributions typical of scientific co-authorship networks, where transient, single-publication authors dominate~\cite{fowler_legislative_2006, newman_scientific_2001a, newman_scientific_2001b}; legislators, by contrast, typically serve several terms (Supplementary Section~\ref{si:descriptive}).

The distribution of hyperedge sizes, measured as the number of legislators sponsoring each bill, is far from uniform and exhibits clear patterns rooted in institutional constraints (\Cref{fig:ho_description}\textbf{d}-\textbf{e} (\textit{right})).
As expected, smaller bills tend to be more common than larger bills, as larger coalitions become increasingly harder to sustain due to inherent political constraints. Coalition sizes span more than two orders of magnitude, from a single sponsor to more than four hundred legislators in the House.
The size distributions decay smoothly across the majority threshold in both chambers; the slight excess just above it in some Congresses likely reflects institutional practices, such as those around commemorative legislation, rather than a general incentive to reach majority support.
The number of cosponsors has been found to matter mainly for whether a bill receives committee consideration~\cite{wilson_cosponsorship_1997}.

Because the same two legislators can thus meet in a coalition of three or of three hundred, we treat coalition size as an explicit variable in all that follows.

\subsection*{Small coalitions are the most bipartisan}

\begin{figure}[ht!]
\centering
\includegraphics[width=0.95\linewidth]{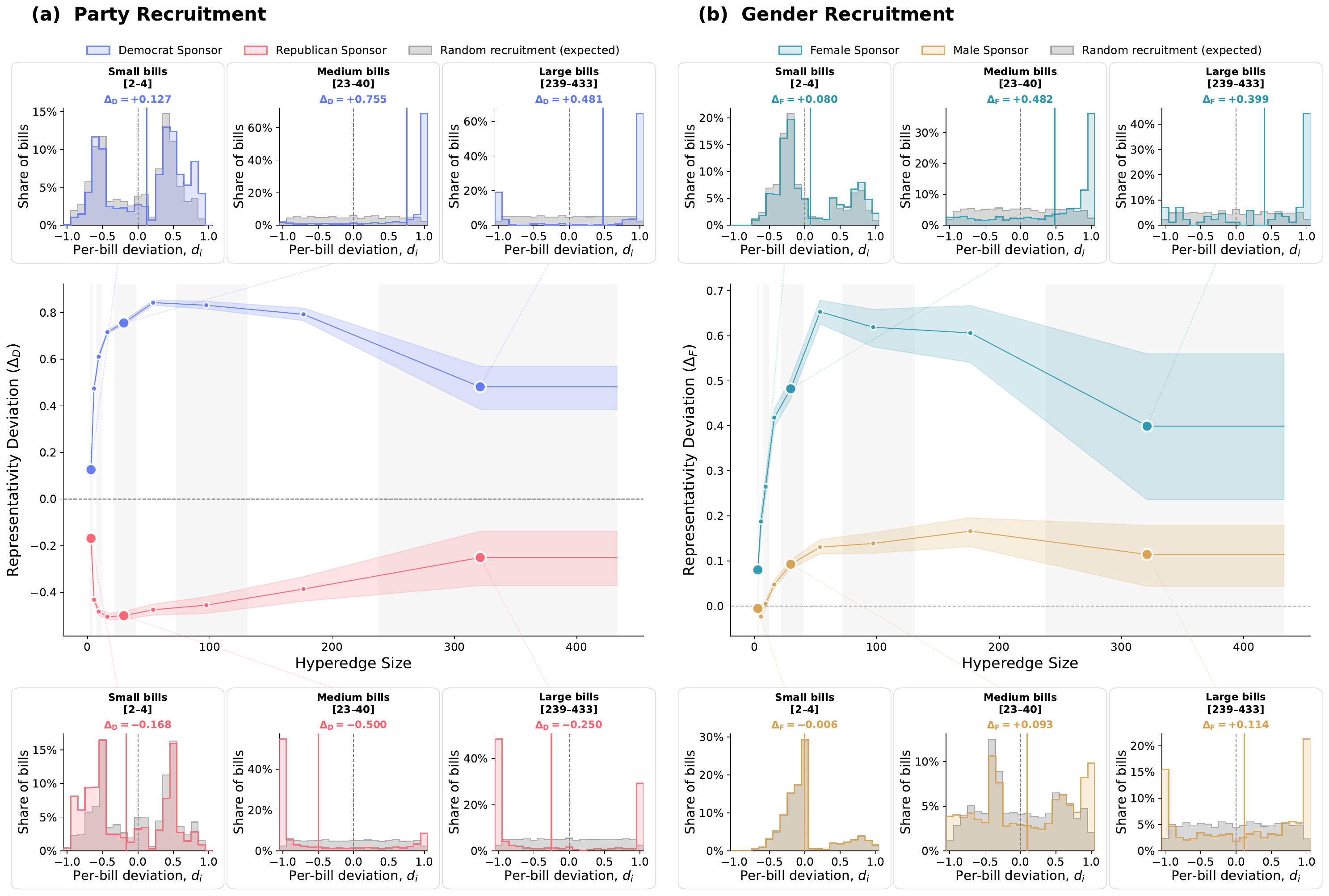}%
\caption{\textbf{House: hyperedge size modulates partisan and gender composition.}
Representativity deviation $\Delta$ as a function of hyperedge size (\textit{center}), and the distributions of the per-bill deviations $d_i$ behind it (\textit{top} and \textit{bottom}).
Hyperedge sizes, sponsor included, were divided into 9 log-spaced bins; bands are $95\%$ bootstrap confidence intervals (1000 resamples of the bills in each bin).
\textbf{(a)} Representativity deviation of Democrats ($\Delta_D$) across bill sizes, in Democratic- and Republican-sponsored bills (\textit{center}). On \textit{top} (Democratic sponsor) and \textit{bottom} (Republican sponsor), for the smallest, middle and largest bins: the empirical distribution of the bills' $d_i$ (colored) and the distribution expected under random recruitment for the same bills (grey); solid and dashed vertical lines mark their means, $\Delta_D$ and $0$.
\textbf{(b)} Representativity deviation of women ($\Delta_F$) across bill sizes, in female- and male-sponsored bills, with the same panels as in (a).}
\label{fig:hyperedge_size_homophily_effects}
\end{figure}

Because coalition sizes span more than two orders of magnitude (\Cref{fig:ho_description}\textbf{d}-\textbf{e}), the same co-appearance can mean different things across this range; we therefore first ask whether bill composition is uniform across sizes or whether size modulates its magnitude.

To evaluate compositional patterns across scales, we analyze the representativity deviation score ($\Delta$; positive values indicate over-representation relative to random draws from the chamber, and negative values the opposite, see Methods) for partisan and gender attributes across bill hyperedge sizes spanning the 93rd to 118th Congresses. 
Because the null preserves chamber composition, $\Delta$ measures homophily in excess of what that composition alone would produce (``inbreeding'' homophily in the terminology of McPherson et al.~\cite{mcpherson_birds_2001}), but it does not separate preference from structural sources such as shared party, committees or delegations~\cite{kossinets_origins_2009}, a distinction we return to when modeling individual joins.

Partisan affiliation emerges as the dominant structuring force of bill composition, with legislators heavily overrepresented in co-partisan bills and underrepresented in opposing-party bills (\Cref{fig:hyperedge_size_homophily_effects}\textbf{a}). 
In smaller bills, where coalition formation may center on narrow legislative objectives, cross-party collaboration is most feasible, and party composition remains closest to random expectation ($\Delta_D=+0.13$ and $-0.17$ for coalitions of 2--4 legislators under Democratic and Republican sponsors). 
As bills grow, cross-party participation declines and coalitions become increasingly pure, with the strongest sorting among coalitions of a few dozen members ($\Delta_D$ up to $+0.84$ and down to $-0.50$). 
Towards the largest hyperedge sizes, however, bills become more mixed again (239--433 members; $+0.48$ and $-0.25$). On a share scale they are close to random mixing, partly because coalitions this large must include the other party; $\Delta$, which saturates at these sizes, ranks them as more partisan than the smallest (Supplementary Section~\ref{si:cross_party_by_congress}). Across Congresses, small coalitions retained their cross-party share while mid-size coalitions lost much of theirs (Supplementary Figs.~\ref{fig:suppmat_cross_party_house}--\ref{fig:suppmat_cross_party_senate}), and the return of mixing at the largest sizes is consistent with accounts in which broad bipartisan support is typically needed for enactment~\cite{curry_limits_2020}. Bills in commemorative and symbolic policy areas have the largest typical coalitions (Supplementary Fig.~\ref{fig:suppmat_policy_area}\textbf{a,c}), which may contribute to the renewed mixing, although health, international affairs and taxation remain the most common areas among coalitions of 20 or more members (Supplementary Section~\ref{si:analysis}). Party is thus the main organizer of cosponsorship, as party-centered accounts of Congress lead one to expect~\cite{Cox_McCubbins_2005}, but how strongly it organizes a coalition depends on the coalition's size.

Gender composition follows a different pattern: women are over-represented ($\Delta > 0$) at nearly every size, regardless of the primary sponsor's gender, but much more under female sponsors (\Cref{fig:hyperedge_size_homophily_effects}\textbf{b}).
In female-sponsored bills, women's over-representation follows a clear scale dependence: small bills sit closest to random mixing ($\Delta_F=+0.08$), coalitions of 40--72 members show the peak ($+0.65$), and the largest bills see this over-representation diminish slightly ($+0.40$).
Male-sponsored bills show a modest initial rise in female participation from small to medium scales ($\Delta_F$ from $-0.01$ to between $+0.09$ and $+0.17$) before plateauing across the largest hyperedge sizes.

Party affiliation modifies this pattern.
The over-representation of women is driven predominantly by coalitions led by Democratic women, which show the strongest overall excess of both women and Democrats. In contrast, Republican female-sponsored bills display female underrepresentation ($\Delta < 0$) in small and medium sizes, rising only to random-mixing levels in large bills (Supplementary Fig.~\ref{fig:suppmat_homophily_cross_labels_house}\textbf{b}).

Female participation in male-sponsored bills similarly reflects this partisan gap, as coalitions led by Republican men include women at rates substantially lower than those led by Democratic men across all scales. This agrees with evidence that gender differences in cosponsorship depend on party~\cite{gagliarducci_gender_nodate, bagues_gender_2026}, and leaves open whether women's over-representation reflects a preference for same-gender partners or the party composition of the coalitions they join, a question we return to when modeling individual joins. 
While these findings uncover robust structural patterns across 26 congressional terms, pooling data across the full historical span may smooth over term-specific political events and shifting institutional constraints.

\subsection*{Coalitions are built from recurring groups}
\label{sec:simpliciality_multiplicity}

\begin{figure}[!ht]
\centering
\includegraphics[width=0.99\linewidth]{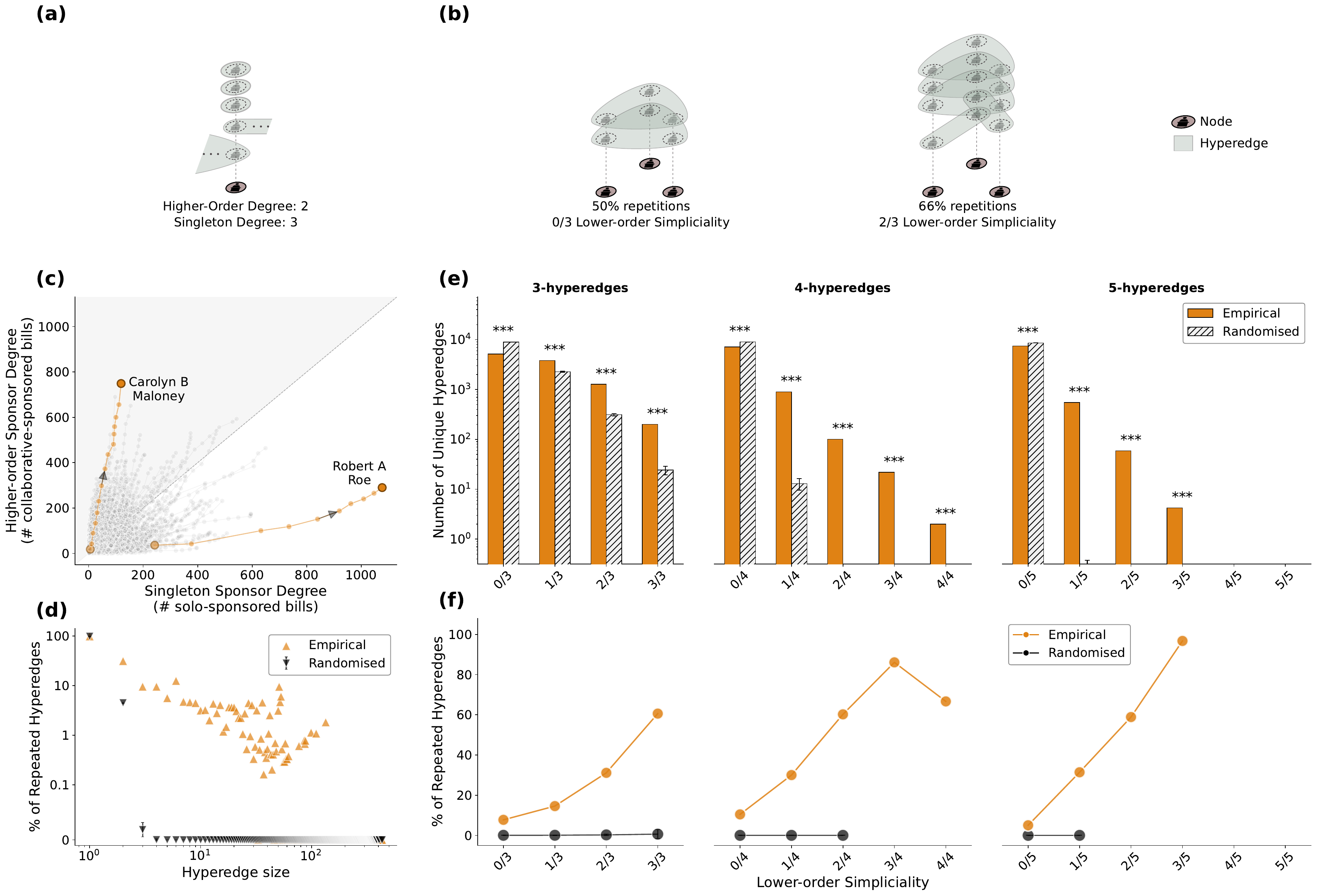}
\caption{\textbf{House: structural properties of cosponsorship hypergraphs reveal pronounced multiplicity and simpliciality.} All panels report on the aggregated hypergraph of House bills from the 93rd to 118th Congresses. \textbf{(a)} Schematic of a legislator's singleton sponsor degree (size-1 bills) and higher-order sponsor degree (bills with $>1$ sponsors). \textbf{(b)} Schematics of hyperedge multiplicity and lower-order simpliciality. \textbf{(c)} Singleton and higher-order sponsor degrees accumulated over the Congresses in which each legislator participates; Carolyn B.\ Maloney and Robert A.\ Roe are highlighted. Most members ($70.1\%$) predominantly engage in multi-sponsor coalitions in their last Congress recorded. \textbf{(d)} Share of repeated hyperedges by hyperedge size in the empirical and randomized hypergraphs. Small coalitions are frequently repeated, but this multiplicity decays rapidly as hyperedge size increases; randomization effectively eliminates repetition for all sizes above three. \textbf{(e)} Counts of unique hyperedges of size 3, 4, and 5 (left to right) by lower-order simpliciality. Compared with a randomized configuration model that preserves degree and hyperedge size distributions (hashed bars; error bars indicate one standard deviation), the empirical data exhibit an excess of highly simplicial coalitions. Significance asterisks: the observed counts always lie outside all corresponding 100 randomized hypergraphs ($***:  p \leq 0.01$). \textbf{(f)} Share of repeated hyperedges by lower-order simpliciality in the empirical hypergraph (orange) and randomized hypergraphs (black; error bars, almost indiscernible, indicate one standard deviation). Hyperedges containing denser lower-order subsets are substantially more likely to be repeated.}

\label{fig:ho_structure}
\end{figure}

Composition, however, does not reveal whether the same legislators recur together, since party sorting alone could produce the patterns of \Cref{fig:hyperedge_size_homophily_effects}. 
We therefore ask whether coalitions are built from recurring individuals and subgroups. 
Studies of team assembly find that collaborators tend to repeat earlier collaborations~\cite{guimera_team_2005}, and that the subsets of a new team have often worked together already~\cite{juul_hypergraph_2024}; in Congress, caucuses and state delegations provide standing groups that could play this role, although caucuses tend to reinforce---rather than cut across---party lines~\cite{victor_social_2009}.
\Cref{fig:ho_structure}\textbf{c} compares legislators' singleton sponsor degree, i.e., the number of bills introduced without cosponsors, with their higher-order sponsor degree, i.e., the number of bills introduced with at least one cosponsor, throughout Congresses.
This distinction separates legislators who are predominantly coalition builders from those who act as ``lone wolves''.
These contrasting profiles are well exemplified by Representative Carolyn B. Maloney (D) and Representative Robert A. Roe (D) (Supplementary Section~\ref{si:builders}).
Part of this contrast reflects the era: Roe's tenure overlaps the period in which singleton bills dominated (\Cref{fig:ho_description}\textbf{c}) and members adapted slowly to cosponsorship~\cite{koger_position_2003}.
Both legislators display clear, consistent trends of evolution throughout the Congresses in which they served, remaining within the same characteristic region from term to term.
These examples therefore suggest that legislators' propensity to gather support for their bills may reflect stable individual legislative strategies or behavioral tendencies, rather than being determined solely by their formal institutional roles (Supplementary Section~\ref{si:builders}). 

We find that entire coalitions recur far more often than expected by chance. In \Cref{fig:ho_structure}\textbf{d} we report the share of hyperedges of each size that are multiples of another hyperedge (bills cosponsored by the same set of legislators).

Although the share of repeated hyperedges decreases with hyperedge size, it remains as high as $10\%$ for hyperedges of size $51$: one in ten bills sponsored by 51 legislators shares its complete sponsoring set with at least one other bill. In randomized hypergraphs, such repetition essentially vanishes above three members.

To investigate the interplay between group cohesiveness and lower-order structure, we relate \emph{lower-order simpliciality} and multiplicity for House hyperedges of size 3, 4, and 5, across the 93rd–118th Congresses.
Here, we employ the notion of \emph{lower-order simpliciality}~\cite{landry_simpliciality_2024}: the fraction of a coalition's subgroups with one member fewer that have also sponsored a bill of their own.
As an example, for a triplet of legislators who cosponsored a bill together, the corresponding hyperedge is \emph{maximally lower-order simplicial} if each pair of legislators has also cosponsored at least one bill together. In contrast, it has no lower-order simpliciality if none of the three pairs has cosponsored a bill together. Co-appearance on the three-person bill itself does not count: each pair must also appear as a separate two-person bill.
\Cref{fig:ho_structure}\textbf{e} shows, for each hyperedge size $k \in \{3, 4, 5\}$, the number of unique size-$k$ hyperedges across all possible levels of lower-order simpliciality.

These empirical counts are compared with the corresponding mean counts obtained from 100 randomized hypergraphs with the same degree and hyperedge size distributions.
In the merged empirical House hypergraph, we observe an over-expression of hyperedges with positive lower-order simpliciality and a slight under-expression of non-simplicial hyperedges across all three hyperedge sizes; three-member coalitions whose three pairs all exist as two-person bills are about an order of magnitude more common than expected.

When relating simpliciality with multiplicity (\Cref{fig:ho_structure}\textbf{f}), a clear and consistent pattern emerges in the House data: the higher the lower-order simpliciality, the higher the multiplicity (among three-member coalitions, from under $10\%$ to about $60\%$ repeated). The corresponding Senate analysis (Supplementary Fig.~\ref{fig:suppmat_simpliciality_multiplicity_senate}) shows the same qualitative association.
This pattern is consistent with a simple account: a group of legislators who repeatedly collaborate in smaller coalitions (which constitute the lower-order hyperedges that make the larger hyperedge simplicial) is more likely to cosponsor multiple bills as the full group, thereby generating repeated realizations of the same hyperedge.
In other words, legislators who frequently work together in subgroups also tend to work together repeatedly as a larger coalition, in line with the transitivity reported for cosponsorship networks~\cite{bratton_networks_2011} and the simplicial structure of other collaboration systems~\cite{benson_simplicial_2018, juul_hypergraph_2024}.
Simpliciality and multiplicity are therefore two observable faces of repeated collaboration among stable groups of legislators. Part of the exact repetition also reflects batch introductions by the same group, such as the 2023 episode discussed below.

\subsection*{Large coalitions are introduced early in a Congress}

The structures above are aggregated over entire careers, yet legislators' incentives change within each Congress~\cite{mayhew2004congress}; we therefore ask whether coalitions of different sizes are built at different points of the term.
The timing of interactions plays a central role in social systems, influencing how relationships form and how collaborative activities emerge, not just in face-to-face settings \cite{cencetti_temporal_2021} but also in networks of scientific coauthorship~\cite{newman_scientific_2001a, wuchty_increasing_2007} and digital communications~\cite{kossinets_empirical_2006}.
To study when coalitions of different sizes are formed, we use the introduction date of each bill.
By examining the daily rate of bill introduction across the full span of the dataset, from the first day of the 93rd Congress to the last day of the 118th Congress, and by superimposing the within-congress trajectories of all terms, we identify the recurring patterns that characterize the life cycle of a typical Congress.

\begin{figure}[ht]
\centering
\includegraphics[width=0.999\linewidth]{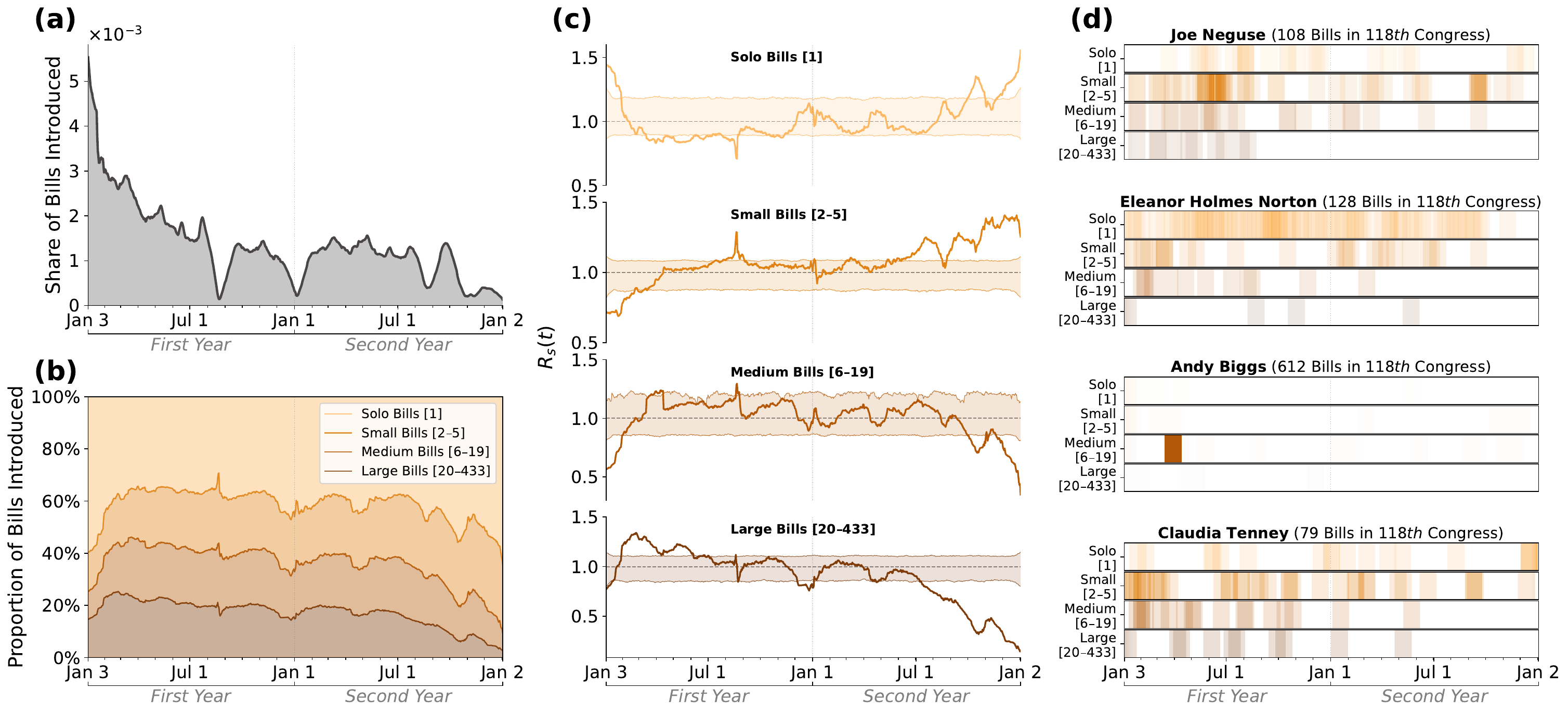}
\caption{\textbf{House: temporal evolution of legislative activity over a congressional term.} Since all Congresses follow an approximately identical calendar structure, we aggregate daily activity across all studied congressional terms and smooth the resulting time series using a 30-day rolling window. 
\textbf{(a)} Overall frequency of bill introductions by day. The solid line denotes the share of bills introduced on each day across all Congresses, showing that most bills are introduced early in the term. 
\textbf{(b)} Proportion of bills introduced each day by bill-size category. Bills of size $2+$ are divided into small (2--5), medium (6--19) and large (20+) categories, while bills of size $1$ form a separate \emph{solo} category.
\textbf{(c)} Relative daily prevalence of each bill-size category over a standardized congressional term, measured as the ratio between the daily proportion of bills in a given size category and that category's overall proportion across the full term. Values above (below) $1$ indicate that the category is overrepresented (underrepresented) on that day. Shaded areas are the $95\%$ confidence intervals obtained by randomly shuffling introduction dates among hyperedges $1000$ times and recomputing the relative shares.
\textbf{(d)} Relative daily activity of individual legislators by bill-size category, showcasing heterogeneous activity patterns across legislators. Highlighted legislators are the most prolific bill introducers in the $118$th Congress within Democratic men (\textit{top}), Democratic women (\textit{middle top}), Republican men (\textit{middle bottom}), and Republican women (\textit{bottom}).}
\label{fig:temporal_bill_introduction}
\end{figure}
When the daily bill-introduction trajectories of all Congresses are temporally aligned and normalized to a common Congress duration, a seasonal pattern emerges, with a peak at the beginning of the Congress, a minimum towards the end, and accentuated falls corresponding to recess periods in August and late December/ early January (\Cref{fig:temporal_bill_introduction}\textbf{a}).
Separating bills by size allows us to identify how different coalition-size categories are distributed across the legislative calendar.
Large and medium-size bills are distributed mostly in the first months of Congress, whereas smaller bills are more evenly distributed across the legislative calendar (\Cref{fig:temporal_bill_introduction}\textbf{b,c}).
Solo bills, by contrast, are introduced at higher rates at the very beginning and the very end of a Congress, the periods closest to elections. 
This pattern, however, is attenuated for small bills, which, although less common early in the terms and slightly more at the end, are more uniformly distributed throughout a Congress.
These patterns are present in both chambers, although they are particularly clear in the House (Senate results in Supplementary Fig.~\ref{fig:suppmat_temporality_senate}).

One interpretation is that legislators' incentives change over the course of a Congress.
At the start of a new Congress, legislators arrive with new mandates.
This early surge may be intensified by pressure to deliver on high-profile campaign promises, consistent with the traditional ``first 100 days'' window, during which both the President and the concurrent Congress face heightened media scrutiny and public expectation \cite{predicting_legislative_first_100, hyperpartisanship_first_100}.
Broad coalitions may also be easier to assemble in this period, before positions on the term's agenda have settled.
As the Congress approaches its end, however, legislative attention shifts: reelection campaigns become more salient, especially in the House, where every seat is contested every two years, and legislators have less reason to invest in the negotiation needed to assemble large coalitions~\cite{mayhew2004congress}.
Instead, introductions in this period tend to be smaller; this is consistent with the position-taking account of sponsorship and cosponsorship~\cite{mayhew2004congress, koger_position_2003}, in which bills signal positions to constituents and agenda setters rather than build broad coalitions, although we do not observe bill content or intent directly.
Two simpler explanations could produce the same pattern: 
(i) bills introduced at the start of a Congress include reintroductions of bills from the previous one, which may arrive with their earlier supporters;
or (ii) bills introduced late in the term have less time to gather cosponsors before the Congress ends, so their final coalitions may be smaller for that reason alone. 
The second explanation concerns how coalitions grow after introduction, which we examine in the next section.

The most active sponsors also differ in when, and with how many cosponsors, they introduce bills.
As representative examples, we examine Joe Neguse (most active Democratic man), Eleanor Holmes Norton (most active Democratic woman), Andy Biggs (most active Republican man), and Claudia Tenney (most active Republican woman) in the House during the 118th Congress.
While these lawmakers display varying overall introduction volumes, clear contrasts emerge in their preferred coalition sizes.
Eleanor Holmes Norton relies predominantly on solo introductions and small cosponsoring coalitions with continuous activity across the session.
In contrast, Joe Neguse and Claudia Tenney exhibit a broader distribution across small and medium bands, occasionally forming large coalitions.
Andy Biggs, by contrast, introduced almost all of his bills in the medium band and within a few days.

On 29 March 2023, during the debt-ceiling negotiations, he and five Freedom Caucus colleagues introduced more than 500 spending-cut bills~\cite{azcentral_biggs_2024}.
Legislators with similar levels of activity can thus follow very different calendars.

\subsection*{Coalition recruitment carries a social memory of earlier groups}

\begin{figure}[ht!]
\centering
\includegraphics[width=\linewidth]{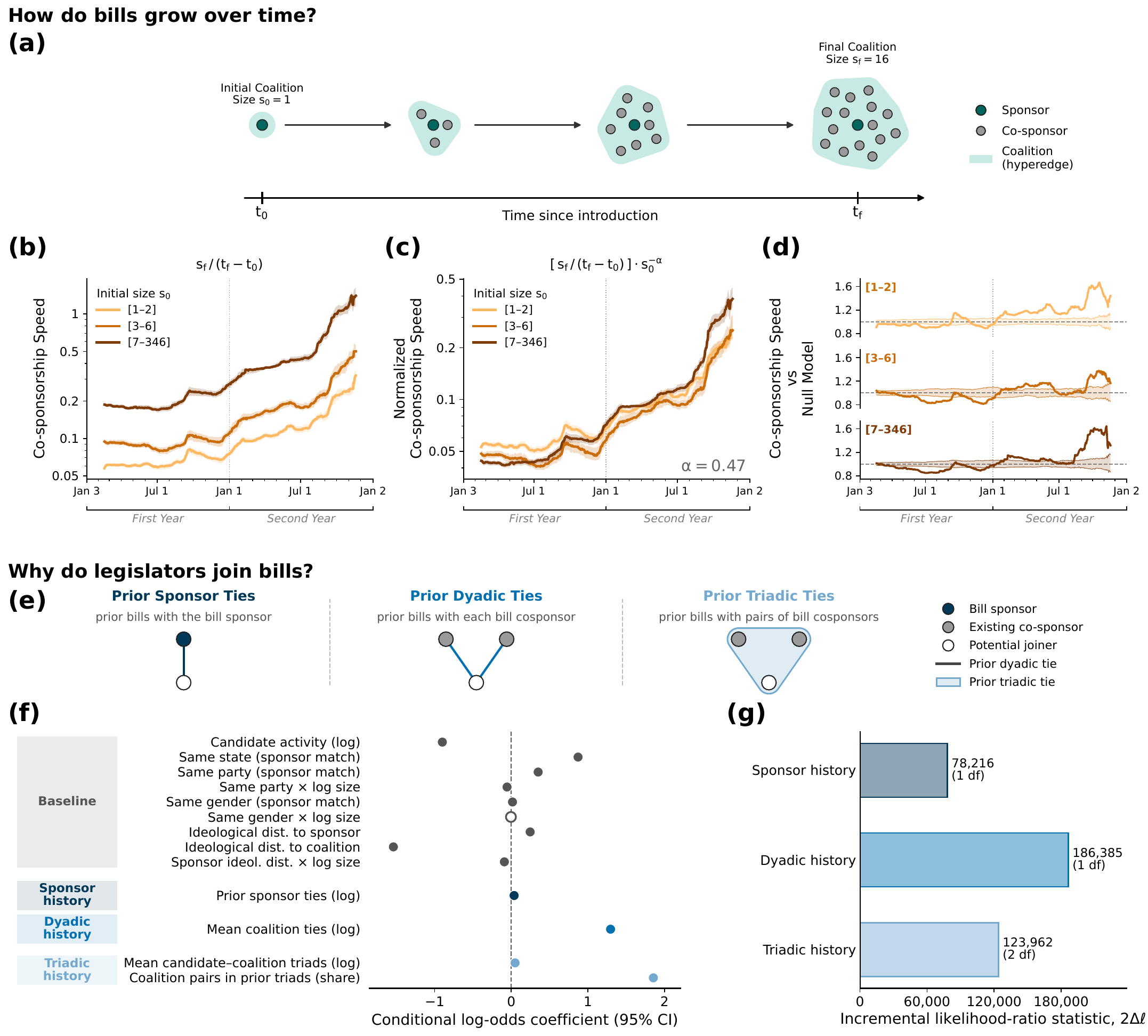}
\caption{\textbf{House: coalition assembly and the relational context of recruitment.} \textbf{(a)} A coalition grows from its initial size $s_0$ at introduction ($t_0$) to its final size $s_f$ ($t_f$). \textbf{(b)} Median cosponsorship speed $s_f/(t_f-t_0+1)$ of the bills introduced in rolling 90-day windows, by introduction date, stratified by initial size. \textbf{(c)} The same speed rescaled by $s_0^{-\alpha}$, with $\alpha=0.47$. \textbf{(d)} Empirical cosponsorship speed relative to a null model in which introduction days are shuffled among the bills of each initial-size class, keeping every bill's assembly time within the Congress. \textbf{(e)} The three relational histories a candidate can have with a forming coalition: prior bills with the sponsor, with individual cosponsors, and with pairs of cosponsors. \textbf{(f)} Conditional-logit coefficients and 95\% confidence intervals of the full model; coefficients are conditional on all other terms and are on different covariate scales. \textbf{(g)} Likelihood-ratio statistic $2\Delta\ell$ for adding each history block, on its own, to the baseline model; the sequential (nested) increments are reported in Supplementary Section~\ref{si:recruit_ladder}.}

\label{fig:recruitment}
\end{figure}

Our third question joins the two previous results: cosponsors are recruited through personal contacts and ``Dear Colleague'' letters~\cite{campbell_cosponsoring_1982, fowler_connecting_2006} and join over weeks and months~\cite{kessler1996dynamics}, so we ask, as studies of group formation and team assembly do~\cite{backstrom_group_2006, guimera_team_2005, juul_hypergraph_2024}, whether the subgroup structure of \Cref{fig:ho_structure} and the calendar of \Cref{fig:temporal_bill_introduction} reappear in who joins next and how fast.
Indeed, legislative coalitions are group-valued structures assembled sequentially through individual decisions to cosponsor bills across the lifespan of a congressional term. 
Rather than emerging fully formed at introduction, non-singleton coalitions undergo a dynamic accretion process driven both by macro-level calendar position and micro-level relational history.

At the macroscopic level, coalitions grow after introduction at rates that depend on both their size and their introduction date (\Cref{fig:recruitment}\textbf{a}). As a Congress progresses, bills introduced later in the term face an abbreviated operational window before adjournment, inducing a phenomenon of temporal compression.

When tracking the median rate of cosponsor additions against the introduction date (\Cref{fig:recruitment}\textbf{b}), bills grouped into distinct initial-size classes exhibit structured variation in recruitment velocity. This heterogeneity is substantially reduced when accounting for coalition scale. By normalizing the cosponsorship accretion speed by a power of the bill's initial size ($s_0^{\alpha}$, with $\alpha=0.47$; \Cref{fig:recruitment}\textbf{c}), the speed curves across size strata become closely aligned, though not identical. 

For all sizes, normalized accretion speed rises toward the end of the Congress. Compared with a null model, this speed-up is not visible over periods of days or weeks, but it emerges when growth is measured over two to three months (\Cref{fig:recruitment}\textbf{d}; Supplementary Section~\ref{si:assembly}): over these periods, bills introduced late in a Congress gain cosponsors faster than the shorter time remaining alone would produce. What limits late coalitions is therefore the end of the Congress rather than a lower willingness to join, which qualifies the incentive account: as the term ends, legislators do not become less ready to support colleagues' bills.

Beneath these aggregate accretion curves lies a discrete sequence of prospective choice events. At each recruitment step $t_i \to t_{i+1}$, a candidate legislator encounters a bill already supported by a forming coalition. To isolate the relational features guiding who joins next, each event compares the observed joiner against a risk set of members of the chamber in that Congress who have not yet joined.
Comparing conditional choice models that add each relational block to a common baseline reveals the relative importance of individual attributes, pairwise ties, and group context.

\begin{itemize}
\item Baseline Controls \& Dyadic Ties: Individual activity, geographic matching, party alignment, and ideological proximity to the sponsor establish a baseline choice probability. The small positive coefficient on ideological distance (measured as the absolute difference between DW-NOMINATE first dimension scores) to the sponsor, which weakens as coalitions grow, is conditional on distance to the coalition (\Cref{fig:recruitment}\textbf{f}) and should not be read as attraction to ideologically distant sponsors.

We find that ties to the coalition matter more than ties to the sponsor. Added on its own to the baseline, mean dyadic ties to the current coalition improve the fit more than prior ties to the primary sponsor (\Cref{fig:recruitment}\textbf{g}), and once both are included, the sponsor-tie coefficient is small (\Cref{fig:recruitment}\textbf{f}); since the current coalition includes the sponsor, part of the sponsor tie is already contained in the coalition ties. Recruitment is thus organized around the coalition already assembled rather than around the sponsor alone.

\item Higher-Order Relational History: Pairwise ties do not exhaust the relational context. On its own, triadic history (candidate co-appearance history with pairs of current coalition members, $x_{\text{triad}}$) improves the baseline by less than coalition ties and by more than sponsor ties (\Cref{fig:recruitment}\textbf{g}); added after both pairwise blocks, it still improves the fit (Supplementary Section~\ref{si:recruit_ladder}). In the full model, the effect is carried by the share of current coalition pairs with whom the candidate has co-appeared before, rather than by the number of such co-appearances, although the two terms are correlated: having worked with members of the coalition together, and not only with each of them separately, predicts joining. In this sense, the formation of a new coalition retains a social memory of earlier groups, a legislative counterpart of the higher-order memory found in other temporal group interactions~\cite{cencetti_temporal_2021, gallo_higher-order_2024}. Because the blocks are correlated, these single-block gains overlap and should not be summed.

\end{itemize}

Neither same-gender matching with the sponsor nor its interaction with coalition size has a detectable effect at the moment of joining (\Cref{fig:recruitment}\textbf{f}), so the over-representation of women in completed coalitions (\Cref{fig:hyperedge_size_homophily_effects}\textbf{b}) is more consistent with induced than with choice homophily~\cite{mcpherson_birds_2001, kossinets_origins_2009}: conditional on party, ideology and prior ties, the same-gender coefficient is close to zero. Because prior ties may themselves reflect earlier same-gender choices, this does not rule out choice homophily, nor matching to the gender composition of the coalition itself. Party matching with the sponsor weakens as coalitions grow (\Cref{fig:recruitment}\textbf{f}), although large completed coalitions are more partisan (\Cref{fig:hyperedge_size_homophily_effects}\textbf{a}); in large coalitions, party sorting may therefore depend more on proximity to the coalition than on the sponsor's party. The models include interactions with coalition size but not with the date of introduction, so whether these relationships differ between bills introduced early and late in a Congress remains to be tested.

Together, these results suggest that the recurring groups of \Cref{fig:ho_structure} matter for recruitment through candidates' earlier collaboration with parts of the current coalition, 
a higher-order analogue of triadic closure~\cite{kossinets_empirical_2006, bratton_networks_2011, carpenter2004friends, benson_simplicial_2018} and of the finding that joining a group depends on how one's contacts within it are connected~\cite{backstrom_group_2006}. Because these models are predictive and observational, this relational history may also reflect shared committees, caucuses, or state delegations that we do not observe~\cite{kossinets_origins_2009}.

\section*{Discussion}
Roll-call voting in Congress polarized over the five decades we study, yet cross-party cosponsorship persisted~\cite{harbridge_bipartisanship_2015}.
Analyzing each bill as a group of legislators, we find that this cooperation is greatest in the smallest coalitions, that coalitions are repeatedly built from the same subgroups, and that who joins a forming coalition depends more on the members already on it than on its sponsor.
Coalition size changes the strength, and in some cases the direction, of patterns that pairwise network measures average across all bills.

Who coalesces with whom depends on coalition size. Party organizes coalitions of every size, as party-centered accounts of Congress lead one to expect~\cite{Cox_McCubbins_2005}, but its hold is weakest in the smallest coalitions, strongest among coalitions of a few dozen members, and weaker again among the largest, consistent with the broad support that enactment typically requires~\cite{curry_limits_2020} and with the large coalitions typical of commemorative and symbolic bills. The bipartisanship that survived polarization therefore appears mainly where coalitions are small and least visible, while the most sorted coalitions are of a size at which bills may serve as statements of party position~\cite{lee_insecure_2016, gelman_losing_2020}. Whether small cross-party coalitions reflect shared constituencies, such as the state delegations for which the recruitment model shows a strong same-state effect, or collaboration on policy across party lines, remains open.

The gender results point to a process that operates through groups. Women are over-represented chiefly in coalitions led by Democratic women, yet candidates show no preference for sponsors of their own gender when they join, which suggests that the gender composition of coalitions arises through party and through the coalitions legislators join more than through the choice of individual partners~\cite{mcpherson_birds_2001, kossinets_origins_2009}. Whether women are more likely to join coalitions in which other women are already present, as accounts of tokenism and critical mass in group composition would suggest~\cite{kanter_effects_1977}, remains open.

Coalitions are also built from one another. Entire coalitions recur, and those whose subgroups have sponsored bills of their own recur most often, consistent with standing working groups such as caucuses, state delegations and committees~\cite{victor_social_2009}. Which of these organizations supply the recurring groups remains to be established, as does whether coalitions built from them fare better than coalitions that reach beyond them: weak, bridging ties between legislators have been linked to legislative success~\cite{kirkland_relational_2011}, whereas recruitment in our data draws on strong, repeated ones.

The calendar adds a second constraint. Large coalitions appear early in a Congress, and bills introduced late gain cosponsors faster than the time remaining would predict, yet end with smaller coalitions: the end of the term limits how large coalitions become without reducing legislators' readiness to join them. This qualifies accounts in which approaching elections shift effort from coalition building to position taking~\cite{mayhew2004congress, koger_position_2003}. The shift may concern which bills legislators introduce, as the late rise in solo bills suggests, and late joining may itself be a form of position taking, since cosponsoring a bill that cannot advance before adjournment costs little. Because Senate terms are staggered, comparing senators who face reelection with those who do not would test the electoral reading directly.

Recruitment, finally, depends on the coalition already assembled. Whether a legislator joins is predicted more by prior ties to the members already on a bill than by ties to its sponsor, and further by prior collaboration with pairs of those members, so that coalition formation retains a social memory of earlier groups and not only of earlier pairs. This bears on how cosponsorship is explained. If cosponsorship were mainly an exchange of support between pairs of legislators, ties to the sponsor would be expected to dominate; instead, the relevant reference appears to be the set of members already on the bill. Members already on a bill may serve as cues for later joiners, as colleagues do in accounts of roll-call voting~\cite{kingdon_congressmens_1989}, and sponsors' efforts to attract support~\cite{campbell_cosponsoring_1982} may work through them. Lobbyists and staff, whose communication is itself often brokered by third parties~\cite{carpenter2004friends}, are another possible channel; linking cosponsorship to records of lobbying contacts would help distinguish these explanations.

These results also bear on measurement. Network measures of connectedness, influence and partisan structure~\cite{fowler_legislative_2006, fowler_connecting_2006, waugh_party_2011, neal_homophily_2022, sotoudeh_network-based_2024, lo_statistical_2025, zhang_community_2008} count a co-appearance on a three-member bill like one on a bill with three hundred cosponsors, and so combine close coordination with broad endorsement. Because cross-party cooperation and party sorting sit in coalitions of different sizes, polarization measured from cosponsorship networks that combine coalitions of all sizes~\cite{neal_sign_2020, zhang_community_2008} may track mid-size coalitions more than small ones; the size gradient itself steepened as Congress polarized, because mid-size coalitions became more partisan while small ones did not. For questions in which group size carries information, a representation that keeps size explicit (a hypergraph or an equivalent two-mode encoding) is needed; pairwise projections remain adequate for questions that do not condition on size.

A central contribution of this work is methodological as much as empirical: by extending the most widely used higher-order cosponsorship dataset from the 108th to the 118th Congress and enriching it with demographic, policy, and outcome metadata, we provide a resource that should support a substantially broader range of analyses than was previously possible.

Several limitations and directions for future work follow naturally from our approach.
First, we treat all participants in a hyperedge symmetrically, aside from annotating the primary sponsor, hence simplifying the procedurally distinct roles of sponsors and cosponsors; richer directed or weighted hypergraph formalisms could capture this distinction more faithfully.
Second, our hypergeometric null models condition on chamber-level demographic composition but do not account for other potential confounders, such as committee assignments, seniority, or geographic delegation, which may themselves correlate with both coalition size and the demographic attributes we study; the recruitment models are predictive and observational, and the relational histories they use may partly reflect these shared settings.
Third, while we have focused on the U.S. Congress due to its data availability and outsized influence on the comparative literature, the framework we propose, treating coalition size as an explicit dimension of analysis, is directly applicable to cosponsorship data from other legislatures~\cite{briatte_network_2016, lee_time-dependent_2016}, where rules on cosponsorship and the strength of party control differ; such comparisons would show how much of what we report reflects the open cosponsorship rules of the U.S. Congress.
Fourth, the timing results classify bills by their completed coalition size, which depends on the time left in a Congress, and the recruitment models let relational effects vary with coalition size but not with the date of introduction, so whether recruitment works differently early and late in a term remains open.
Finally, while we provide a preliminary analysis of chamber passage probabilities across hyperedge sizes and sponsor attributes (Supplementary Fig.~\ref{fig:suppmat_legislative_success} and Section~\ref{si:suppmat_legislative_success}), we do not exhaustively map legislative outcomes against the full higher-order structural and temporal dynamics of cosponsorship networks. 
Detailed evaluations of downstream success, including passage in the secondary chamber and enactment into law, are left for future work using this resource.

Legislatures record the membership, the order of joining and the outcome of group efforts, which few other social settings do. Our results suggest that cooperation in Congress is organized around groups whose size, composition and history all matter, and that this record offers a setting in which theories of how social groups form~\cite{backstrom_group_2006, guimera_team_2005} can be tested.
We anticipate that the dataset and analytical framework introduced here will support continued work in this direction, both within the study of the U.S. Congress and in the broader study of higher-order structure in social and political systems~\cite{preti_higher-order_2024}.

\section*{Methods}
\label{sec:methods}

\subsection*{Hypergraph Construction}

We model bill cosponsorship data as hypergraphs where nodes represent legislators and hyperedges represent bills. A hypergraph is a pair $H = (V, E)$ where $V$ is a finite set of nodes and $E=(e_b)_{b\in\mathcal{B}}$ is a collection of subsets of $V$ indexed by bills, so that distinct bills with identical cosponsor sets are kept as repeated hyperedges (multiplicity). We treat all participating legislators as members of a single sponsoring set and distinguish the primary sponsor by an annotation of the hyperedge (\Cref{fig:ho_description}\textbf{b}), although sponsorship and cosponsorship play procedurally distinct roles. The number of legislators in a Congress can exceed the number of seats (435 in the House, 100 in the Senate) because deaths and resignations are followed by special elections and, in the House, because non-voting delegates are included.
After the initial data collection, detailed in Supplementary Section~\ref{si:construction}, we construct hypergraphs from the bill-sponsorship data, with minor modifications tailored to each particular analysis.
First, we discard bills from the 104th House of Representatives with more than one sponsor, since these would not be straightforwardly tractable when analyzing bills based on their sponsor class.
Second, we exclude bills without cosponsors (i.e., bills with only a sponsor) from the analysis of homophilic patterns, as these bills do not incorporate information about group interactions.
Finally, in the gender and party analysis, we discard legislators (and the bills they participate in) who are not affiliated with either the Republican or Democratic Party.

\subsection*{Hypergeometric Distribution as a Null-model for Bill Composition}

A recent survey discusses the problem of defining and measuring higher-order homophily~\cite{laber_guide_2026}.
One of the most well-known approaches to this problem is by Veldt et al. \cite{veldt_combinatorial_2023}, where homophily can be summarized as an affinity score $h_{n,k}(X)$ that accounts for how much nodes of class $X$ tend to participate in hyperedges of size $n$ where exactly $k$ members are of class $X$.
This work uses a hypergeometric-based baseline to normalize the affinity score with respect to a random mixing scenario.
Inspired by their definition, we also evaluate attribute homophily as a deviation from an expected hypergeometric distribution.
To assess whether bills by a given sponsor class include more cosponsors from a target class $\mathcal{C}$ (Democrats, or women) than expected, we define a baseline null model in which cosponsors are drawn at random from the members of the chamber in the bill's Congress.
For a specific bill $i \in \{1, \cdots, B\}$, let $N_i$ denote the number of cosponsors (all except the original sponsor), $M_i$ the number of members of the chamber in that Congress, and $n_i$ the number of those members of class $\mathcal{C}$, each adjusted to exclude the bill's primary sponsor.
If cosponsors are drawn uniformly at random, the number of cosponsors $X_i$ belonging to class $\mathcal{C}$ follows a hypergeometric distribution:
\begin{center}
$
P(X_i=k) = \text{Hypergeom}(k; M_i,n_i,N_i)= \frac{\binom{n_i}{k}\binom{M_i-n_i}{N_i-k}}{\binom{M_i}{N_i}}
$
\end{center}
Bills differ in size and occur in Congresses with different compositions, so we compare each bill with its own null distribution rather than pooling bills before the comparison.
Inspired by Cliff's delta, for bill $i$ with $k_i^{obs}$ cosponsors of class $\mathcal{C}$ we compute
\begin{center}
$d_i = P(X_i < k_i^{obs}) - P(X_i > k_i^{obs}), \qquad X_i \sim \text{Hypergeom}(M_i, n_i, N_i)$,
\end{center}
i.e. the Cliff's delta between the single observation $k_i^{obs}$ and its null distribution.
$d_i \in [-1,1]$ is positive when bill $i$ has more class-$\mathcal{C}$ cosponsors than
expected by chance and negative when it has fewer; under random recruitment its expected value
is zero.
The representativity deviation of class $\mathcal{C}$ is the average over the $B$ bills,
\begin{center}
$\Delta = \frac{1}{B} \sum_{i=1}^{B} d_i \in [-1,1]$,
\end{center}
with $\Delta>0$ indicating over-representation relative to chance and $\Delta<0$
under-representation. Confidence intervals are obtained by bootstrapping the $d_i$ (1000
resamples, 95\% percentile interval).

To show how the bills of a set deviate, and not only their average, we compare the empirical distribution of the $d_i$ with the distribution expected under random recruitment.
We denote these distributions over values $x \in [-1, 1]$ of $d$ by $\rho$, to distinguish them from the hypergeometric probabilities $P(X_i = k)$ over counts $k$.
The empirical distribution is the relative frequency of the observed values,
\begin{center}
$\rho_{\texttt{obs}}(x) = \frac{1}{B} \sum_{i=1}^{B} \mathbb{I}(d_i = x)$,
\end{center}
whose mean is $\Delta$.
Under random recruitment, bill $i$ would take each possible count $k$ with probability $P(X_i = k)$, and with it the value $d_i(k) = P(X_i < k) - P(X_i > k)$.
Averaging over the same $B$ bills gives the expected distribution
\begin{center}
$\rho_{\texttt{null}}(x) = \frac{1}{B} \sum_{i=1}^{B} \sum_{k} P(X_i = k)\, \mathbb{I}\big(d_i(k) = x\big)$,
\end{center}
a finite mixture of the bills' own null distributions of $d$, which is a valid probability distribution (Supplementary Section~\ref{si:null}) and has mean zero, since every bill's null has mean zero.
Because each possible count is converted into $d$ through its own bill's null before averaging, $\rho_{\texttt{obs}}$ and $\rho_{\texttt{null}}$ describe the same per-bill quantity: their means differ by exactly $\Delta$, and differences in shape show how the deviation is spread across bills (for instance, a few strongly over-represented bills, or coalitions polarized at both ends).
In \Cref{fig:hyperedge_size_homophily_effects}, both are shown as histograms with bars of width $0.1$.
Small bills can take only a few values of $d$, so both distributions are concentrated on a few values for them.

\subsection*{Lower-order Simpliciality}
Inspired by Landry et al.~\cite{landry_simpliciality_2024}, we define \emph{lower-order simpliciality} as the fraction of immediately lower-order sub-hyperedges that are contained in a given hyperedge. A hyperedge is a \emph{simplex} whenever all of its sub-hyperedges are also present in the hypergraph, and \emph{simpliciality} refers to the fraction of hyperedges that are simplices.
More precisely, given a hypergraph $H = (V, E)$, the \emph{lower-order simpliciality} of a hyperedge $h$ of size $k$ is the ratio $\sigma_{k-1} = i/k$ where $i = \left|\{e \in E : |e|=k-1 \wedge e \subseteq h\}\right|$.
Since a hyperedge of size $k$ contains exactly $k$ possible sub-hyperedges of size $k-1$, $\sigma_{k-1}(h)$ ranges from 0 to 1.
For example, for a hyperedge of size 3, \emph{lower-order simpliciality} can take the values $0$, $1/3$, $2/3$ or $1$, depending on how many of the three possible pairwise hyperedges among the nodes in the triplet are present in the hypergraph.
To test the statistical significance of simpliciality counts and hyperedge duplicates in the pooled chamber hypergraphs (Figure ~\ref{fig:ho_structure}), we devised randomized null samples preserving node degrees and hyperedge sizes using a Markov Chain sampling procedure.
Given an initial hypergraph, we iteratively performed double-hyperedge swaps.
At each step, we selected two distinct hyperedges and one node from each hyperedge; then we exchanged the two node memberships.
The swap was accepted only when it preserved the hyperedge sizes and did not introduce repeated nodes within either hyperedge.
We performed $1000 \times |E|$ swaps, where $|E|$ is the number of hyperedges.
After discarding the first $20\%$ of swaps as an initial transient, we collected $100$ randomized hypergraph samples along the chain.
We estimated chain mixing by monitoring the convergence of degree assortativity and the share of hyperedges with multiplicity greater than one.

\subsection*{Disentanglement of Temporal Patterns Across Different Hyperedge Sizes}

To analyze the temporal evolution of legislative activity across the two-year congressional session ($\sim$ 730 calendar days, standardizing leap years), we aggregated bill introductions from the 93rd to the 118th Congresses for both the House of Representatives and the Senate. 
Bills were partitioned into discrete size categories ($s$) using three log-spaced bins for bill sizes $\geq 2$ (2--5, 6--19 and 20 or more legislators in the House; 2--3, 4--7 and 8 or more in the Senate) and one bin for solo bills.
Let $N_{c,s}(t)$ denote the number of bills introduced in size category $s$ on relative day $t \in [1, 730]$ of Congress $c$. The total pooled daily volume for category $s$ across all analyzed Congresses is $N_s(t) = \sum_{c} N_{c,s}(t)$, and the total pooled volume across all bill sizes on day $t$ is $N(t) = \sum_{s} N_s(t)$.
We calculated the aggregate daily introduction density $\rho(t)$ as the proportion of all historical bills introduced on day $t$:
\begin{center}
$
\rho(t) = \frac{N(t)}{\sum_{t'=1}^{730} N(t')}
$
\end{center}
Within each day $t$, the conditional proportion of introduced bills belonging to size category $s$ is:
\begin{center}
$
P_s(t) = \frac{N_s(t)}{N(t)} = \frac{\sum_{c} N_{c,s}(t)}{\sum_{c} \sum_{s'} N_{c,s'}(t)}
$
\end{center}
We smoothed $\rho(t)$ and $P_s(t)$ across time using a centered 30-day moving average.
To identify size-specific seasonal preferences independent of overall introduction volume, we evaluated the normalized temporal ratio $R_s(t)$ (\Cref{fig:temporal_bill_introduction}\textbf{c}):
\begin{center}
$
R_s(t) = \frac{P_s(t)}{P_s^{\text{global}}}
$
\end{center}
where $P_s^{\text{global}}$ is the time-invariant baseline proportion for size category $s$ across the entire two-year session:
\begin{center}
$
P_s^{\text{global}} = \frac{\sum_{t=1}^{730} N_s(t)}{\sum_{t=1}^{730} N(t)}
$
\end{center}
We assessed statistical significance against a non-parametric null model with $1000$ permutations.
For each iteration, daily bill counts within each Congress were independently shuffled across the 730-day timeline to destroy temporal correlations while preserving total bill volume and aggregate size distributions.
$R_s^{(b)}(t)$ was recalculated following the same 30-day centered smoothing.
We constructed a 95\% confidence band using the $2.5^{\text{th}}$ and $97.5^{\text{th}}$ percentiles of the permuted distributions.

To illustrate the diversity of bill introduction timing across legislators, we identified the most active primary sponsor within each of four demographic groups (Democratic/Republican $\times$ man/woman) in the most recent Congress in our dataset ($118$th).
Daily sponsorship activity $n_{i,s}(t)$ for legislator $i$ in bill category $s$ was normalized by their total introduced bill count $N_i$:
\begin{center}
$
\tilde{A}_{i,s}(t) = \frac{n_{i,s}(t)}{N_i}
$
\end{center}
and later smoothed over a 30-day window.
Profiles were scaled relative to the legislator’s overall maximum activity level, $\hat{A}_{i,s}(t) = \frac{\tilde{A}_{i,s}(t)}{\max_{s',t'} \tilde{A}_{i,s'}(t')}$.

\subsection*{Sequential Coalition Assembly and Choice Modeling}

The assembly speed analysis uses the bills with dated cosponsorships (Congresses 97--118, all bill types, at least one cosponsor). To compare assembly profiles across the congressional term, each bill's introduction date $t_0$ was expressed as its day in the term, counted from January~3 of the term's first year, on a common calendar of 730 days. Bills were stratified into three quantile classes of their initial coalition size $s_0$, which includes the sponsor and the cosponsors who joined on the day of introduction (\Cref{fig:recruitment}\textbf{b}--\textbf{d}). We measured cosponsorship speed as $s_f/(t_f-t_0+1)$, where $s_f$ is the final coalition size, sponsor included, and $t_f$ the date of the last recorded cosponsorship; bills whose whole coalition joined on the day of introduction were left out. For each class, we took the median speed of the bills introduced in rolling 90-day windows shifted by one day. To test for size scaling across heterogeneous coalition sizes, we rescaled speed by $s_0^{-\alpha}$, with $\alpha$ chosen to minimise the mean, over windows, of the variance across classes of the log median rescaled speed, and compared the size-stratified trajectories. In the null model (\Cref{fig:recruitment}\textbf{d}), introduction days were shuffled among the bills of each class, uniformly among the assignments that keep every bill's assembly time within the Congress (1000 shuffles; Supplementary Section~\ref{si:assembly}).

In the micro-level recruitment analysis we use an event-resolved history of sponsorship transitions. Each dated, non-original cosponsorship defined a discrete choice event $e = (b, t, i)$, representing the transition $S_b(t^-) \to S_b(t^-) \cup \{i\}$ from the pre-event coalition $S_b(t^-)$, which includes the sponsor. Legislators joining on the same date were evaluated against the identical pre-day coalition state to prevent artificial within-day ordering.
For each event, we formed a choice set $A_e$ by combining the observed joiner $i$ with a risk set of 10 legislators sampled uniformly from the members of the chamber in that Congress who had not joined bill $b$. Prior collaborations were counted over all earlier bills of the chamber from the 93rd Congress onwards; legislators without a DW-NOMINATE score in a Congress were given the chamber's median score in that Congress.

Recruitment probabilities were estimated using conditional logit choice models: 

\begin{center}
$
\text{Pr}(j \text{ joins} \mid A_e) = \frac{\exp(\boldsymbol{\beta}^T \mathbf{x}_{ej})}{\sum_{\ell \in A_e} \exp(\boldsymbol{\beta}^T \mathbf{x}_{e\ell})}
$
\end{center}

which condition out event-specific intercepts, holding constant all institutional, temporal, and bill-level characteristics common to choice set $A_e$.
The contribution of each history block was assessed in two ways, on the same choice events. First, each block was added on its own to the baseline controls (\Cref{fig:recruitment}\textbf{g}). Second, the blocks were added in sequence, so that each is evaluated given the previous ones:
\begin{enumerate}[start=0]
\item Baseline controls (activity, shared state, shared party, shared gender, ideology, and coalition-size interactions);
\item Prior sponsor ties;
\item Mean prior coalition dyadic ties;
\item Triadic history (mean prior candidate--pair co-occurrences and share of coalition pairs with at least one prior co-occurrence);
\end{enumerate}
Both comparisons use likelihood-ratio statistics $2\Delta\ell$ and AIC; for the sequential ladder, $2\Delta\ell = 2(\ell_k - \ell_{k-1})$ (Supplementary Section~\ref{si:recruit_ladder}). 
Full mathematical formulations for all covariate blocks are provided in Supplementary Section~\ref{si:features}; the sample, the nested specifications and their fit statistics are in Supplementary Sections~\ref{si:recruit_spec} and~\ref{si:recruit_ladder}.

\section*{Data Availability}
The complete raw data and preprocessed hypergraphs used in this work are available at \url{https://doi.org/10.5281/zenodo.18337053}.

\section*{Code Availability}
The code used to construct the hypergraphs and produce the reported analyses will be made 
available in a versioned public repository before publication.

{\renewcommand{\addcontentsline}[3]{}\bibliography{refs}}

\section*{Acknowledgements}

B.P. and G. Petri acknowledge support from the “BeyondTheEdge: Higher-Order Networks and Dynamics” project (European Union, REA Grant Agreement No. 101120085). A.S. acknowledges support from the European Union’s Horizon 2020 research and innovation programme under the Marie Sk\l{}odowska-Curie grant agreement no. 101208090.

\section*{Author contributions statement}

\textit{B.P.: Conceptualization, Visualization, Methodology, Investigation, Data Curation, Software, Writing - original draft. G.d.F.M.: Conceptualization, Methodology, Writing - Review \& Editing. G. Preti: Software, Investigation, Dataset Curation, Conceptualization, Writing - Review \& Editing. G.Petri: Dataset Curation, Conceptualization, Methodology, Writing - Review \& Editing. A.S.: Visualization, Methodology, Writing - Review \& Editing.}

\section*{Competing interests}

The authors declare no competing interests.

\clearpage
\immediate\write\csname @auxout\endcsname{\string\newlabel{MainLastPage}{{}{\the\numexpr\value{page}-1\relax}{}{}{}}}
\setcounter{page}{1}
\hypersetup{pageanchor=false} %
\rfoot{\small\sffamily\bfseries\thepage/\pageref*{LastPage}}
\setcounter{figure}{0}
\setcounter{table}{0}
\renewcommand{\thefigure}{S\arabic{figure}}
\renewcommand{\thetable}{S\arabic{table}}
\renewcommand{\theHfigure}{S\arabic{figure}}
\renewcommand{\theHtable}{S\arabic{table}}

\vspace*{2em}
\begin{center}
{\LARGE \textbf{Supplementary Information:}\\
How coalition size shapes bill cosponsorship in the U.S. Congress\par}
\vskip 1.5em
{\large Bernardo Pereira, Giulia Preti, Andrea Santoro,
Gianmarco De Francisci Morales and Giovanni Petri\par}
\end{center}
\vskip 1.5em

\setcounter{tocdepth}{2}
\setlength{\tocsep}{2pc} %
\titlecontents{section}[\tocsep]
  {\addvspace{1pt}\small\bfseries\sffamily}
  {\contentslabel[\thecontentslabel]{\tocsep}}
  {}
  {\hfill\thecontentspage}
  []
\titlecontents{subsection}[\tocsep]
  {\addvspace{0pt}\small\sffamily}
  {\contentslabel[\thecontentslabel]{\tocsep}}
  {}
  {\ \titlerule*[.5pc]{.}\ \thecontentspage}
  []
\vspace*{-1.5em}
\tableofcontents
\clearpage

\section{Data}\label{si:data}

This section documents the dataset analyzed throughout the main text: how it relates to earlier releases of the congressional cosponsorship data, how it was collected and cleaned, and the legislative process it records. The hypergraph construction itself is described in the main-text Methods.

\subsection{Related datasets}\label{si:related}
Although numerous works have independently assembled and analyzed bill cosponsorship data from the U.S. Congress, using datasets of varying scope, temporal coverage, and metadata richness, Fowler's data~\cite{fowler_legislative_2006, fowler_connecting_2006} has had the most lasting impact on the network science literature. 
Beyond cosponsorship, political cooperation in legislative bodies has also been studied through the lens of committee memberships and roll-call votes~\cite{porter_community_2007, waugh_party_2011, porter_network_2005}, each offering a complementary window onto the structure of legislative alliances.
In its network form, Fowler's cosponsorship dataset has been made accessible through repositories such as Netzschleuder~\cite{peixoto2020netzschleuder}; in its higher-order form, it has been processed and released by Benson et al.~\cite{benson_simplicial_2018}, made available through Benson's data repository~\cite{benson_cornell_data}, and subsequently incorporated into the XGI data library~\cite{landry_xgi_2023}. 
In this form, it has become a standard benchmark in the higher-order networks literature~\cite{iacopini_group_2022, mancastroppa_hyper-cores_2023, hood_broad_2025, preti_higher-order_2024}.
This trajectory reflects a broader landscape in which repositories of higher-order network data, such as XGI~\cite{landry_xgi_2023}, HypergraphX Data\cite{lotito_hypergraphx-data_2026}, the Aachen repository~\cite{ahorn_repository}, and Benson's collection~\cite{benson_cornell_data}, are actively growing, but remain considerably sparser than their pairwise counterparts. 
Established databases for conventional networks, including the Colorado Index of Complex Networks (ICON)~\cite{clauset2016icon}, the Stanford Network Analysis Project (SNAP)~\cite{snapnets}, Netzschleuder~\cite{peixoto2020netzschleuder}, SocioPatterns~\cite{sociopatterns}, KONECT~\cite{konect}, and the Network Repository~\cite{networkrepository}, offer a breadth and variety of datasets that the higher-order community has yet to match. 
Enriching this ecosystem with high-quality, richly annotated hypergraph datasets is therefore an important priority for the field.

The existing higher-order releases of the congressional cosponsorship data, both in Benson's repository and in the XGI library, cover the period from 1973 to 2005, encompassing the 93rd through the 108th Congress.
Our dataset extends this temporal scope to the 118th Congress (2023-2025), adding two decades of legislative activity and bringing the record up to the most recently completed Congress.
Beyond this temporal extension, the existing releases provide node-level information limited to legislator name, state, district, and party, and edge-level information consisting of bill timestamps.
While this has been sufficient to support a range of structural analyses, it leaves a significant portion of the available legislative record unexploited. Our dataset extends this foundation substantially.
At the node level, we additionally provide each legislator's gender, birth year, and a consistent identifier linking records across Congresses.
At the hyperedge level, we include the bill's primary sponsor, its date of introduction, its type, its policy area, and its legislative outcome, whether it passed the House, the Senate, and was ultimately signed into law.
Together, these additions enable a class of analyses on gender dynamics, temporal evolution, policy-area specialization, and legislative success that are not supported by the datasets currently available in the higher-order networks community.

\subsection{Dataset construction}\label{si:construction}
To construct a comprehensive dataset of U.S. Congressional bill cosponsorship hypernetworks, we developed a multi-stage pipeline that integrates data from the \textbf{Congress.gov API}~\footnote{\url{https://api.congress.gov}}, the \textbf{GovInfo Bulk Data Repository}~\footnote{\url{https://www.govinfo.gov/bulkdata/BILLSTATUS}}, the \textbf{Biographical Directory of the United States Congress}~\footnote{\url{https://www.congress.gov}}, and the \textbf{Congressional Bills Project}~\footnote{\url{http://www.congressionalbills.org}}. 
Our pipeline consists of a collection phase followed by a harmonization and cleaning phase.

All data were collected for the $93$rd through the $118$th Congresses (1973–2025).
This range reflects a structural limitation of the Congress.gov API, which does not provide sponsor or cosponsor information for earlier Congresses.

\subsubsection{Data Acquisition}

We developed a custom scraper to query the Congress.gov API~\footnote{An API key is required for access.} and parse GovInfo XML files.
The acquisition workflow proceeds in four stages:

\begin{enumerate}
\item \textbf{Bill Enumeration.} For each Congress, we retrieved the complete list of bills via the \texttt{/bill/\{congress\}} endpoint, utilizing pagination to ensure complete coverage. 
Each bill is uniquely identified by its \emph{(congress, bill type, bill number)}. 
The bill types include:
    \begin{itemize}
        \item \textit{hconres}, \textit{sconres}: House and Senate Concurrent Resolutions. Must be passed by both chambers but not signed by the President. They do not have the force of law and are used to make or amend rules that apply to both chambers.
        \item \textit{hjres}, \textit{sjres}: House and Senate Joint Resolutions. Must be passed by both chambers and signed by the President. They have the force of law and are used for continuing or emergency appropriations, declarations of war, and proposing amendments to the Constitution.
        \item \textit{hres}, \textit{sres}: House and Senate Simple Resolutions. They do not require the approval of both chambers. They do not have the force of law and are used for internal rules, procedures, or other matters within the prerogative of one chamber.
        \item \textit{hr}, \textit{s}: House and Senate Bills. They deal with domestic and foreign issues and programs. Public bills pertain to matters that affect the general public, while private bills affect just certain individuals and organizations.
    \end{itemize}

\item \textbf{Bill Metadata Retrieval.} For each enumerated bill, we queried the \texttt{/bill/\{congress\}/\{bill type\}/\{bill number\}} endpoint to extract the sponsor’s biographical data (name, state, district, party, \emph{bioguideId}) and the bill’s metadata (title, introduction date, policy area, number of cosponsors, and number of cosponsors including the withdrawn cosponsors). We retrieved partial information on the legislative outcomes (passed House/Senate, enacted as Public Law) from the Congressional Bill Project data dump and from the GovInfo XML files.

\item \textbf{Cosponsor Retrieval.} For each enumerated bill, we obtained detailed information on all cosponsors (if any) using the \texttt{/cosponsors} endpoint. Each record includes the same legislator fields captured for the bill's sponsor.

\item \textbf{Legislator Metadata Retrieval.} To enrich nodes, we scraped the \texttt{/member/\{bioguideId\}} endpoint to obtain additional legislator attributes: birth year, party-membership history, district-assignment history, number of cosponsored bills, and number of sponsored bills.
\end{enumerate}

Given the large volume of requests required by the pipeline, we implemented a robust request handler compliant with the API's rate limits (5000 requests/hour). The handler uses exponential backoff with jitter to recover from HTTP 429 (Too Many Requests) and intermittent 500-level server errors, ensuring data completeness without violating service terms.

\subsubsection{Data Harmonization and Cleaning}

Following acquisition, we harmonized records with the Congressional Bills Project dataset (covering the 93rd–114th Congresses) to fill metadata gaps.
The cleaning workflow, implemented via custom Python modules, addressed several inconsistencies:

\begin{itemize}
\item \textbf{Entity Resolution.} We used the \emph{bioguideId} as the primary key for legislators. When IDs were missing, we applied a fuzzy matching procedure that cross-referenced legislators’ names, states, and active Congresses against the Biographical Directory of the U.S. Congress.
\item \textbf{Name Normalization.} We normalized names using Unicode NFKD decomposition to remove diacritics and resolved common nickname–formal-name variants via a curated lookup table (e.g., Chuck → Charles).
\item \textbf{Party and District Alignment.} Historical numerical party codes were mapped to standard character codes (D, R, I). We also corrected inconsistencies arising from Congress party switches or redistricting events.
\item \textbf{Imputation of Missing Attributes.} Missing demographic information, such as gender, was imputed using a verified dictionary of legislators.
\item \textbf{Cosponsor Deduplication.} Occasional duplicate cosponsor entries in the Congress.gov API were removed. Minor inconsistencies in cosponsor counts may remain due to limitations in the source data.
\end{itemize}

\begin{table*}[ht!]
    \centering
    \begin{tabular}{lll}
        \toprule
        \textbf{Field Name} & \textbf{Field Type} & \textbf{Description} \\
        \midrule
        \textbf{BillNum} & Int & Bill number. \\
        \textbf{BillType} & Cat & Bill type (e.g., \emph{hr}, \emph{s}, \emph{hres}). \\
        \textbf{Chamber} & Int & Chamber code ($0$=House, $1$=Senate). \\
        \textbf{CongressNum} & Int & Congress number. \\
        CosponsorNum & Int & Count of cosponsors. \\
        CosponsorWWNum & Int & Count of cosponsors (including withdrawn). \\
        PassH & Int & $1$ if passed House, $0$ otherwise. \\
        PassS & Int & $1$ if passed Senate, $0$ otherwise. \\
        PLaw & Int & $1$ if became Public Law, $0$ otherwise. \\
        PolicyArea & String & Primary policy subject area. \\
        IntroducedDate & Date & Date of introduction. \\
        Title & String & Official title of the bill. \\
        \textbf{bioguideId} & String & Legislator's unique ID. \\
        \bottomrule
    \end{tabular}
    \caption{Fields in the cleaned bill-sponsor dataset. The fields in bold are present in the cleaned bill-cosponsor dataset.}
    \label{tbl:cleaned_data_fields}
\end{table*}

\begin{table*}[ht!]
    \centering
    \begin{tabular}{lll}
        \toprule
        \textbf{Field Name} & \textbf{Field Type} & \textbf{Description} \\
        \midrule
        \textbf{bioguideId} & String & Legislator ID from the Biographical Directory. \\
        \textbf{birthYear} & Int & Birth Year of the legislator. \\
        \textbf{NameFirst} & String & First name of the legislator. \\
        \textbf{NameLast} & String & Last name of the legislator. \\
        \textbf{directOrderName} & String & Full name of the legislator.\\
        \textbf{Gender} & Int & Gender of the legislator ($0$=Male, $1$=Female). \\
        \textbf{cosponsoredLegislation} & Int & Number of cosponsored bills (up to October 2025). \\
        \textbf{sponsoredLegislation} & Int & Number of sponsored bills (up to October 2025). \\
        Party & Cat & Party code (D, R, I). \\
        StartYearInParty & Int & First year of party membership. \\
        EndYearInParty & Int & Final year of party membership ($-1$ if ongoing). \\
        Chamber & Int & Chamber served ($0$=House, $1$=Senate). \\
        CongressNum & Int & Congress number. \\
        StartTerm & Int & Start year of district representation. \\
        \multirow{2}{*}{EndTerm} & \multirow{2}{*}{Int} & Last year of district representation ($-1$ if ongoing).\\
        & & ($-1$ if still a member). \\
        State & Cat & State code.\\
        District & Int & District code (House members only).\\
        \bottomrule
    \end{tabular}
    \caption{Fields in the cleaned legislator-level datasets. The fields in bold represent static attributes.}
    \label{tbl:cleaned_member_data_fields}
\end{table*}

\subsubsection{Data Quality}

To ensure reliability, we cross-validated our scraped dataset against the Congressional Bills Project for the overlapping sessions (93rd–114th). We verified that sponsor IDs and party affiliations matched across sources. Discrepancies were manually reviewed and resolved using the Biographical Directory. The final dataset accounts for edge cases such as mid-session resignations, party switches, and special elections.

\subsubsection{Access and Availability}
The data are available at \url{https://doi.org/10.5281/zenodo.18337053}. The code used to construct the hypergraphs and produce the analyses will be released in a versioned public repository before publication (main-text Code Availability).

\subsection{The legislative process in the U.S. Congress}
Legislative activity in the U.S. Congress is divided between two chambers: the House of Representatives, with 435 members apportioned among the states according to population, and the Senate, with 100 senators, two per state. A bill follows these steps: (1) a legislator, the primary sponsor, drafts the bill, which other legislators may join as cosponsors; (2) the sponsor introduces it in their chamber; (3) it is referred to the appropriate committee for analysis and evaluation; (4) it returns to the chamber for debate and a vote; (5) if passed, it is sent to the other chamber, where steps (3) and (4) are repeated; (6) if passed there as well, it is sent to the President, who can sign it into law or veto it. Cosponsorship can occur at any point until the bill is passed by the chamber, which is why coalitions grow after introduction (main-text Fig.~5a). %

\section{Coalitions across five decades}\label{si:descriptive}

This section supports the main-text section ``Cosponsorship coalitions span two orders of magnitude'' (main-text Fig.~1). The main text summarizes each of the first three observations below in one sentence; here we give them in full. We then relate coalition size to bill type and policy area, and set out the analysis of coalition sizes near the majority threshold.

\subsection{Chamber size}

The House comprises 435 seats, apportioned among states and their congressional districts according to population, while the Senate comprises 100 seats, with each state holding exactly two regardless of population. In practice, the number of nodes in any given Congress hypergraph may deviate slightly from these nominal values, as the death or resignation of a sitting legislator during a term necessitates a special election to fill the vacant seat, resulting in more legislators being recorded in that Congress than expected.

\subsection{Bill counts over time}

Main-text Fig.~1\textbf{c} summarizes the number of bills (hyperedges) recorded in each chamber across the full period covered by our dataset. The House hypergraphs are considerably larger, both in nodes and hyperedges, reflecting the greater number of seats and the higher volume of legislative activity in the lower chamber. Variation in the number of bills introduced across Congresses is closely linked to the broader socio-political context of each legislative period. After the 95th Congress (1977--79), the number of bills introduced in the House declined steeply following the lifting of the 25-cosponsor cap per bill, which ended the introduction of many duplicate bills that served as political endorsements of previously introduced ones \cite{congressgov2026sponsorship}. Moreover, in the 104th Congress (1995-97), the Republican Party gained control of the House for the first time in more than 40 years and regained control of the Senate after 8 years, in what was called the \textit{Republican Revolution}, coinciding with the lowest polyadic bill (with more than one legislator supporting it) introduction rates observed in both chambers over the 1973-2025 period. Finally, at the onset of the 2008 Global Financial Crisis, legislative activity increased in both chambers, reaching the highest bill-introduction rates observed in the recent past. The pattern of evolution for singleton bills versus polyadic ones is analogous in both chambers. After an initial period where singleton bills dominated legislative activity, there was an inflection during 1981-83 in the Senate and 1983-85 in the House, and the preponderance of polyadic bills has risen ever since.

\subsection{Degree distributions}

Main-text Fig.~1\textbf{d}--\textbf{e} (\textit{left}) depicts the distribution of node degrees, i.e., the number of bills in which each legislator participates across all Congresses. We also report a pooled distribution obtained by aggregating node degrees across Congresses. These distributions display a characteristic uneven profile in both chambers. A minority of legislators are involved in very few bills, the bulk of the chamber participates in a moderate number of cosponsorship events, and relatively few legislators accumulate an exceptionally large number of bill involvements. This asymmetric profile, with most legislators concentrated around a single peak of a few hundred bills, is consistent with and extends the findings of Fowler~\cite{fowler_legislative_2006}, who reported analogous patterns in earlier data; our dataset confirms that this distributional shape is robust over a substantially longer temporal window that encompasses more recent years. Notice that this profile contrasts with the right-skewed, heavy-tailed degree distributions typical of scientific co-authorship networks, where transient, single-publication authors dominate \cite{fowler_legislative_2006, newman_scientific_2001a, newman_scientific_2001b}. As Fowler first observed, this divergence stems from fundamentally different boundary dynamics. While academic networks are shaped by continuous turnover (e.g., students permanently exiting the field), legislators typically serve multiple terms with sustained activity, directly truncating the long tail of low-degree nodes characteristic of scientific co-authorship networks.

\clearpage
\section{Composition across coalition sizes}\label{si:composition}

This section supports the main-text section ``Small coalitions are the most bipartisan'' (main-text Fig.~2). It reports the Senate results, establishes the properties of the representativity deviation score $\Delta$ defined in the main-text Methods, breaks the results down by the party and gender of the sponsor, and describes the temporal dependency of coalition size based on sponsorship gender and party.

\subsection{Senate}

\begin{figure}[!ht]
\centering
\includegraphics[width=\linewidth]{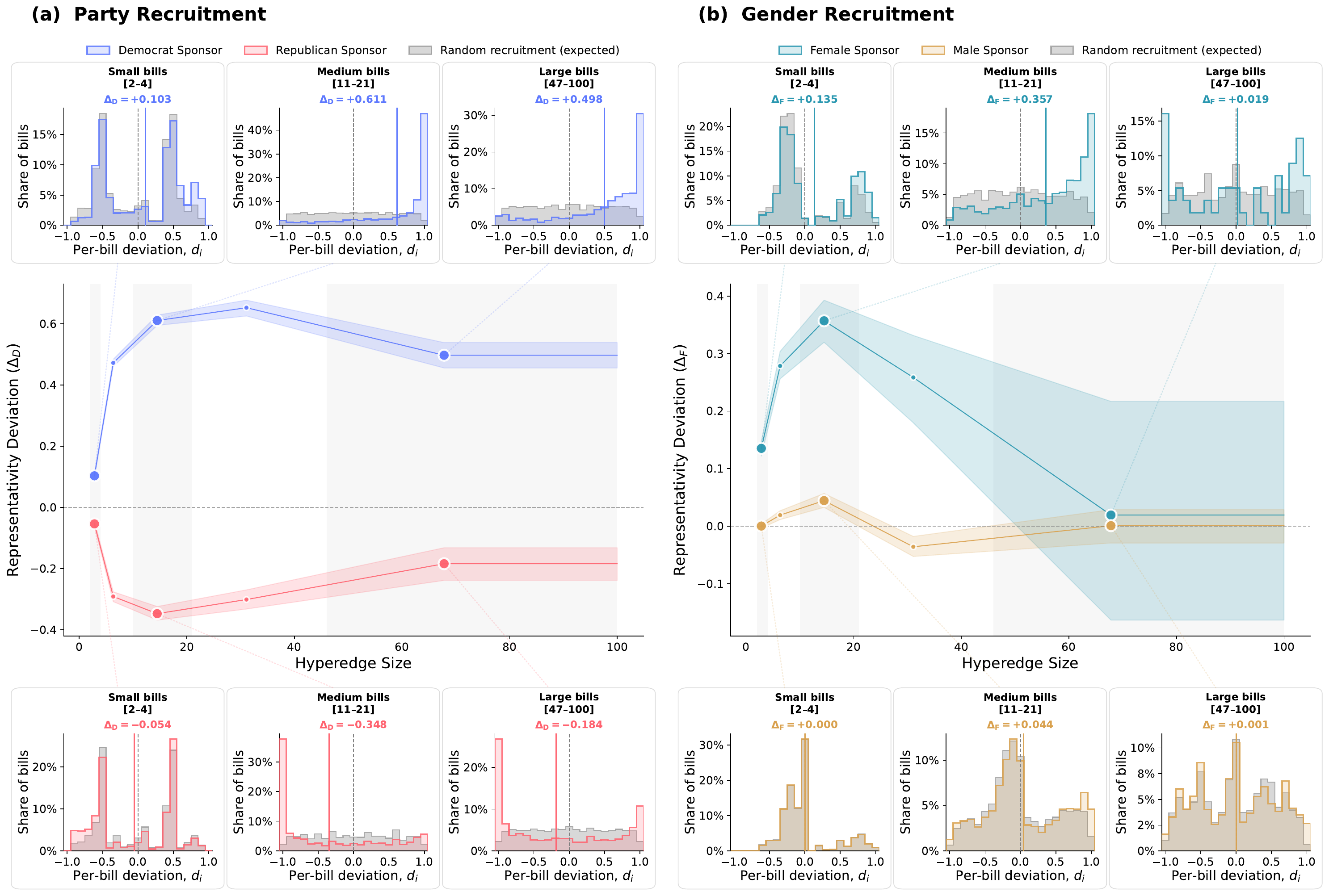}
\caption{\textbf{Senate: hyperedge size modulates partisan and gender composition.} Representativity deviation $\Delta$ as a function of hyperedge size (\textit{center}), and the distributions of the per-bill deviations $d_i$ behind it (\textit{top} and \textit{bottom}), as in main-text Fig.~2. Hyperedge sizes, sponsor included, were divided into 5 log-spaced bins; bands are $95\%$ bootstrap confidence intervals (1000 resamples of the bills in each bin). \textbf{(a)} Representativity deviation of Democrats ($\Delta_D$) across bill sizes, in Democratic- and Republican-sponsored bills (\textit{center}). On \textit{top} (Democratic sponsor) and \textit{bottom} (Republican sponsor), for the smallest, middle and largest bins: the empirical distribution of the bills' $d_i$ (colored) and the distribution expected under random recruitment for the same bills (grey); solid and dashed vertical lines mark their means, $\Delta_D$ and $0$. \textbf{(b)} Representativity deviation of women ($\Delta_F$) across bill sizes, in female- and male-sponsored bills, with the same panels as in (a).}
\label{fig:suppmat_homophily_senate}
\end{figure}

The Senate shows the same size gradient for party as the House (Supplementary Fig.~\ref{fig:suppmat_homophily_senate}\textbf{a}). Party composition is closest to random mixing in the smallest coalitions ($\Delta_D=+0.10$ and $-0.05$ for coalitions of 2--4 senators under Democratic and Republican sponsors), strongest among coalitions of a few dozen senators ($+0.61$ and $-0.35$ for 11--21 senators), and weaker among the largest ($+0.50$ and $-0.18$ for 47--100). Women are over-represented under female sponsors in small and medium coalitions ($\Delta_F=+0.14$ and $+0.36$) but not in the largest ($+0.02$), and coalitions led by men are close to random mixing at every size (Supplementary Fig.~\ref{fig:suppmat_homophily_senate}\textbf{b}).

\subsection{The null model is a distribution}\label{si:null}

For bill $i$, the null probabilities $P(X_i = k)$ of its possible counts $k$ of class-$\mathcal{C}$ cosponsors are non-negative and sum to one, $\sum_k P(X_i = k) = 1$.
Each possible count gives a value of the deviation score, $d_i(k) = P(X_i < k) - P(X_i > k)$, so the null distribution of $d$ for bill $i$ assigns probability $P(X_i = k)$ to the value $d_i(k)$.
The expected distribution shown in the panels of main-text Fig.~2 averages these over the $B$ bills of a set:

\begin{center}
$
\rho_{\texttt{null}}(x) = \frac{1}{B} \overset{B}{\underset{i=1}{\sum}} \sum_{k} P(X_i = k)\, \mathbb{I}\big(d_i(k) = x\big).
$
\end{center}

\paragraph{It is a probability distribution.} Every term is non-negative, so $\rho_{\texttt{null}}(x) \geq 0$. Summing over all values $x$, each pair $(i, k)$ is counted exactly once:

\begin{center}
$
\begin{aligned}
\sum_{x} \rho_{\texttt{null}}(x)
&= \frac{1}{B} \sum^{B}_{i=1} \sum_{k} P(X_i = k) \sum_{x} \mathbb{I}\big(d_i(k) = x\big) \\
&= \frac{1}{B} \sum^{B}_{i=1} \sum_{k} P(X_i = k)
 = \frac{1}{B} \sum^{B}_{i=1} 1 = 1,
\end{aligned}
$
\end{center}

\noindent which also implies $\rho_{\texttt{null}}(x) \leq 1$.

\paragraph{Its mean is zero.} For a single bill, let $X_i$ and $X_i'$ be two independent draws from its null distribution. The expected deviation score is

\begin{center}
$
\sum_{k} P(X_i = k)\, d_i(k)
= \sum_{k} P(X_i = k) \big[ P(X_i' < k) - P(X_i' > k) \big]
= P(X_i' < X_i) - P(X_i' > X_i) = 0,
$
\end{center}

\noindent since $X_i$ and $X_i'$ are identically distributed, so the two probabilities are equal by symmetry.
The mean of $\rho_{\texttt{null}}$ is the average of these per-bill means, and is therefore zero for any set of bills, whatever their sizes and Congresses.
The empirical distribution $\rho_{\texttt{obs}}$ has mean $\Delta$, so the two means differ by exactly $\Delta$.

\subsection{Bounds of the representativity deviation score $\Delta$}\label{si:delta_bounds}

To correctly assess the meaning of high or low values of $\Delta$, it is important to remark how this parameter is bounded. We start by rewriting the definition of $d_i$ for each bill $i$ as: 

\begin{center}
$
d_i = p_{\text{less}}(i) - p_{\text{greater}}(i)
$
\end{center}
where 
$
p_{\text{less}}(i) = P(X_i < k^{obs}_i)$ and $p_{\text{greater}}(i) = P(X_i > k^{obs}_i)
$, with $X_i \sim \text{Hypergeom}(M_i, n_i, N_i)$ the null distribution of bill $i$.

Next, we note that 
$
0 \leq p_{\text{less}}(i), p_{\text{greater}}(i) \leq 1
$
and 
$
p_{\text{less}}(i) +  p_{\text{greater}}(i) \leq 1
$
and thus we conclude
\begin{center}
$
-1 \leq p_{\text{less}}(i) - p_{\text{greater}}(i) = d_i \leq 1
$.
\end{center}

Since $\Delta$ is an average of the $d_i$ values, it follows that $-1 \leq \Delta \leq 1$.

\subsubsection{Asymmetry between Democratic and Republican sponsors}

In the House, coalitions led by Democratic sponsors over-represent Democrats more strongly ($\Delta_D$ up to $+0.84$) than coalitions led by Republican sponsors under-represent them ($\Delta_D$ down to $-0.50$; main-text Fig.~2\textbf{a}). For bills with few cosponsors, the attainable range of $d_i$ depends on the class share: for a bill with a single cosponsor, drawn from a chamber in which a share $p$ of members belongs to the target class, $d_i$ can only take the values $1-p$ (if the cosponsor belongs to the class) or $-p$ (if not). At small sizes, values of $\Delta$ are therefore not directly comparable between classes of different size; for party the difference is small, since the shares of Democrats and Republicans are both close to one half, but it matters for gender, where women are a small minority. These bounds widen toward $[-1,1]$ as the number of cosponsors grows, and are already close to $[-1,1]$ at the sizes, a few dozen members, where the asymmetry is largest, so they cannot explain it there. The asymmetry may instead reflect majority status, which changed several times over the period, or differences in how the two parties build coalitions; separating these explanations requires analyzing Congresses by majority party, which we leave to future work.

\subsection{Cross-party participation across Congresses}\label{si:cross_party_by_congress}

At the largest coalition sizes, the bill-specific null distribution is narrow, so most observed deviations lie close to either end of the $d_i$ scale. In the House class of 239--433 members, 81--88\% of bills have $|d_i|>0.9$. At these sizes, $\Delta$ therefore records how consistently bills fall above or below random mixing more directly than how far their cross-party share lies from it. We consequently complement $\Delta$ with the mean opposite-party share of cosponsors.

\begin{figure}[!ht]
\centering
\includegraphics[width=\linewidth]{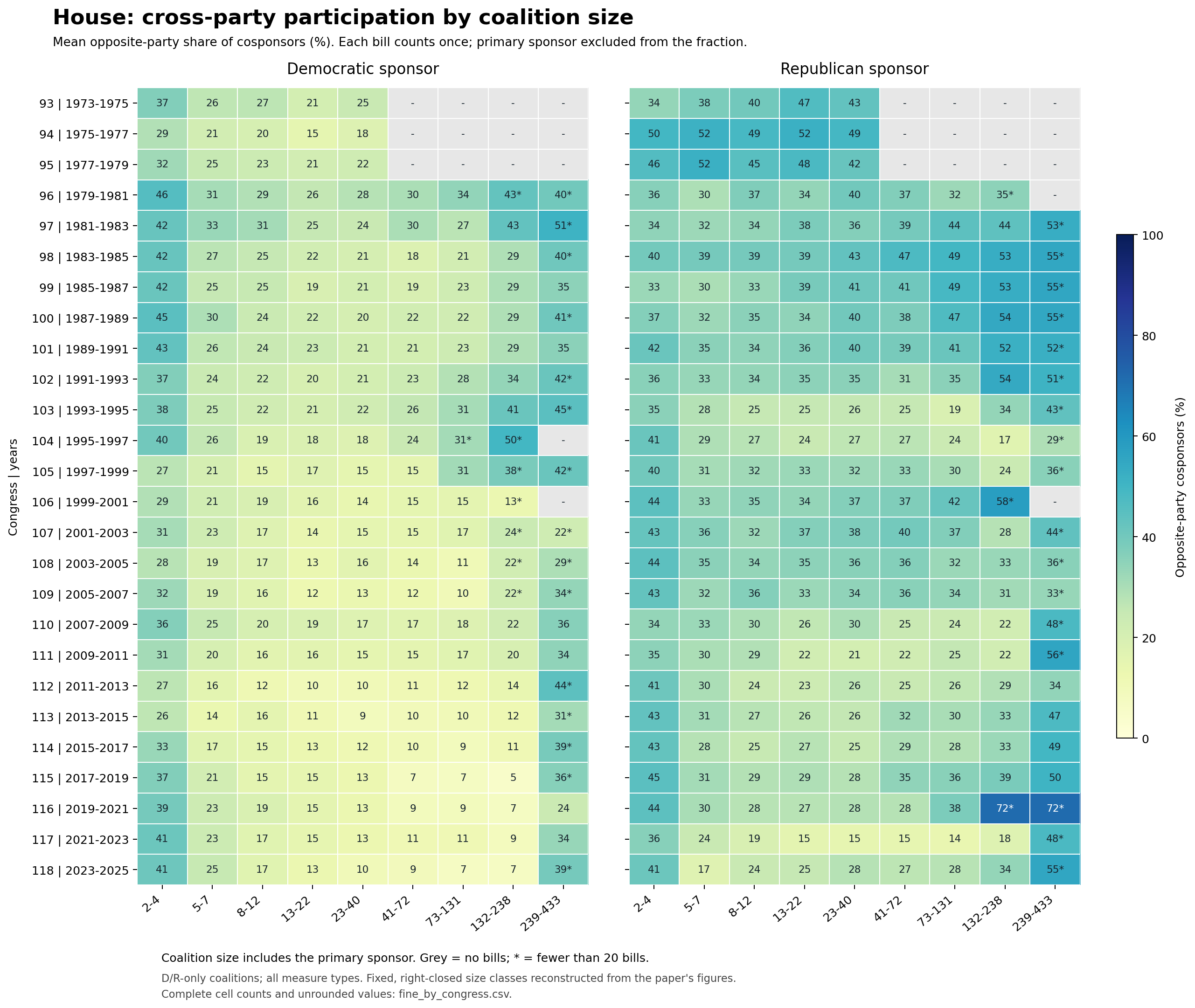}
\caption{\textbf{House: opposite-party share of cosponsors by Congress and coalition size.} Each cell shows the mean share of a bill's cosponsors who belong to the party opposite that of its sponsor, separately for Democratic- and Republican-sponsored bills. Coalition size includes the sponsor. Grey cells contain no bills; an asterisk marks cells with fewer than 20 bills.}
\label{fig:suppmat_cross_party_house}
\end{figure}

\begin{figure}[!ht]
\centering
\includegraphics[width=\linewidth]{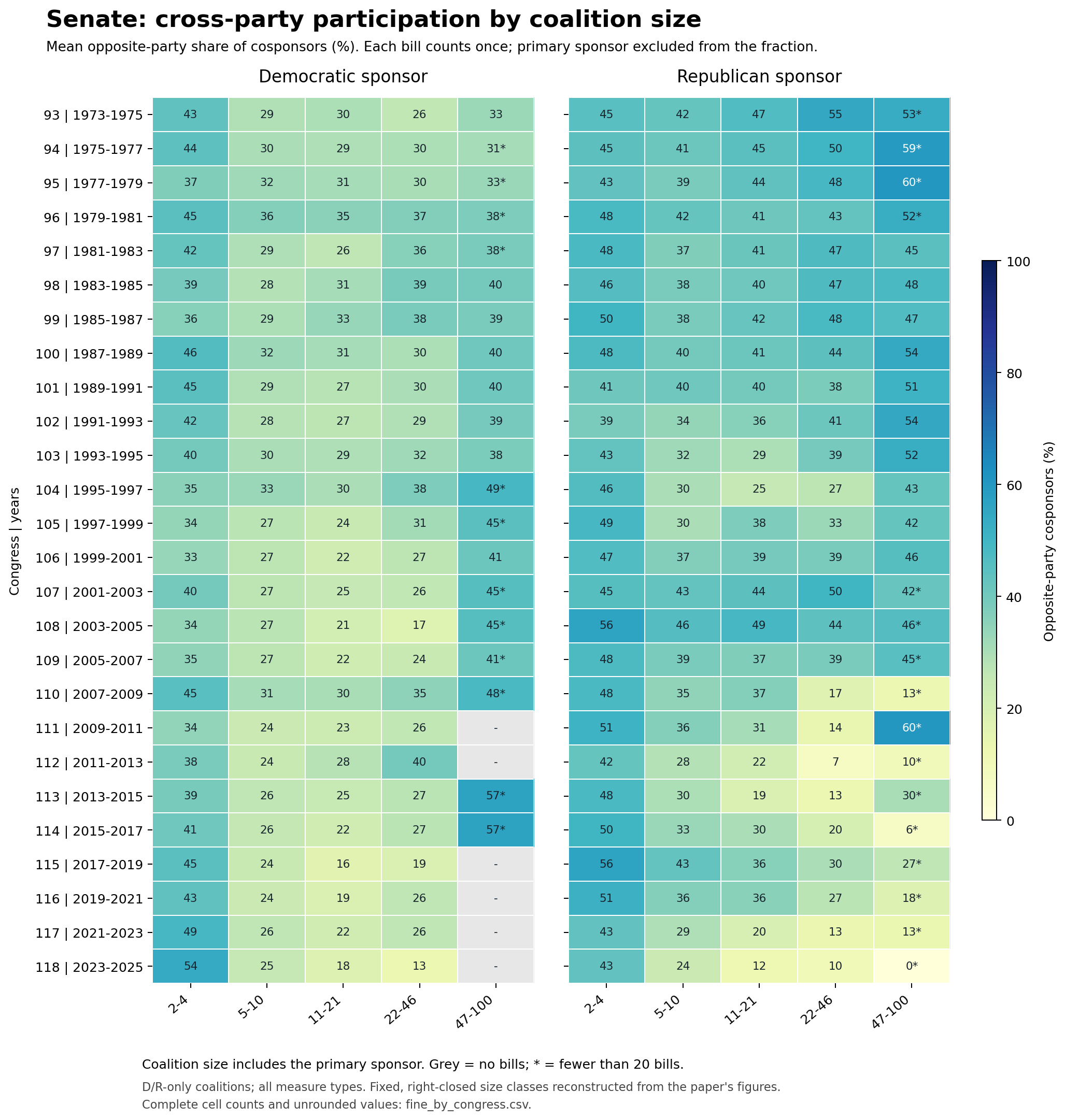}
\caption{\textbf{Senate: opposite-party share of cosponsors by Congress and coalition size.} As in Supplementary Fig.~\ref{fig:suppmat_cross_party_house}; grey cells contain no bills and an asterisk marks cells with fewer than 20 bills.}
\label{fig:suppmat_cross_party_senate}
\end{figure}

Supplementary Figs.~\ref{fig:suppmat_cross_party_house}--\ref{fig:suppmat_cross_party_senate} show this share by Congress and coalition-size class. Across chambers and sponsor parties, small coalitions retain a comparatively high cross-party share across the period, whereas the decline is concentrated in coalitions of intermediate size. The return of cross-party participation in the largest classes is partly mechanical, because coalitions covering most of a chamber must include members of the other party; sparse cells are marked and should be interpreted cautiously.

\subsection{Symmetry of the representativity deviation score $\Delta$}
For generality, assume we are dealing with a class type that can take only two values: $a$ or $b$.
In this section we explain the symmetry of the representativity deviation score $\Delta_a$ for a target class $a$ with respect to its complement class $b$. 
That is, in essence, why $\Delta_F = -\Delta_M$ and $\Delta_D = -\Delta_R$.
For a given bill, let $N$ be the number of cosponsors (draws) and $M$ the pool size (legislators of the chamber in that bill's Congress, sponsor excluded).
If the target class is $a$, the observed count is $k_a$ and the pool has $n_a$ successes, so:
$
X_a \sim \text{Hypergeom}(M,\, n_a,\, N).
$
If instead the target class is $b$, since every cosponsor is either $a$ or $b$, the observed count is simply the complement:
$
k_b = N - k_a
$
and because the pool is the same, the number of $b$ candidates is $n_b = M - n_a$. This holds in our data for both class types: for party, because only Democrats and Republicans are kept in the hypergraphs, and for gender, because every legislator is recorded as male or female. Thus, a standard hypergeometric identity gives:

\begin{center}
$
X_b = N - X_a \sim \text{Hypergeom}(M, M-n_a,\, N) = \text{Hypergeom}(M,\, n_b,\, N)
$
\end{center}

So $X_b$ is exactly the complement hypergeometric variable with the same $M$, the same $N$, just $n_b = M - n_a$.

Following this identity, we can rewrite the tail distributions of $b$ as a function of the tail distributions of $a$:
\begin{center}
$
p_{\text{less}}^{(b)} = P(X_b < k_b) = P(N - X_a < N - k_a) = P(X_a > k_a) = p_{\text{greater}}^{(a)}
$
\end{center}

\begin{center}
$
p_{\text{greater}}^{(b)} = P(X_b > k_b) = P(N - X_a > N - k_a) = P(X_a < k_a) = p_{\text{less}}^{(a)}
$
\end{center}

Since $d_i = p_{\text{less}} - p_{\text{greater}}$, for each bill:

\begin{center}
$
d_i^{(b)} = p_{\text{less}}^{(b)} - p_{\text{greater}}^{(b)} = p_{\text{greater}}^{(a)} - p_{\text{less}}^{(a)} = -\,d_i^{(a)}
$
\end{center}

This holds per bill, so it also holds for the mean:

\begin{center}
$
\Delta_b = \frac{1}{B}\sum_i d_i^{(b)} = -\frac{1}{B}\sum_i d_i^{(a)} = -\Delta_a
$
\end{center}

The same holds for every possible count, $d_i^{(b)}(N - k) = -\,d_i^{(a)}(k)$, so the distributions of $d$ are mirror images of each other: $\rho^{(b)}_{\texttt{obs}}(x) = \rho^{(a)}_{\texttt{obs}}(-x)$ and $\rho^{(b)}_{\texttt{null}}(x) = \rho^{(a)}_{\texttt{null}}(-x)$.

Since the bootstrap distribution is obtained by resampling and averaging the same $d_i$ values, the confidence intervals flip sign:

\begin{center}
$
\text{CI}_b = \big(-\text{CI}_a^{\text{upper}},\ -\text{CI}_a^{\text{lower}}\big)
$
\end{center}

This flip is exact in our implementation, which uses the same bootstrap resamples for both classes (a fixed random seed and the same bills in the same order).

\subsection{Representativity deviation by coalition size}\label{si:delta_tables}

Supplementary Tables~\ref{tab:house_delta_d}--\ref{tab:senate_delta_f} report the values of $\Delta$ behind the curves of main-text Fig.~2 and of its Senate counterpart, for party ($\Delta_D$, Democrats as the target class) and gender ($\Delta_F$, women as the target class), with all Congresses pooled.
Rows are the log-spaced size bins of the figures (sizes count the sponsor), and the two blocks of columns correspond to the two sponsor groups.
For each group and bin, we report the number of bills, $\Delta$ with its 95\% bootstrap confidence interval (1000 resamples of the bills), and the range $\Delta$ can attain for that set of bills: $\Delta(\min)$, its value if no bill had a cosponsor of the target class, and $\Delta(\max)$, its value if every bill had as many as possible.
This range, rather than $[-1, 1]$, is the reference for reading $\Delta$.
It is close to $[-1, 1]$ once bills have a few dozen members, but narrower for small bills, and asymmetric when the target class is a minority: for House bills of 2--4 members sponsored by women, $\Delta_F$ can only lie between $-0.31$ and $0.88$ (Supplementary Section~\ref{si:delta_bounds}).
A value of $\Delta$ near one end of its range therefore indicates near-maximal over- or under-representation for bills of that size, even when it is well below 1 in absolute value.

\begin{table*}[htbp]
\centering
\caption{House: representativity deviation of Democrats ($\Delta_D$) by coalition size for bills sponsored by Democrats and by Republicans (Congresses 93--118).}
\label{tab:house_delta_d}

\setlength{\tabcolsep}{3pt} %
\small

\begin{tabular}{l|rrrrrr|rrrrrr}
\toprule
 & \multicolumn{6}{c}{Democrat Sponsor} &
   \multicolumn{6}{c}{Republican Sponsor} \\
Size Bin
& \# bills & $\Delta$ & CI (low) & CI (high) & $\Delta(\min)$ & $\Delta(\max)$
& \# bills & $\Delta$ & CI (low) & CI (high) & $\Delta(\min)$ & $\Delta(\max)$ \\
\midrule
{[2--4]} & 25598 & 0.127 & 0.120 & 0.133 & -0.686 & 0.580 & 18733 & -0.168 & -0.176 & -0.161 & -0.672 & 0.618 \\
{[5--7]} & 9723 & 0.474 & 0.463 & 0.486 & -0.970 & 0.935 & 8212 & -0.431 & -0.445 & -0.418 & -0.964 & 0.948 \\
{[8--12]} & 10202 & 0.611 & 0.601 & 0.622 & -0.998 & 0.990 & 7811 & -0.482 & -0.497 & -0.467 & -0.997 & 0.993 \\
{[13--22]} & 12557 & 0.717 & 0.708 & 0.726 & -1.000 & 0.999 & 8590 & -0.504 & -0.519 & -0.490 & -1.000 & 1.000 \\
{[23--40]} & 11069 & 0.755 & 0.747 & 0.765 & -1.000 & 1.000 & 7126 & -0.500 & -0.516 & -0.483 & -1.000 & 1.000 \\
{[41--72]} & 5873 & 0.843 & 0.832 & 0.853 & -1.000 & 1.000 & 3966 & -0.474 & -0.497 & -0.450 & -1.000 & 1.000 \\
{[73--131]} & 2870 & 0.831 & 0.815 & 0.849 & -1.000 & 1.000 & 1990 & -0.455 & -0.490 & -0.419 & -1.000 & 1.000 \\
{[132--238]} & 1622 & 0.792 & 0.766 & 0.819 & -1.000 & 1.000 & 1087 & -0.385 & -0.438 & -0.332 & -1.000 & 1.000 \\
{[239--433]} & 277 & 0.481 & 0.385 & 0.570 & -1.000 & 1.000 & 242 & -0.250 & -0.371 & -0.138 & -1.000 & 1.000 \\
\bottomrule
\end{tabular}
\end{table*}
 
\begin{table*}[htbp]
\centering
\caption{House: representativity deviation of women ($\Delta_F$) by coalition size for bills sponsored by women and by men (Congresses 93--118).}
\label{tab:house_delta_f}

\setlength{\tabcolsep}{3pt} %
\small

\begin{tabular}{l|rrrrrr|rrrrrr}
\toprule
 & \multicolumn{6}{c}{Female Sponsor} &
   \multicolumn{6}{c}{Male Sponsor} \\
Size Bin
& \# bills & $\Delta$ & CI (low) & CI (high) & $\Delta(\min)$ & $\Delta(\max)$
& \# bills & $\Delta$ & CI (low) & CI (high) & $\Delta(\min)$ & $\Delta(\max)$ \\
\midrule
{[2--4]} & 6186 & 0.080 & 0.068 & 0.092 & -0.312 & 0.883 & 38145 & -0.006 & -0.009 & -0.002 & -0.191 & 0.929 \\
{[5--7]} & 2772 & 0.187 & 0.165 & 0.209 & -0.648 & 0.999 & 15163 & -0.024 & -0.031 & -0.015 & -0.471 & 0.999 \\
{[8--12]} & 2951 & 0.265 & 0.245 & 0.286 & -0.821 & 1.000 & 15062 & 0.005 & -0.003 & 0.013 & -0.591 & 1.000 \\
{[13--22]} & 3272 & 0.418 & 0.398 & 0.438 & -0.913 & 1.000 & 17875 & 0.048 & 0.039 & 0.057 & -0.686 & 1.000 \\
{[23--40]} & 3011 & 0.482 & 0.459 & 0.505 & -0.959 & 1.000 & 15184 & 0.093 & 0.083 & 0.102 & -0.780 & 1.000 \\
{[41--72]} & 1916 & 0.653 & 0.626 & 0.679 & -0.989 & 1.000 & 7923 & 0.131 & 0.116 & 0.147 & -0.956 & 1.000 \\
{[73--131]} & 978 & 0.619 & 0.575 & 0.658 & -0.998 & 1.000 & 3882 & 0.139 & 0.117 & 0.163 & -0.991 & 1.000 \\
{[132--238]} & 444 & 0.606 & 0.541 & 0.667 & -1.000 & 1.000 & 2265 & 0.166 & 0.132 & 0.196 & -0.999 & 1.000 \\
{[239--433]} & 86 & 0.399 & 0.236 & 0.560 & -1.000 & 1.000 & 433 & 0.114 & 0.044 & 0.179 & -1.000 & 0.999 \\
\bottomrule
\end{tabular}
\end{table*}
 
\begin{table*}[htbp]
\centering
\caption{Senate: representativity deviation of Democrats ($\Delta_D$) by coalition size for bills sponsored by Democrats and by Republicans (Congresses 93--118).}
\label{tab:senate_delta_d}

\setlength{\tabcolsep}{3pt} %
\small

\begin{tabular}{l|rrrrrr|rrrrrr}
\toprule
 & \multicolumn{6}{c}{Democrat Sponsor} &
   \multicolumn{6}{c}{Republican Sponsor} \\
Size Bin
& \# bills & $\Delta$ & CI (low) & CI (high) & $\Delta(\min)$ & $\Delta(\max)$
& \# bills & $\Delta$ & CI (low) & CI (high) & $\Delta(\min)$ & $\Delta(\max)$ \\
\midrule
{[2--4]} & 21883 & 0.103 & 0.096 & 0.110 & -0.628 & 0.602 & 16540 & -0.054 & -0.062 & -0.045 & -0.624 & 0.616 \\
{[5--10]} & 8672 & 0.473 & 0.461 & 0.485 & -0.975 & 0.969 & 7200 & -0.292 & -0.308 & -0.276 & -0.973 & 0.972 \\
{[11--21]} & 4297 & 0.611 & 0.595 & 0.629 & -1.000 & 1.000 & 3896 & -0.348 & -0.369 & -0.324 & -1.000 & 1.000 \\
{[22--46]} & 1956 & 0.653 & 0.626 & 0.678 & -1.000 & 1.000 & 2324 & -0.302 & -0.333 & -0.270 & -1.000 & 1.000 \\
{[47--100]} & 731 & 0.498 & 0.456 & 0.539 & -1.000 & 0.995 & 751 & -0.184 & -0.239 & -0.132 & -1.000 & 0.985 \\
\bottomrule
\end{tabular}
\end{table*}
 
\begin{table*}[htbp]
\centering
\caption{Senate: representativity deviation of women ($\Delta_F$) by coalition size for bills sponsored by women and by men (Congresses 93--118).}
\label{tab:senate_delta_f}

\setlength{\tabcolsep}{3pt} %
\small

\begin{tabular}{l|rrrrrr|rrrrrr}
\toprule
 & \multicolumn{6}{c}{Female Sponsor} &
   \multicolumn{6}{c}{Male Sponsor} \\
Size Bin
& \# bills & $\Delta$ & CI (low) & CI (high) & $\Delta(\min)$ & $\Delta(\max)$
& \# bills & $\Delta$ & CI (low) & CI (high) & $\Delta(\min)$ & $\Delta(\max)$ \\
\midrule
{[2--4]} & 5482 & 0.135 & 0.122 & 0.148 & -0.279 & 0.867 & 32941 & 0.000 & -0.003 & 0.004 & -0.182 & 0.831 \\
{[5--10]} & 2161 & 0.278 & 0.256 & 0.304 & -0.685 & 0.995 & 13711 & 0.019 & 0.012 & 0.027 & -0.468 & 0.929 \\
{[11--21]} & 1009 & 0.357 & 0.320 & 0.393 & -0.898 & 0.991 & 7184 & 0.044 & 0.033 & 0.057 & -0.602 & 0.901 \\
{[22--46]} & 337 & 0.258 & 0.180 & 0.331 & -0.927 & 0.958 & 3943 & -0.036 & -0.053 & -0.018 & -0.643 & 0.836 \\
{[47--100]} & 56 & 0.019 & -0.163 & 0.217 & -0.975 & 0.942 & 1426 & 0.001 & -0.029 & 0.029 & -0.852 & 0.758 \\
\bottomrule
\end{tabular}
\end{table*}

\clearpage

\subsection{Pooled Counts Distributions}

In addition to the empirical and null distributions of $d_i$'s, we show the pooled distributions of hypergeometric counts in the empirical case and in the null mixture. Each panel compares the share of bills with exactly $k$ cosponsors of class $\mathcal{C}$, $q_{\texttt{obs}}(k) = \frac{1}{B}\sum_{i} \mathbb{I}(k_i^{obs} = k)$, with the average of the bills' hypergeometric null distributions, $q_{\texttt{null}}(k) = \frac{1}{B}\sum_{i} P(X_i = k)$, a probability distribution with the same argument as in Supplementary Section~\ref{si:null}. Because these curves pool bills of different sizes and Congresses before comparing them, their overlap is not a measure of $\Delta$, which compares each bill with its own null distribution; however, they account for the global pattern exhibited in coalition formation.

\begin{figure}[!ht]
\centering
\includegraphics[width=0.85\linewidth]{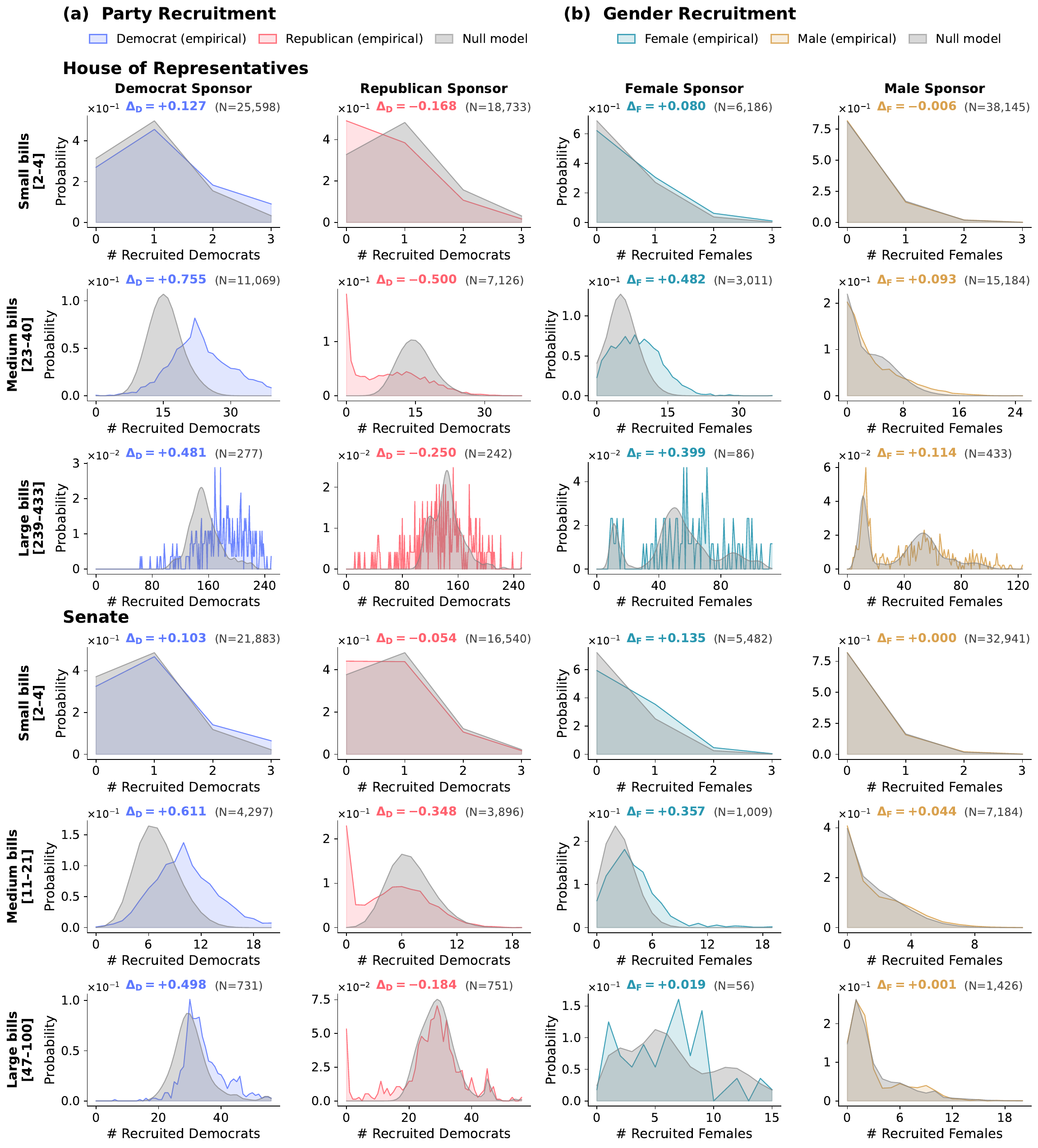}
\caption{\textbf{House \& Senate: pooled count distributions.} House (top block) and Senate (bottom block); rows, the smallest, middle and largest log-spaced size bins of main-text Fig. 2 (sizes include the sponsor). \textbf{(a)}~Party: number of Democratic cosponsors, in bills sponsored by Democrats and by Republicans. \textbf{(b)}~Gender: number of female cosponsors, in bills sponsored by women and by men. Each panel shows the share of bills with exactly $k$ cosponsors of the target class, $q_{\texttt{obs}}(k)$ (coloured, empirical), and the average of the bills' hypergeometric null distributions, $q_{\texttt{null}}(k)$ (grey). Above each panel, the representativity deviation $\Delta$ of those bills and their number $N$; $\Delta$ compares each bill with its own null and is not the overlap of the two curves.}
\label{fig:suppmat_pooled_counts}
\end{figure}

\subsection{Party and gender of the sponsor jointly}

\begin{figure}[!ht]
\centering
\includegraphics[width=\linewidth]{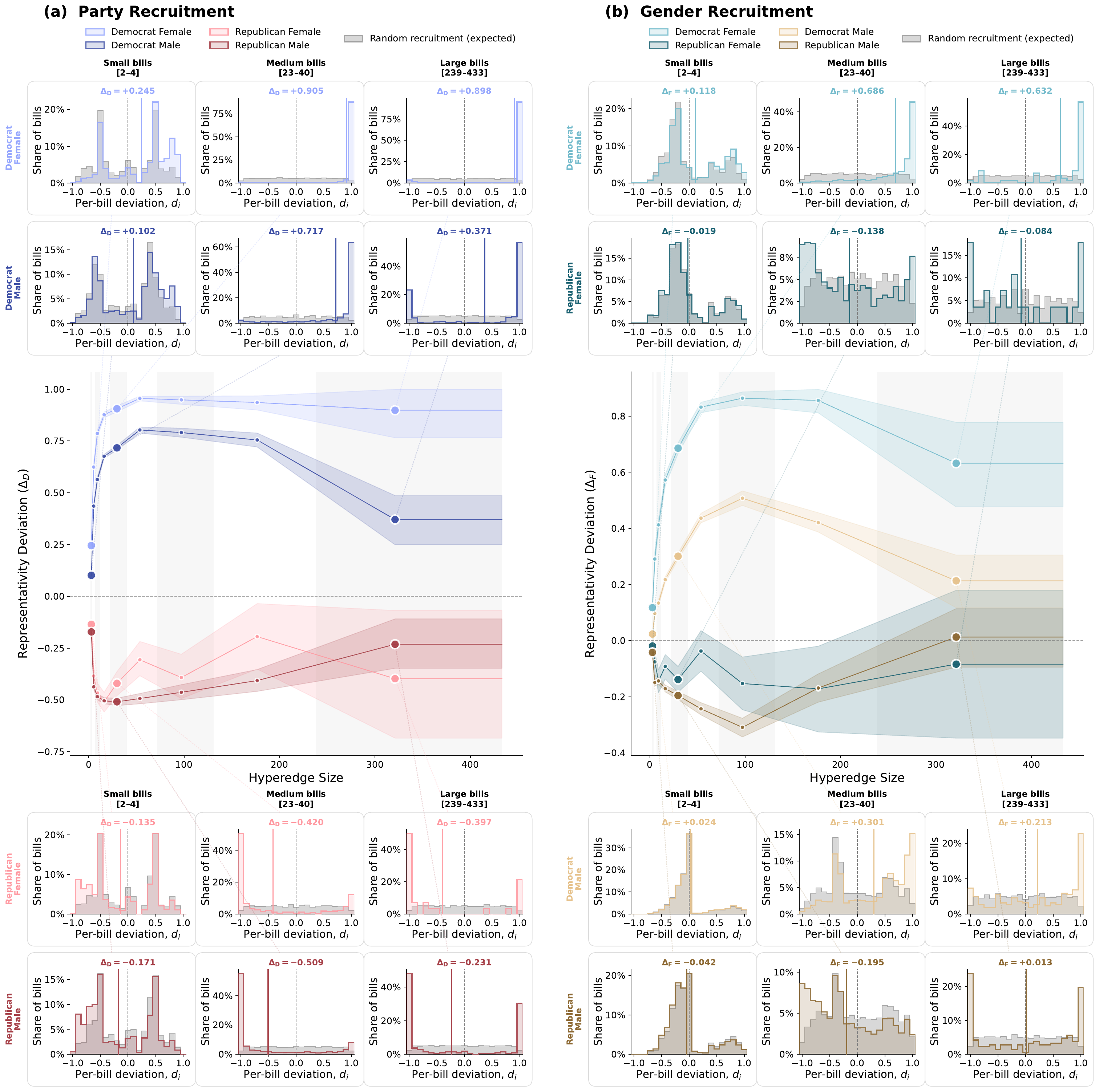}
\caption{\textbf{House: party and gender composition by the party--gender class of the sponsor.} Bills sponsored by Democratic women, Democratic men, Republican women and Republican men; sizes include the sponsor and are divided into the log-spaced bins of main-text Fig. 2.
\textbf{(a)}~Representativity deviation of Democrats ($\Delta_D$) across bill sizes, one curve per sponsor class (\textit{centre}); bands are $95\%$ bootstrap confidence intervals (1000 resamples of the bills in each bin). Above (Democratic sponsors) and below (Republican sponsors), for the smallest, middle and largest bins: the empirical distribution of the bills' per-bill deviations $d_i$ (coloured) and the distribution expected under random recruitment for the same bills (grey); solid and dashed vertical lines mark their means, $\Delta_D$ and $0$. \textbf{(b)}~The same for the representativity deviation of women ($\Delta_F$), with female sponsors above and male sponsors below. Colours as in the legends.}
\label{fig:suppmat_homophily_cross_labels_house}
\end{figure}

\begin{figure}[!ht]
\centering
\includegraphics[width=\linewidth]{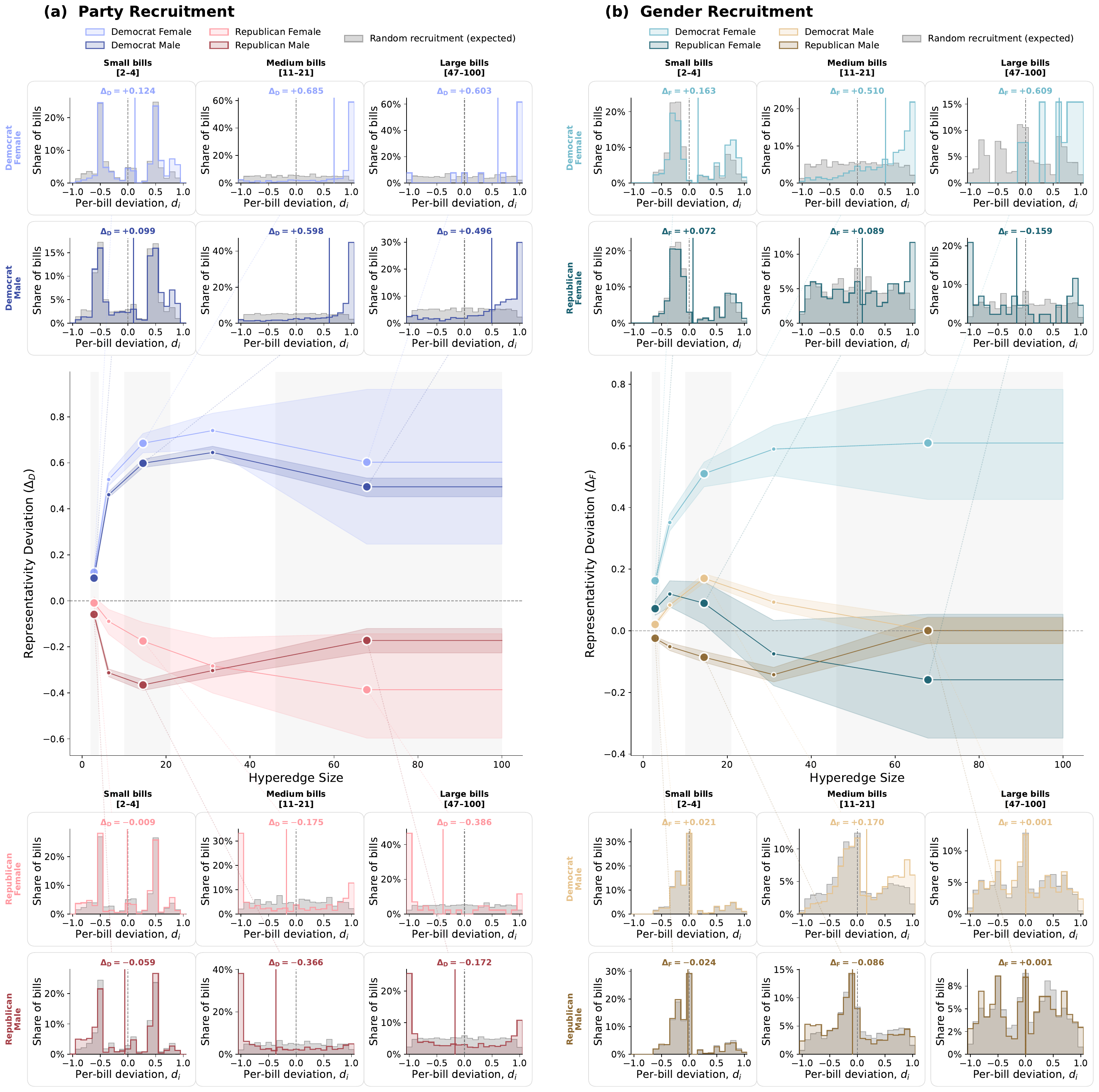}
\caption{\textbf{Senate: party and gender composition by the party--gender class of the sponsor.} As Supplementary Fig.~\ref{fig:suppmat_homophily_cross_labels_house}, for the Senate.}
\label{fig:suppmat_homophily_cross_labels_senate}
\end{figure}

Supplementary Figs.~\ref{fig:suppmat_homophily_cross_labels_house} and~\ref{fig:suppmat_homophily_cross_labels_senate} split the composition results by the sponsor's party and gender. In the House, the over-representation of women is driven by coalitions led by Democratic women: for coalitions of 23--40 members, $\Delta_F=+0.69$ under Democratic women and $+0.30$ under Democratic men, but $-0.14$ under Republican women and $-0.20$ under Republican men (Supplementary Fig.~\ref{fig:suppmat_homophily_cross_labels_house}\textbf{b}). The Senate shows the same ordering (Supplementary Fig.~\ref{fig:suppmat_homophily_cross_labels_senate}\textbf{b}): for coalitions of 11--21 members, $\Delta_F=+0.51$ under Democratic women and $+0.17$ under Democratic men, against $+0.09$ under Republican women and $-0.09$ under Republican men, with wider confidence intervals for female sponsors, who are few in the Senate. Within each party, coalitions led by women include more women than those led by men; in the House the difference is much larger among Democrats, while in the Senate it appears in both parties. In both chambers, coalitions led by Democratic women are also the most Democratic (Supplementary Figs.~\ref{fig:suppmat_homophily_cross_labels_house}\textbf{a} and~\ref{fig:suppmat_homophily_cross_labels_senate}\textbf{a}). These are the patterns summarized in the main text. Together with the absence of same-gender matching at joining time (main-text Fig.~5f), they suggest that the gender composition of coalitions follows party, with the sponsor's gender adding to it mainly among Democrats.

\subsection{Coalition size by gender and party across terms}

\begin{figure}[!ht]
\centering
\includegraphics[width=0.63\linewidth]{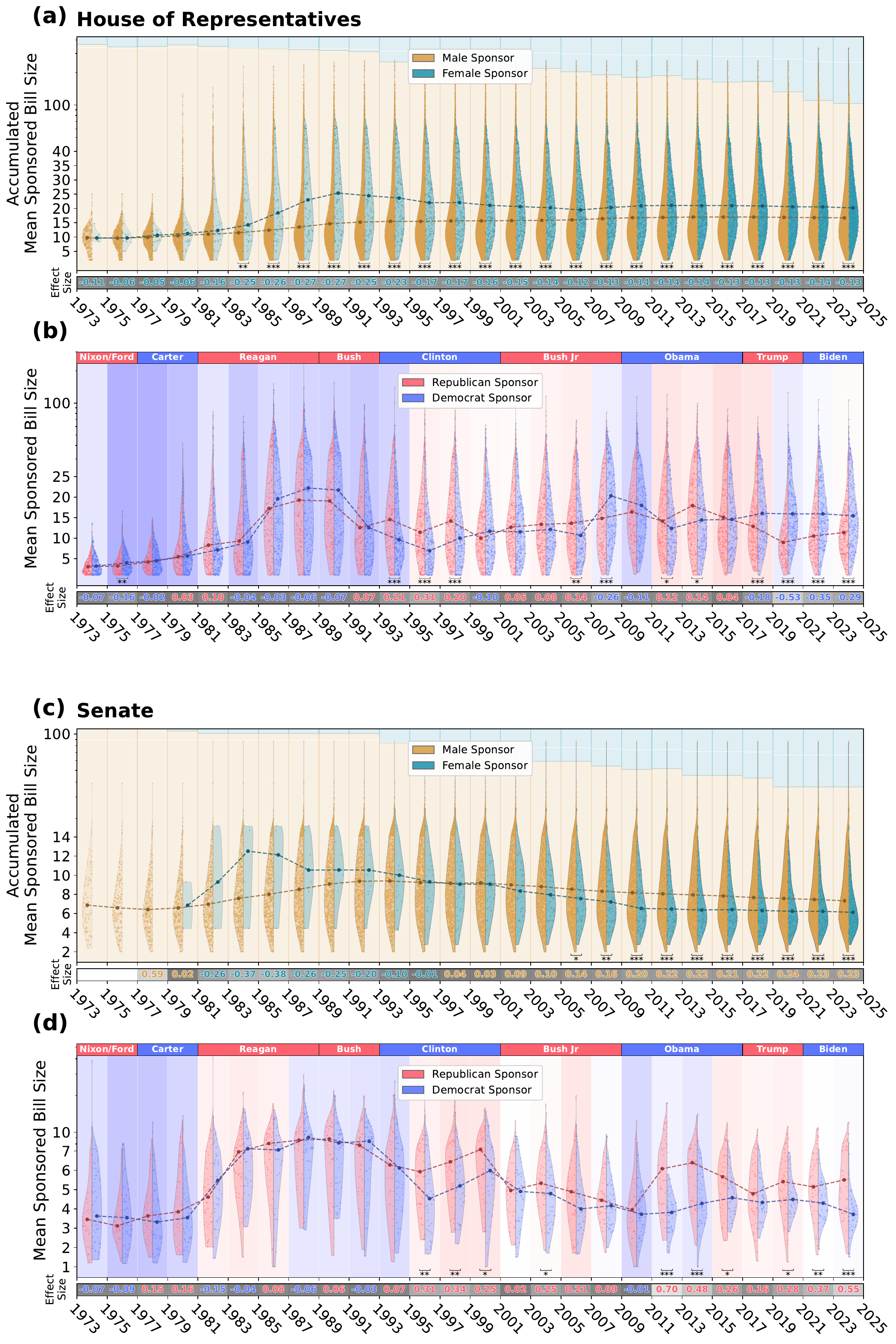}
\caption{\textbf{House \& Senate: mean coalition size of sponsored bills by gender and party.}
\label{fig:suppmat_mean_bill_sizes_across_terms}
\textbf{(a,c)} Distributions of mean size of sponsored bills by each legislator, male or female. Since women's representation increases over time, differences between male and female legislators are evaluated using accumulated distributions that include all Congresses up to and including the current one. One point in each distribution refers to a legislator in a specific term. Background color is divided according to the chamber gender composition. Dashed lines represent the evolution of the medians. Significance levels are obtained using the Mann--Whitney U test (asterisks denote statistical significance, with $^*p<0.05$, $^{**}p<0.01$, and $^{***}p<0.001$). Effect size (measured through Cliff's delta) is reported at the bottom of each term column; positive values indicate larger mean coalitions for bills sponsored by men, negative values for bills sponsored by women.
\textbf{(b,d)} Distributions of mean size of sponsored bills by each legislator, Republican and Democrat. Dashed lines represent the evolution of the medians. One point in each distribution refers to a legislator in that term. Background colors indicate the majority party in each Congress, with color intensity proportional to the majority margin. Significance levels and effect sizes are measured the same as in the panel above, with positive values indicating larger mean coalitions for bills sponsored by Republicans and negative values for bills sponsored by Democrats. An oscillating pattern is observed in which, for many terms, the party holding the majority has the higher median.}
\end{figure}

Supplementary Fig.~\ref{fig:suppmat_mean_bill_sizes_across_terms} describes sponsors rather than coalitions, showing the mean size of the coalitions they lead, by gender and party, Congress by Congress. In the House, women sponsor bills with larger mean coalition size than men from the 1980s onward (Supplementary Fig.~\ref{fig:suppmat_mean_bill_sizes_across_terms}\textbf{a}), in line with evidence that women engage more in coalition building~\cite{jackie_jill_2011, when_women_2013}. Differences between the parties are not completely explained but seem to be mostly driven by majority status (Supplementary Fig.~\ref{fig:suppmat_mean_bill_sizes_across_terms}\textbf{b,d}).

This section supports the main-text section ``Coalitions are built from recurring groups'' (main-text Fig.~3). It discusses the two legislators used as examples, the Senate results, and the checks of the recurrence analysis.

\subsection{Coalition builders and solo sponsors}\label{si:builders}

Main-text Fig.~3\textbf{c} compares legislators' singleton sponsor degree, i.e., the number of bills introduced without cosponsors, with their higher-order sponsor degree, i.e., the number of bills introduced with at least one cosponsor, throughout Congresses. This distinction separates legislators who are predominantly coalition builders from those who act as ``lone wolves.'' Coalition builders may consistently and deliberately pursue legislative agendas that overlap with those of their colleagues or actively seek peer support for their bills, whereas ``lone wolf'' behavior may reflect either difficulty in attracting peer support or a deliberate preference for signaling political stances without building broader agreement. These contrasting profiles are well exemplified by Representative Carolyn B. Maloney (D) and Representative Robert A. Roe (D). Maloney, who served in the House from 1993 to 2023, introduced almost 800 bills with at least one cosponsor, compared with fewer than 200 bills with no cosponsors. Roe, by contrast, served from 1969 to 1993 and introduced more than 1000 bills without cosponsors, compared with more than 200 bills with at least one other sponsor. Part of this contrast reflects era: Roe's tenure overlaps the period in which singleton bills dominated (main-text Fig.~1\textbf{c}) and members adapted slowly to cosponsorship~\cite{koger_position_2003}. Both legislators display clear, consistent trends of evolution throughout the Congresses in which they served, remaining within the same characteristic region from term to term. This persistence does not appear to map straightforwardly onto formal leadership status: neither legislator occupied an especially low or high position in their party's chamber hierarchy, and both held their most prominent institutional roles as chairs or co-chairs of House committees. These examples therefore suggest that legislators' propensity to gather support for their bills may reflect stable individual legislative strategies or behavioral tendencies, rather than being determined solely by their formal institutional roles.

\subsection{Senate}

\begin{figure}[!ht]
\centering
\includegraphics[width=\linewidth]{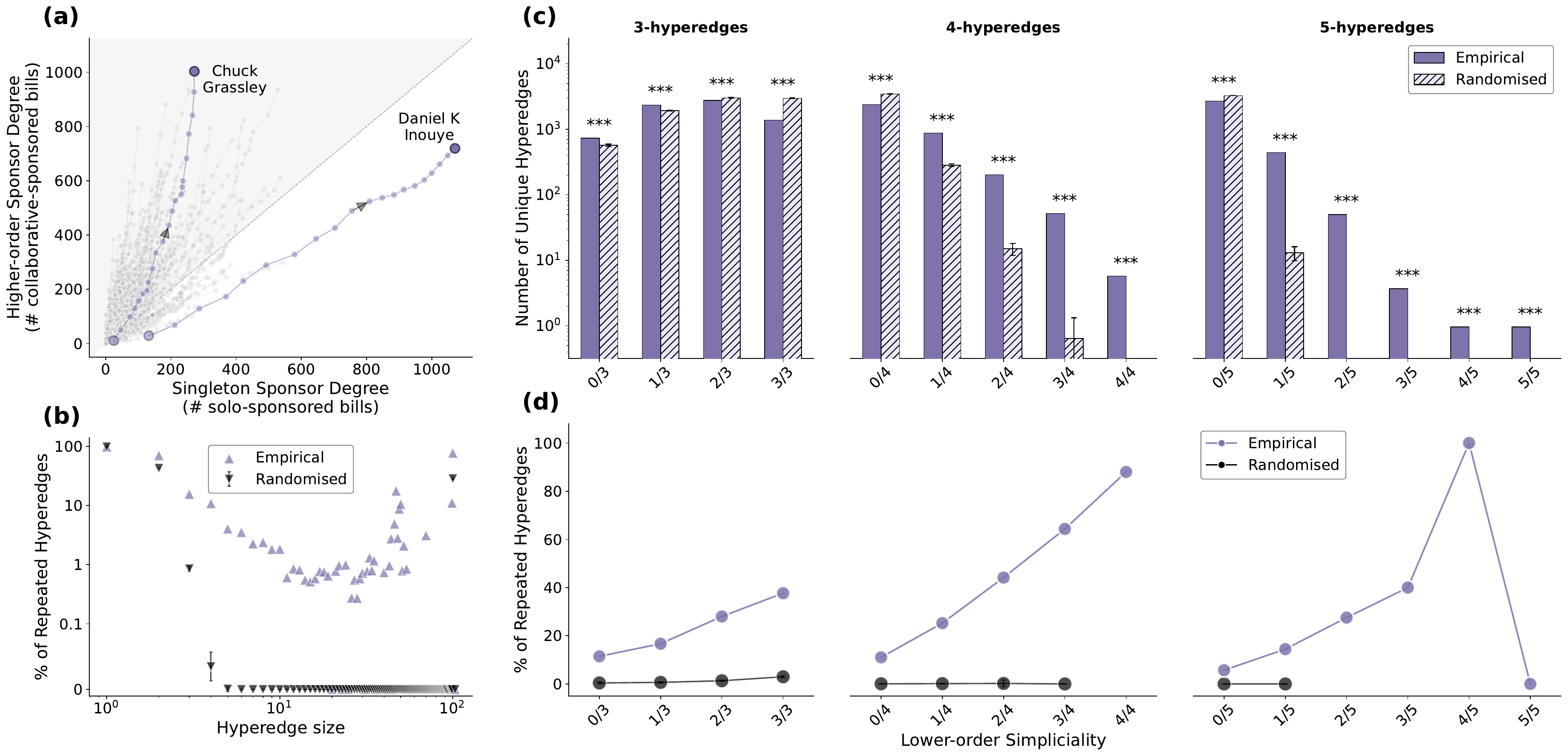}
\caption{\textbf{Senate: structural properties of cosponsorship hypergraphs.} All panels report on the aggregated hypergraph of Senate bills from the 93rd to 118th Congresses. \textbf{(a)} Comparing a legislator's singleton degree (size-1 bills) against their higher-order degree (bills with $>1$ sponsors) accumulated throughout the Congresses in which they participate shows that most senators ($77.9\%$) predominantly engage in multi-sponsor coalitions in their last Congress recorded; Chuck Grassley and Daniel K. Inouye are highlighted. Each legislator's curve serves as a proxy for how their preferences for collaborative and individual legislative work change throughout their participation in Congress. \textbf{(b)} Share of hyperedges of each size whose complete sponsoring set also sponsors another bill. Repetition decays with size and rises again among the largest coalitions; randomizing the hypergraph eliminates it for all sizes above three. \textbf{(c)} Counts of unique hyperedges of size 3, 4, and 5 (left to right) by lower-order simpliciality (\textit{x-axis} icons), compared with a randomized configuration model that preserves degree and hyperedge size distributions (hashed bars; error bars indicate one standard deviation). For sizes 4 and 5 the empirical data show an excess of highly simplicial coalitions; for size 3, fully simplicial triads are less common than in the null. Asterisks indicate empirical p-values from comparison to the randomized null $(*p < 0.05, **p < 0.01, ***p < 0.001)$. \textbf{(d)} Share of repeated hyperedges by lower-order simpliciality: hyperedges whose subgroups have sponsored bills of their own are more likely to be repeated. Purple curves: empirical hypergraph; black curves with error bars (almost indiscernible): mean and standard deviation over the randomized hypergraphs.}
\label{fig:suppmat_simpliciality_multiplicity_senate}
\end{figure}

In the Senate (Supplementary Fig.~\ref{fig:suppmat_simpliciality_multiplicity_senate}), most senators also sponsor more bills with cosponsors than without ($77.9\%$), and the share of repeated coalitions again rises with lower-order simpliciality (panel \textbf{d}), as in the House. Two differences stand out. Exact repetition rises again among the largest Senate coalitions (panel \textbf{b}), and three-member coalitions whose three pairs all appear as two-person bills are less common than in the randomized null (panel \textbf{c}), whereas for four- and five-member coalitions highly simplicial groups are over-represented, as in the House. The first difference may reflect resolutions supported by most of the chamber and introduced more than once; the second may reflect the small size of the Senate, in which pairs of senators often co-appear on two-person bills even in the randomized null.

\clearpage
\section{Timing within a Congress}\label{si:timing}

This section supports the main-text section ``Large coalitions are introduced early in a Congress'' (main-text Fig.~4).

\subsection{Senate}

\begin{figure}[!ht]
\centering
\includegraphics[width=\linewidth]{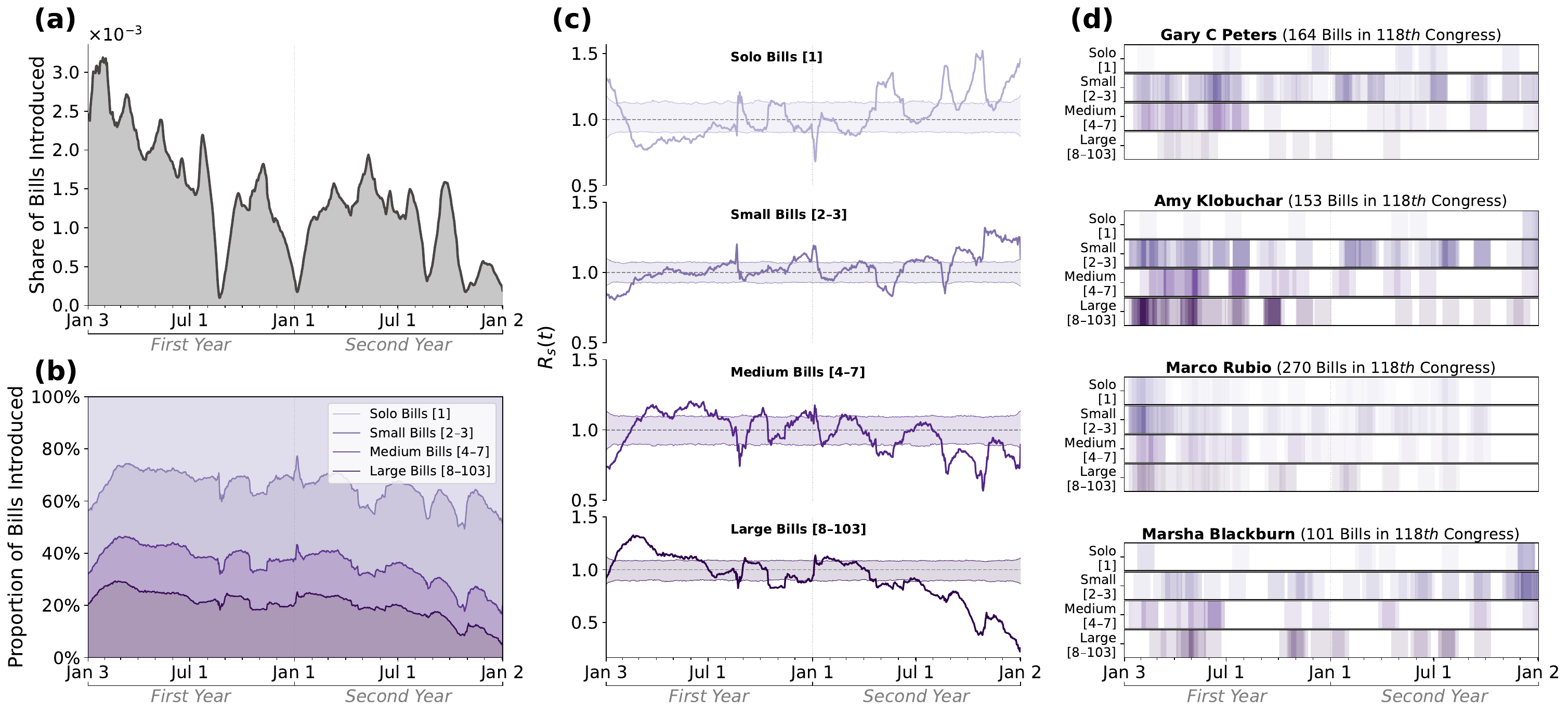}
\caption{\textbf{Senate: temporal evolution of legislative activity over a congressional term.} As in main-text Fig.~4, daily activity is aggregated across all congressional terms and smoothed with a 30-day rolling window. \textbf{(a)} Share of bills introduced on each day across all Congresses; most bills are introduced early in the term. \textbf{(b)} Proportion of bills introduced each day by bill-size category: solo (1), small (2--3), medium (4--7) and large (8+). \textbf{(c)} Relative daily prevalence of each size category over a standardized congressional term, measured as the ratio between the daily proportion of bills in a category and that category's overall proportion across the term; values above (below) $1$ indicate over-representation (under-representation). Shaded areas are the $95\%$ confidence intervals obtained by shuffling introduction dates among hyperedges $1000$ times. \textbf{(d)} Relative daily activity of individual senators by bill-size category. Highlighted are the most prolific bill introducers in the 118th Congress among Democratic men (Gary C. Peters, \textit{top}), Democratic women (Amy Klobuchar, \textit{middle top}), Republican men (Marco Rubio, \textit{middle bottom}) and Republican women (Marsha Blackburn, \textit{bottom}).}
\label{fig:suppmat_temporality_senate}
\end{figure}

The Senate (Supplementary Fig.~\ref{fig:suppmat_temporality_senate}) shows the same calendar as the House: large coalitions are over-represented early in the term and under-represented toward its end, while solo bills are more common at its start and end. Size categories are defined on Senate coalition sizes (2--3, 4--7 and 8 or more senators), which are smaller than in the House.

\clearpage
\section{Recruitment into forming coalitions}\label{si:recruitment}

This section supports the main-text section ``Coalition recruitment carries a social memory of earlier groups'' (main-text Fig.~5) and the main-text Methods on sequential coalition assembly and choice modeling. It defines the assembly-speed statistic and its null, the covariates of the choice models, the sample, the two comparisons of the history blocks, and the validation of the results.

\subsection{Senate}

\begin{figure}[!ht]
\centering
\includegraphics[width=\linewidth]{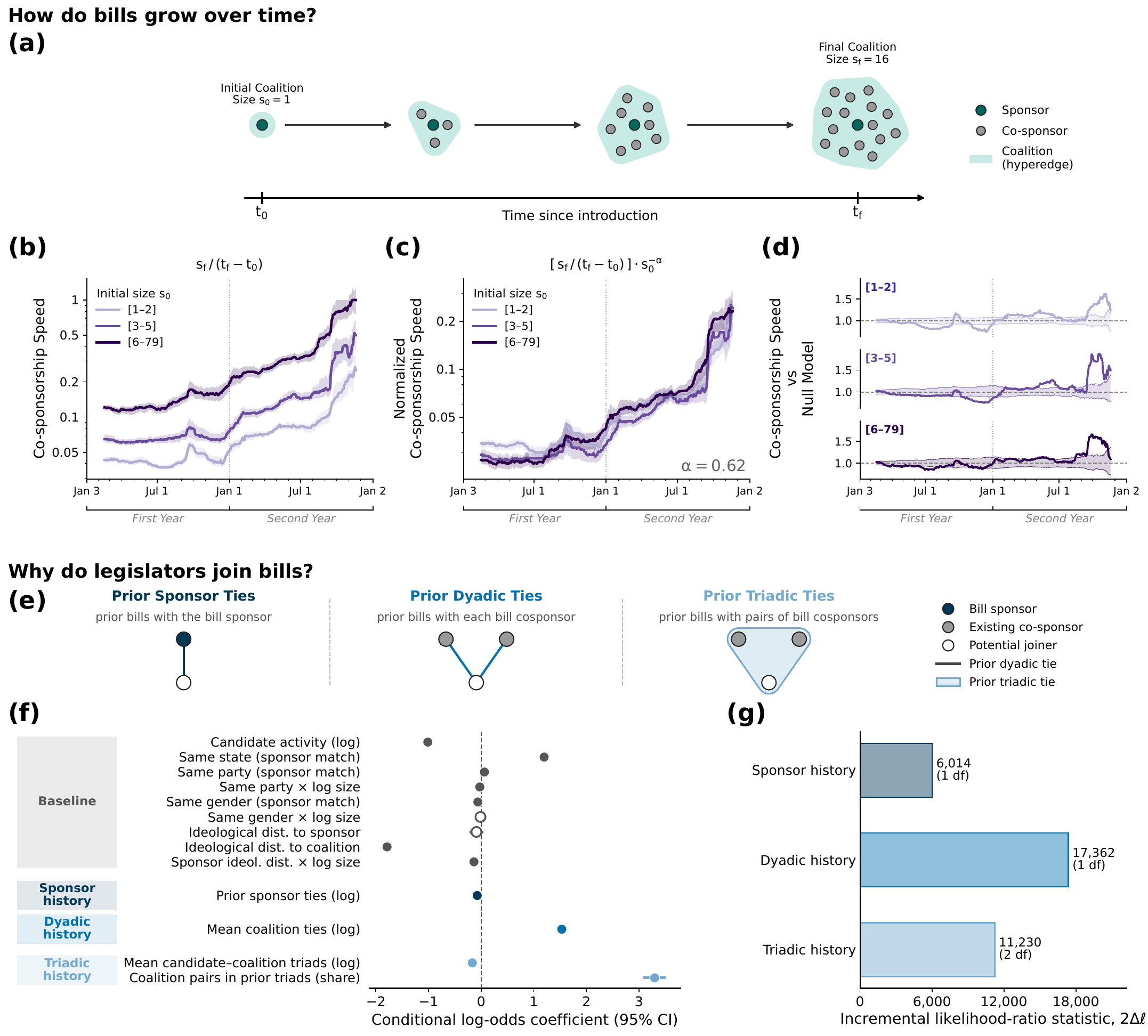}
\caption{\textbf{Senate: coalition assembly and the relational context of recruitment.} Panels as in main-text Fig.~5. (a)~A coalition grows from its initial size $s_0$ at introduction ($t_0$) to its final size $s_f$ ($t_f$). (b)~Median cosponsorship speed $s_f/(t_f-t_0+1)$ of the bills introduced in rolling 90-day windows, by initial-size class; bands, 95\% bootstrap intervals. (c)~The same, with speed multiplied by $s_0^{-\alpha}$. (d)~Observed over null median speed for each initial-size class; band, 2.5--97.5\% of the null. (e)~The three relational histories of a candidate with a forming coalition. (f)~Conditional-logit coefficients of the full model with 95\% confidence intervals. (g)~Likelihood-ratio statistic $2\Delta\ell$ for adding each history block on its own to the baseline model.}
\label{fig:suppmat_recruitment_senate}
\end{figure}

The Senate (Supplementary Fig.~\ref{fig:suppmat_recruitment_senate}) is analyzed exactly as the House in main-text Fig.~5, with the same bills, size classes, null model and choice-model specification. Similarly to the House, we observe a pronounced increase of coalition join speed near the end of the terms across all initial bill sizes (Supplementary Fig.~\ref{fig:suppmat_recruitment_senate}\textbf{d}). Beyond bill growth speed, we also find similar contributions of conditional log-odds coefficients across the features studied, with the coalition pairs in prior triads also being the feature with the largest preponderance.

\subsection{Assembly speed and its null model}\label{si:assembly}

The speed analysis uses the bills with dated cosponsorships (Congresses 97--118; all bill types of the chamber; bills with at least one cosponsor). For a bill introduced on day $t_0$ of its Congress, $t_f$ is the date of its last recorded cosponsorship and $s_f$ its final coalition size, sponsor and original cosponsors included. Cosponsorship speed is $s_f/(t_f-t_0+1)$, in members per day; the $+1$ gives a bill whose coalition is complete on the day of introduction a lifetime of one day. Bills whose whole coalition joined on the day of introduction ($t_f=t_0$) have no assembly phase and are left out. Bills are classified by their initial size $s_0$, the sponsor plus the cosponsors who joined on the day of introduction, into three quantile classes (main-text Fig.~5b--d). For each class and each rolling window of 90 days shifted by one day ($[0,90)$, $[1,91)$, \dots, $[640,730)$), we take the median speed of the bills introduced in the window, and place it at the window's centre.

The exponent $\alpha$ of the normalised speed $s_f/(t_f-t_0+1)\cdot s_0^{-\alpha}$ is chosen to make the size classes as similar as possible: it minimises, over $\alpha\in[-2,2]$, the mean over windows of the variance across classes of the log median normalised speed ($\alpha=0.47$ in the House, $\alpha=0.68$ in the Senate).

The null model of main-text Fig.~5d keeps every bill's size, speed and assembly time, and randomizes when it was introduced. Within each initial-size class, the introduction days are shuffled among the bills, with Congresses pooled on a common two-year calendar, under the constraint that every bill still completes its coalition before the end of the Congress ($t_0+t_f-t_0<730$). Among all such assignments, one is drawn uniformly at random. This preserves each class's distribution of introduction days and of speeds, and removes any association between the two beyond what the constraint imposes. Bills introduced late can only have short assembly times, so the constraint alone makes late windows faster in the null; panel~d therefore shows the observed median relative to the null median (1000 shuffles), and the late-term speed-up is the part not explained by the time left.

\begin{figure}[!ht]
\centering
\includegraphics[width=0.85\linewidth]{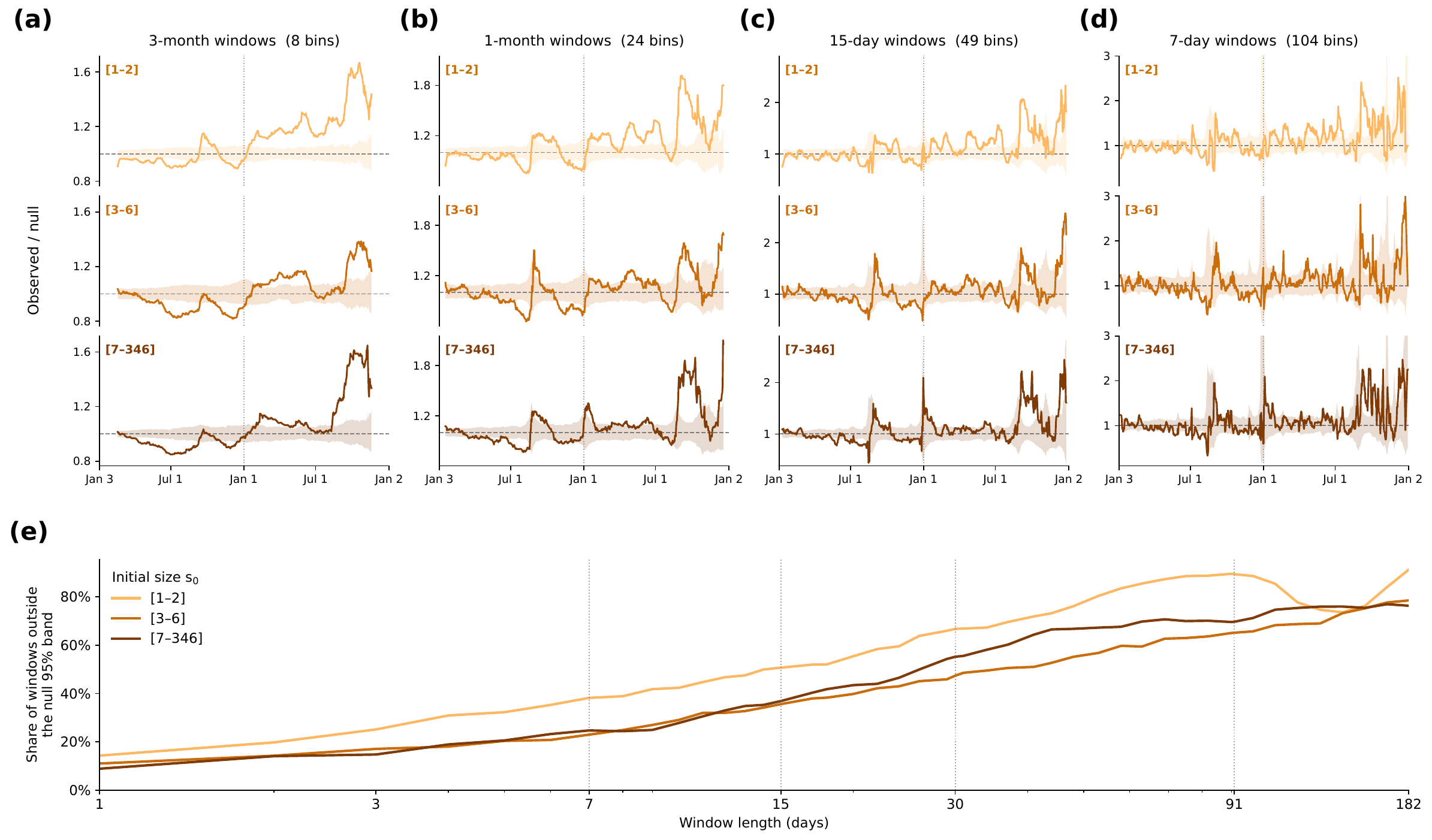}
\caption{\textbf{House: cosponsorship speed against the null model, for several window lengths.} (a)--(d)~Observed over null median speed in rolling windows of 91, 30, 15 and 7 days shifted by one day, for each initial-size class; band, 2.5--97.5\% of 500 shuffles of the null of main-text Fig.~5d. (e)~For every window length from 1 to 182 days, share of windows whose observed median lies outside the null band. Overlapping windows are not independent, so (e) describes the comparison and is not a test.}
\label{fig:suppmat_recruitment_rolling_null_sweep_house}
\end{figure}

\begin{figure}[!ht]
\centering
\includegraphics[width=0.85\linewidth]{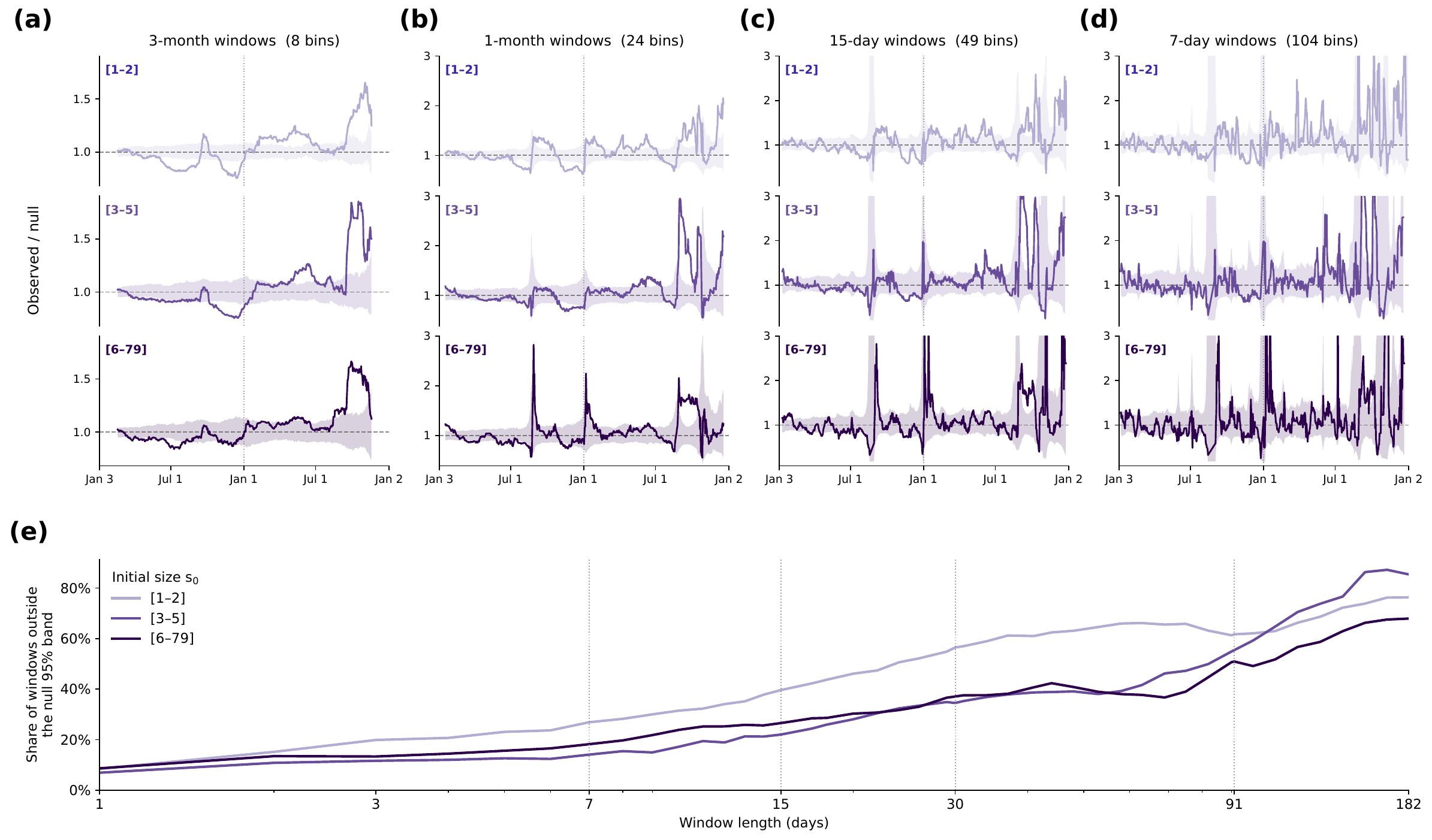}
\caption{\textbf{Senate: cosponsorship speed against the null model, for several window lengths.} As Supplementary Fig.~\ref{fig:suppmat_recruitment_rolling_null_sweep_house}.}
\label{fig:suppmat_recruitment_rolling_null_sweep_senate}
\end{figure}

\begin{figure}[!ht]
\centering
\includegraphics[width=0.9\linewidth]{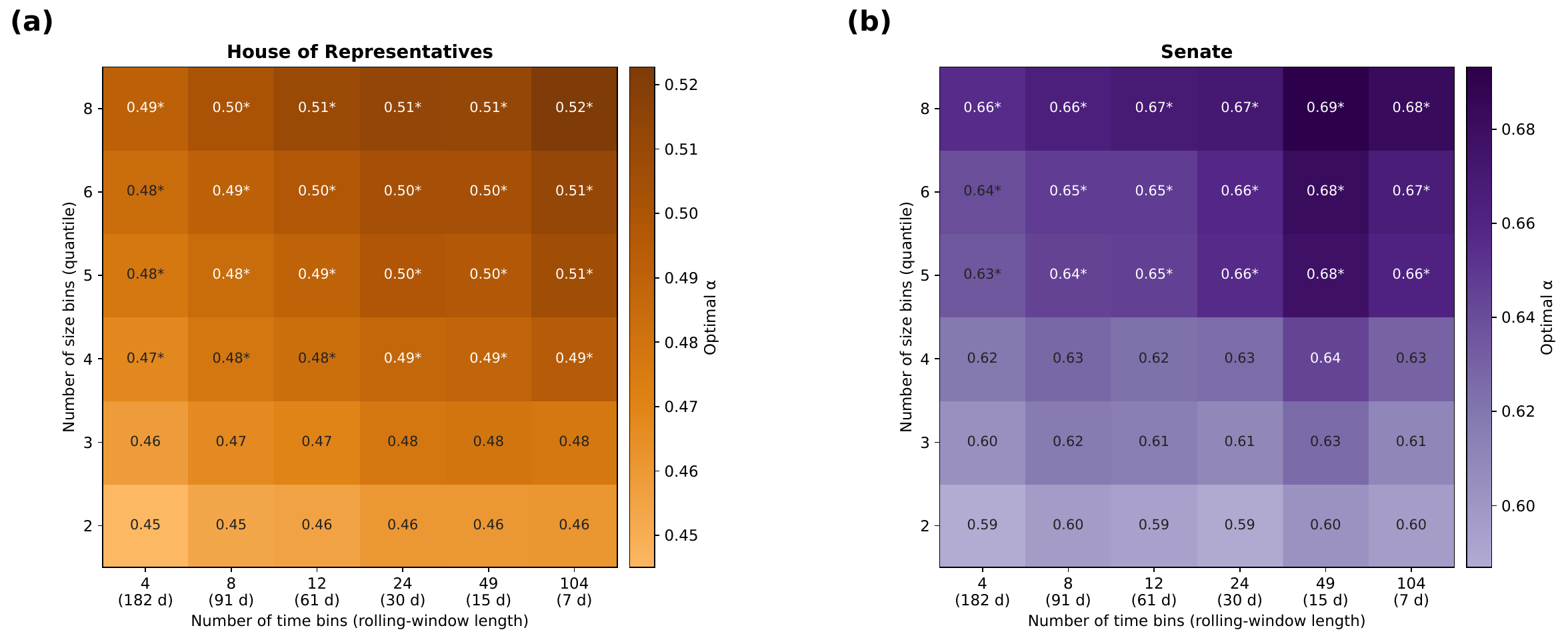}
\caption{\textbf{Normalisation exponent $\alpha$ for different binnings.} (a)~House, (b)~Senate. Each cell is the $\alpha$ that minimises the mean, over rolling windows, of the variance across initial-size classes of the (log median of speed)$\,\cdot\, s_0^{-\alpha}$. Columns: number $n$ of time bins, i.e.\ rolling windows of $730/n$ days (in parentheses); rows: number of quantile classes of $s_0$. An asterisk marks cells in which tied initial sizes left fewer classes than requested.}
\label{fig:suppmat_alpha_matrices}
\end{figure}

Supplementary Figs.~\ref{fig:suppmat_recruitment_rolling_null_sweep_house}--\ref{fig:suppmat_recruitment_rolling_null_sweep_senate} repeat this comparison for windows of 7, 15, 30 and 91 days, and summarise, for every window length from 1 to 182 days, the share of windows whose observed median lies outside the null band. The late-term speed-up appears for windows of two to three months and not for shorter ones. Supplementary Fig.~\ref{fig:suppmat_alpha_matrices} shows how $\alpha$ depends on the number of initial-size classes and on the window length.

\subsection{Covariates}\label{si:features}

Table~\ref{tab:si:features} summarizes the covariates entering the conditional choice models and their definitions. All counts of earlier collaborations are cumulative over every bill of the chamber from the 93rd Congress up to the date $t$ of the event, all bill types included. For bills with dated cosponsorships (Congresses 97--118), a collaboration counts from the date on which the later of the legislators joined the bill; for earlier bills, which have no join dates, every member counts from the introduction date.

\begin{table}[!ht]
\centering
\small
\begin{tabular}{lll}
\toprule
Block & Covariate & Definition (candidate $i$, sponsor $s$, current coalition $S$) \\
\midrule
Baseline & Candidate activity (log) & earlier bill memberships of $i$ \\
 & Same state, same party, same gender & match between $i$ and $s$ \\
 & Same party $\times$ log size; same gender $\times$ log size & matches interacted with $\log(1+|S|)$ \\
 & Ideological distance to sponsor / to coalition & $|\theta_i-\theta_s|$; distance to the coalition's mean \\
 & Sponsor distance $\times$ log size & $|\theta_i-\theta_s|\log(1+|S|)$ \\
Sponsor history & Prior sponsor ties (log) & earlier bills shared by $i$ and $s$ \\
Dyadic history & Mean coalition ties (log) & mean earlier bills shared by $i$ and each $j\in S$ \\
Triadic history & Mean candidate--coalition triads (log) & mean earlier bills shared by $i$ and a pair of $S$ \\
 & Coalition pairs in prior triads (share) & share of pairs of $S$ with at least one such bill \\
\bottomrule
\end{tabular}
\caption{Covariates of the conditional choice models, grouped in the blocks of the nested ladder (main-text Fig.~5f--g). The current coalition $S$ includes the sponsor.}
\label{tab:si:features}
\end{table}

In terms of notation, we denote by $i$ the candidate legislator evaluated at recruitment event $e$; $s$ as the primary sponsor of the bill; $S = S_b(t^-)$ as the current forming coalition immediately before recruitment date $t$: the sponsor, the original cosponsors, and every cosponsor who joined before $t$; $|S|$ as the current coalition size and, finally, $\theta_i$: corresponds to the first-dimension DW-NOMINATE score of legislator $i$ in the Congress of the event (from $-1$, most liberal, to $+1$, most conservative). Legislators without a score in that Congress are given the median score of the chamber in that Congress.

\paragraph{1. Candidate Activity \& Demographic Baseline}.

\begin{itemize}
     \item \textbf{Candidate Activity}: $\log(1 + a_i(t))$, where $a_i(t)$ is the number of bills that candidate $i$ had joined before date $t$.
    \item \textbf{Same State Indicator}: $\mathbb{I}\{\text{State}_i = \text{State}_s\}$
    \item \textbf{Same Party Indicator}: $\mathbb{I}\{\text{Party}_i = \text{Party}_s\}$
    \item \textbf{Same Party $\times$ Log Coalition Size}: $\mathbb{I}\{\text{Party}_i = \text{Party}_s\} \times \log(1 + |S|)$
    \item \textbf{Same Gender Indicator}: $\mathbb{I}\{\text{Gender}_i = \text{Gender}_s\}$
    \item \textbf{Same Gender $\times$ Log Coalition Size}: $\mathbb{I}\{\text{Gender}_i = \text{Gender}_s\} \times \log(1 + |S|)$.
\end{itemize}

The three match indicators are 0 when the attribute is missing for $i$ or $s$.

\paragraph{2. Ideological Proximity Features}

\begin{itemize}
     \item \textbf{Ideological Distance to Sponsor}: $|\theta_i - \theta_s|$
    \item \textbf{Ideological Distance to Coalition}: $\left| \theta_i - \frac{1}{|S|} \sum_{j \in S} \theta_j \right|$
    \item \textbf{Sponsor Ideological Distance $\times$ Log Coalition Size}: $|\theta_i - \theta_s| \times \log(1 + |S|)$
\end{itemize}

\paragraph{3. Dyadic Collaboration Features}

\begin{itemize}
     \item \textbf{Prior Sponsor Ties}: $\log(1 + d_{is}(t))$, where $d_{is}(t)$ is the number of earlier bills on which candidate $i$ and the sponsor $s$ were both members before date $t$, in any role.
    \item \textbf{Mean Coalition Ties}: $\log\left( 1 + \frac{1}{|S|} \sum_{j \in S} d_{ij}(t) \right)$, where $d_{ij}(t)$ is the number of earlier bills on which candidate $i$ and coalition member $j$ were both members before date $t$. Since $S$ includes the sponsor, this mean includes $d_{is}(t)$.
\end{itemize}

\paragraph{4. Higher-Order Memory Features}
\begin{itemize}
     \item \textbf{Log Mean Candidate--Coalition Triads}: $\log\left( 1 + \frac{1}{|P|} \sum_{\{j,k\} \in P} q_{ijk}(t) \right)$, where $q_{ijk}(t)$ is the number of earlier bills on which candidate $i$ and both members of the pair $\{j,k\}$ had all been members before date $t$, and $P$ is the set of pairs of $S$: all $\binom{|S|}{2}$ pairs, or 50 pairs drawn at random (reproducibly, from the bill and the date) when $S$ has more.
    \item \textbf{Coalition Pairs in Prior Triads (Share)}: $\frac{1}{|P|} \sum_{\{j,k\} \in P} \mathbb{I}\{q_{ijk}(t) > 0\}$
\end{itemize}

Both triadic covariates are 0 when $S$ has a single member.

\subsection{Specification and sample}\label{si:recruit_spec}

Choice events are built from the bills with dated cosponsorships (Congresses 97--118; all bill types of the chamber; bills of the 104th House with several sponsors excluded). Each cosponsorship dated after the bill's introduction defines a choice event; cosponsors who joined on the same day form separate events, all evaluated against the coalition present before that day, so no order is imposed among them. The observed joiner is compared with a risk set of 10 legislators drawn uniformly, without replacement, among the members of the chamber in that Congress who were not on the bill and did not join it that day (all of them when fewer remain), so each choice set has up to 11 candidates. Controls are drawn with a fixed seed per Congress.
Models are conditional logits (statsmodels \texttt{ConditionalLogit}) with one stratum per event, estimated by maximum likelihood. Covariates with no variation in the sample are dropped.
The House sample contains 1.459.332 choice events (16.052.652 candidate rows) and the Senate sample 182.700 events (2.009.212 rows), from Congresses 97–118 (Supplementary \Cref{tab:si:recruit_sample} by Congress).

\begin{table*}[htbp]
\centering
\caption{Recruitment sample: choice events, candidate rows (joiner and sampled controls) and bills contributing events, by Congress and chamber.}
\label{tab:si:recruit_sample}

\setlength{\tabcolsep}{3pt} %
\small

\begin{tabular}{l|rrr|rrr}
\toprule
 & \multicolumn{3}{c}{House} &
   \multicolumn{3}{c}{Senate} \\
Congress (years)
& Events & Candidate rows & Bills
& Events & Candidate rows & Bills \\
\midrule
97 (1981--83) & 43{.}916 & 483{.}076 & 2{.}359 & 5{.}786 & 63{.}646 & 944 \\
98 (1983--85) & 50{.}478 & 555{.}258 & 2{.}318 & 6{.}561 & 72{.}153 & 1{.}053 \\
99 (1985--87) & 53{.}813 & 591{.}943 & 2{.}450 & 6{.}689 & 73{.}579 & 1{.}123 \\
100 (1987--89) & 61{.}641 & 678{.}051 & 2{.}613 & 7{.}740 & 85{.}140 & 1{.}168 \\
101 (1989--91) & 68{.}278 & 751{.}058 & 2{.}850 & 8{.}168 & 89{.}848 & 1{.}277 \\
102 (1991--93) & 67{.}849 & 746{.}339 & 2{.}962 & 8{.}620 & 94{.}820 & 1{.}311 \\
103 (1993--95) & 54{.}125 & 595{.}375 & 2{.}611 & 5{.}717 & 62{.}887 & 948 \\
104 (1995--97) & 41{.}690 & 458{.}590 & 2{.}094 & 4{.}347 & 47{.}817 & 813 \\
105 (1997--99) & 56{.}072 & 616{.}792 & 2{.}605 & 5{.}927 & 65{.}197 & 1{.}020 \\
106 (1999--01) & 67{.}945 & 747{.}395 & 3{.}032 & 9{.}402 & 103{.}422 & 1{.}444 \\
107 (2001--03) & 65{.}454 & 719{.}994 & 3{.}185 & 7{.}753 & 85{.}283 & 1{.}359 \\
108 (2003--05) & 63{.}640 & 700{.}040 & 3{.}115 & 8{.}245 & 90{.}695 & 1{.}405 \\
109 (2005--07) & 67{.}244 & 739{.}684 & 3{.}266 & 8{.}908 & 97{.}988 & 1{.}522 \\
110 (2007--09) & 84{.}141 & 925{.}551 & 3{.}913 & 11{.}723 & 128{.}913 & 1{.}795 \\
111 (2009--11) & 72{.}345 & 795{.}795 & 3{.}824 & 9{.}036 & 99{.}396 & 1{.}556 \\
112 (2011--13) & 64{.}443 & 708{.}873 & 3{.}387 & 8{.}554 & 94{.}094 & 1{.}382 \\
113 (2013--15) & 71{.}189 & 783{.}079 & 3{.}637 & 9{.}084 & 99{.}924 & 1{.}442 \\
114 (2015--17) & 72{.}614 & 798{.}754 & 3{.}992 & 9{.}507 & 104{.}577 & 1{.}605 \\
115 (2017--19) & 73{.}339 & 806{.}729 & 4{.}448 & 9{.}376 & 102{.}706 & 1{.}726 \\
116 (2019--21) & 97{.}972 & 1{.}077{.}692 & 5{.}635 & 12{.}478 & 137{.}258 & 2{.}206 \\
117 (2021--23) & 83{.}259 & 915{.}849 & 5{.}764 & 9{.}949 & 109{.}439 & 2{.}116 \\
118 (2023--25) & 77{.}885 & 856{.}735 & 6{.}099 & 9{.}130 & 100{.}430 & 2{.}163 \\
\midrule
\textbf{Total} & 1{.}459{.}332 & 16{.}052{.}652 & 76{.}159 & 182{.}700 & 2{.}009{.}212 & 31{.}378 \\
\bottomrule
\end{tabular}
\end{table*}

\subsection{Nested specifications}\label{si:recruit_ladder}

This section reports two comparisons of the history blocks, both estimated on the same choice events. The first adds each block on its own to the baseline model, as plotted in main-text Fig.~5g. The second adds the blocks in sequence, so that each is evaluated given the ones before it; this measures what triadic history adds beyond pairwise ties. Because the blocks are correlated, the single-block gains overlap and should not be summed; the sequential gains add up to the gain of the full model over the baseline, but depend on the order of the blocks. $\Delta$AIC $= 2\,\Delta k - 2\Delta\ell$, with $\Delta k$ the number of added coefficients; negative values favour the larger model.

\begin{figure}[!ht]
\centering
\includegraphics[width=0.58\linewidth]{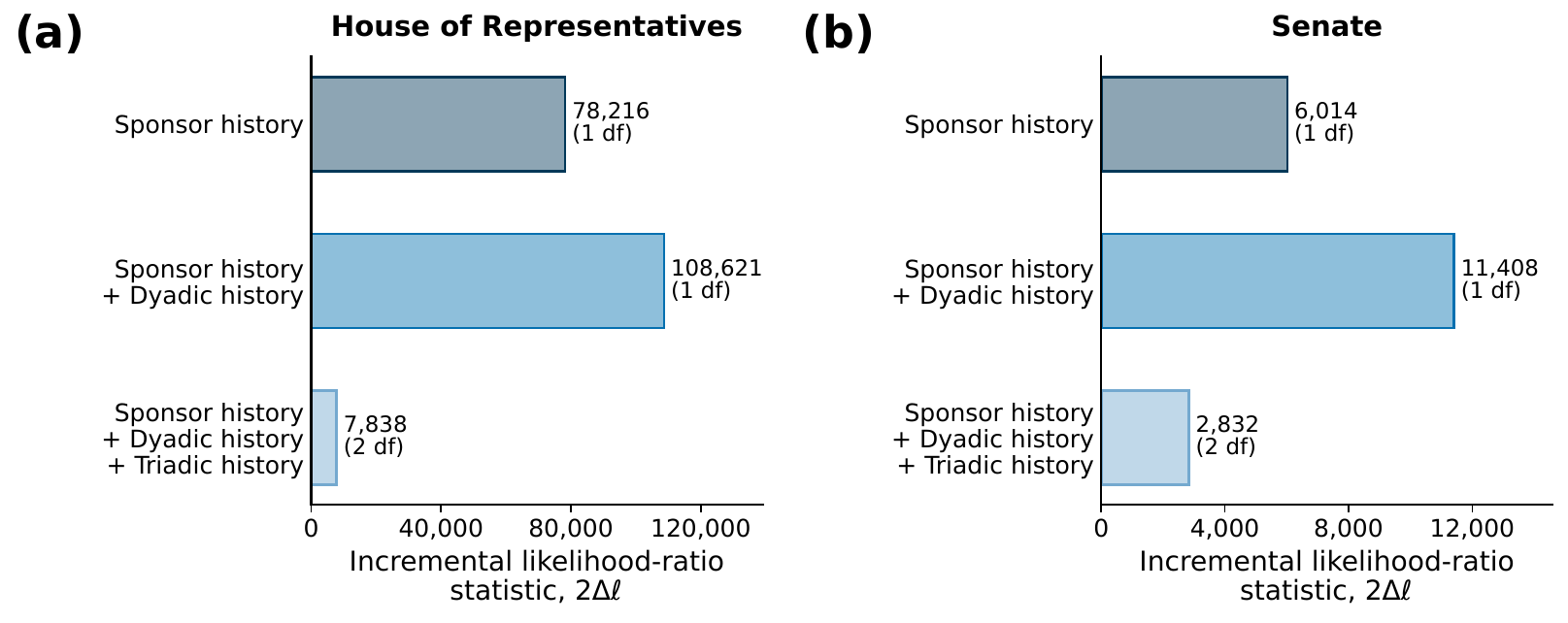}
\caption{\textbf{Likelihood-ratio statistics of the sequential models.} (a)~House, (b)~Senate. Each bar is the statistic $2\Delta\ell$ of a model against the one above it: the baseline plus sponsor history, then plus dyadic history, then plus triadic history. In parentheses, the number of coefficients the last block adds. Each panel has its own scale.}
\label{fig:suppmat_lr_sequential}
\end{figure}

\clearpage

\section{Coalition size across bill types and policy areas}\label{si:analysis}

\begin{figure}[!ht]
\centering
\includegraphics[width=\linewidth]{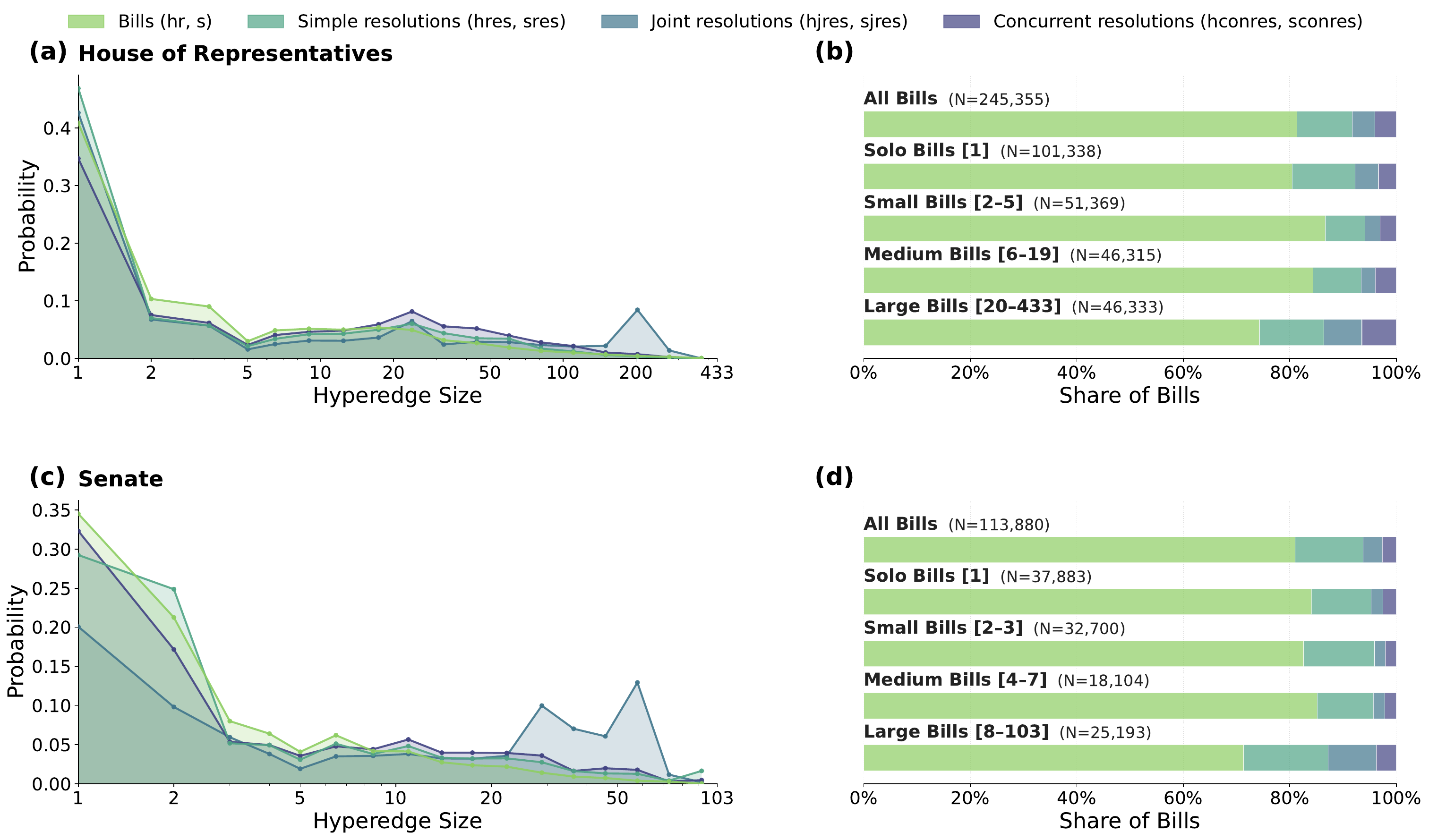}
\caption{\textbf{House \& Senate: coalition size and bill type.} Top row, House; bottom row, Senate; Congresses 93--118, solo bills included. Analogous House and Senate bill types share a colour (e.g.\ bills: H.R., S.). \textbf{(a,c)} Distribution of hyperedge size for each bill type with more than 50 bills: share of the type's bills in each of 20 log-spaced size bins, whose edges lie halfway between whole sizes, plotted at the bin centres; each type is normalised separately. \textbf{(b,d)} Share of each bill type among all bills (\textit{top}) and within each size class: solo bills (size 1), and small, medium and large coalitions. $N$, number of bills in each class.}
\label{fig:suppmat_bill_type}
\end{figure}

\begin{figure}[!ht]
\centering
\includegraphics[width=\linewidth]{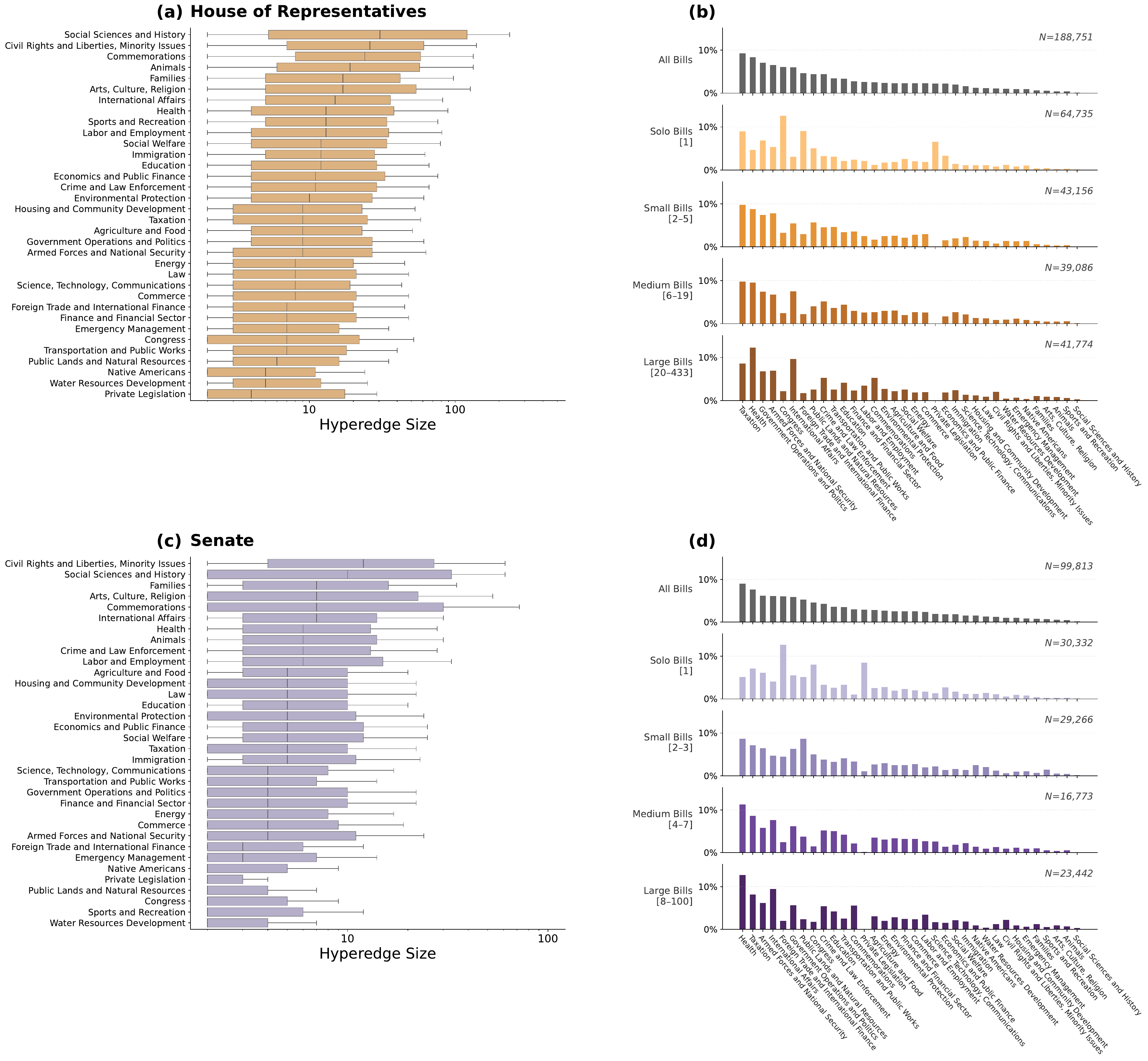}
\caption{\textbf{House \& Senate: coalition size and policy area.} Top row, House; bottom row, Senate; Congresses 96--118 (policy areas are recorded from the 96th Congress), solo bills included, bills without a policy area left out. \textbf{(a,c)} Hyperedge size of the bills with at least one cosponsor in each policy area (box plots; logarithmic size axis), areas ordered by median size. \textbf{(b,d)} Share of each policy area among all bills (\textit{top}) and within each size class: solo bills (size 1), and small, medium and large coalitions, with the same vertical scale in every row; areas ordered from the most to the least frequent. $N$, number of bills in each class.}
\label{fig:suppmat_policy_area}
\end{figure}

Supplementary Figs.~\ref{fig:suppmat_bill_type} and~\ref{fig:suppmat_policy_area} relate coalition size to bill type and policy area, for the House (top rows) and the Senate (bottom rows). Bill sizes count the sponsor, and the size classes are the ones of main-text Fig.4 solo bills (size 1), and small, medium and large coalitions (2--5, 6--19 and 20 or more members in the House; 2--3, 4--7 and 8 or more in the Senate).
In the House, and similarly in the Senate, bills (H.R. and, respectively, S.) make up about four in five of all bills and of small and medium coalitions, but about three in four of the large coalitions, where resolutions, particularly joint and concurrent resolutions, are more common (Supplementary Fig.~\ref{fig:suppmat_bill_type}\textbf{b,d}).

The policy areas whose bills have the largest typical coalitions are Social Sciences and History, Civil Rights and Liberties, Commemorations, Animals, Families, and Arts, Culture and Religion, largely commemorative and symbolic subjects, whereas Private Legislation, Water Resources Development and Native Americans have the smallest (Supplementary Fig.~\ref{fig:suppmat_policy_area}\textbf{a,c}). Because these symbolic areas account for few bills, the most common areas among House coalitions of 20 or more members are nonetheless Health, International Affairs and Taxation (Supplementary Fig.~\ref{fig:suppmat_policy_area}\textbf{b}). The main text uses the first observation as a possible reading of the renewed party mixing among the largest coalitions; testing it requires the policy mix of the largest size bin of main-text Fig.2 (239--433 members).

\section{Legislative success across coalition size}

\begin{figure}[!ht]
\centering
\includegraphics[width=0.8\linewidth]{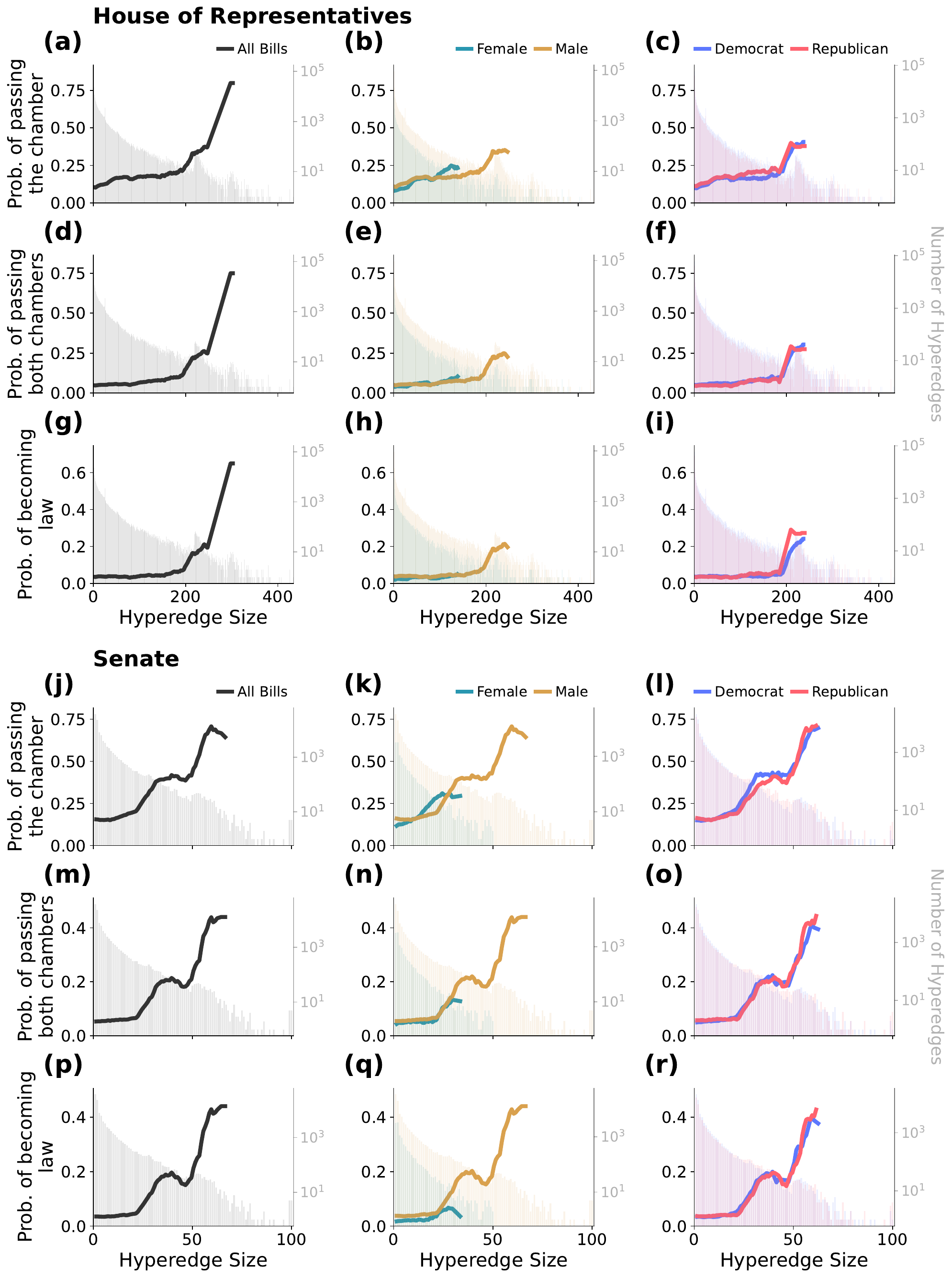}
\caption{\textbf{House \& Senate: legislative success by coalition size, sponsor gender and sponsor party.} House \textbf{(a)--(i)} and Senate \textbf{(j)--(r)}; Congresses 93--118. Rows: probability that a bill passes its chamber, passes both chambers, or becomes law. Columns: all bills, bills by the sponsor's gender, and bills by the sponsor's party. Curves: share of the bills of each size (sponsor included) with that outcome, among sizes with at least 10 bills, smoothed by a centred rolling mean over 44 (House) or 10 (Senate) consecutive sizes. Light bars: number of bills of each size (right axis, logarithmic scale).}
\label{fig:suppmat_legislative_success}
\end{figure}

In both chambers, the probability that a bill passes its chamber increases with the size of its coalition (Supplementary Fig.~\ref{fig:suppmat_legislative_success}\textbf{a,j}). 
Moreover, although success rates differ substantially between sponsors of different gender and party for certain coalition sizes, we report these differences descriptively and do not establish an effect of coalition structure on passage, which we leave for future work.

\subsection{Method}
\label{si:suppmat_legislative_success}

To relate legislative success to coalition size, we pooled the bills of Congresses 93--118 (solo bills included; legislators of other parties than Democrats and Republicans, and their bills, left out) and recorded three outcomes, $L \in \{0, 1\}$: passing the chamber of origin, passing both chambers, and enactment into public law. A bill that never reached a vote counts as not passed.
For each bill size $s$, sponsor included, the empirical success rate is
\begin{center}
$
\bar{y}_{s} = \frac{1}{N_{s}} \underset{i : |i|=s}{\sum}L_i ,
$
\end{center}
where $N_s$ is the number of bills of size $s$ and $L_i$ the outcome of bill $i$; rates are computed for all bills and separately by the gender and the party of the sponsor.
Only sizes with at least 10 bills are used ($N_{s} \ge 10$).
To smooth the curves, each rate is replaced by the mean of the rates of the sizes within a centred window of 10\% of the largest size in the chamber (44 sizes in the House, 10 in the Senate) that meet this threshold.

\end{document}